\documentclass[letterpaper,titlepage,11pt]{article}
\pdfoutput=1

\usepackage{cite}
\usepackage{amsmath,amssymb,amsthm,mathrsfs,bbm}
\usepackage{latexsym,amscd,amsbsy,amsfonts,dsfont}
\usepackage{graphicx}
\usepackage{float}
\usepackage[
      colorlinks=true,
      linkcolor=blue,
      urlcolor=blue,
      filecolor=black,
      citecolor=red,
      pdfstartview=FitV,
      pdftitle={},
        pdfauthor={David Licht, Raimon Luna, Ryotaku Suzuki},
        pdfsubject={},
        pdfkeywords={},
        pdfpagemode=None,
        bookmarksopen=true
      ]{hyperref}\usepackage{caption}

\def\clock{{\count0=\time
           \divide\count0 60
           \ifnum\count0<10 0\fi\the\count0
           \multiply\count0 -60 \advance\count0 \time
           :\ifnum\count0<10 0\fi \the\count0
         }}
\newcommand{\timestamp}{{\small\vbox{\hbox{\tt\jobname.tex}
\hbox{\the\day/\the\month/\the\year, \clock}}}}
\newcommand{\RR}{\mathcal{R}}
\newcommand{\cR}{\mathcal{R}} 

\newcommand{\cL}{{\mathcal L}}
\newcommand{\cS}{{\mathcal S}}
\newcommand{\cX}{{\mathcal X}}
\newcommand{\cI}{{\mathcal I}}
\newcommand{\cJ}{{\mathcal J}}
\newcommand{\cM}{{\mathcal M}}
\newcommand{\cC}{{\mathcal C}}
\newcommand{\cH}{{\mathcal H}}
\newcommand{\veps}{\varepsilon}
\newcommand{\ord}[1]{{\mathcal O}\left(#1\right)}
\newcommand{\fr}[1]{\frac{1}{#1}}
\newcommand{\nonum}{\nonumber\\ }
\newcommand{\hgfunc}[2]{ { \, {}_{#1}  F  {}_{#2} } }
\newcommand{\cout}[1]{}

\newcommand{\pd}{\partial}

\newcommand{\ie}{{\it i.e.,\,}}

\newcommand{\lp}{\left(}
\newcommand{\rp}{\right)}

\newcommand{\sR}{\mathsf{R}}
\newcommand{\fq}{\mathfrak{q}}

\usepackage{bm}
\usepackage{latexsym}
\usepackage[latin1]{inputenc}
\usepackage{amsfonts}
\usepackage{amssymb}
\usepackage{amsmath}
\usepackage{subfigure}

\numberwithin{equation}{section}

\begin{document}

\begin{titlepage}

\vglue 2cm %Formatting
%\vskip 2cm
%
\centerline{\LARGE \bf Black Ripples, Flowers and Dumbbells}
\bigskip
\centerline{\LARGE \bf at large $D$}

\vskip 1.6cm
\centerline{\bf David Licht$^{a}$, Raimon Luna$^{a}$ and Ryotaku Suzuki$^{a,b}$}
\vskip 0.5cm

\centerline{\sl $^{a}$Departament de F{\'\i}sica Qu\`antica i Astrof\'{\i}sica, Institut de
Ci\`encies del Cosmos,}
\centerline{\sl  Universitat de
Barcelona, Mart\'{\i} i Franqu\`es 1, E-08028 Barcelona, Spain}
\centerline{\sl $^{b}$Department of Physics, Osaka City University,}
\centerline{\sl Sugimoto 3-3-138, Osaka 558-8585, Japan}

\smallskip
\vskip 0.5cm
\centerline{\small\tt david.licht@icc.ub.edu,\, raimonluna@icc.ub.edu,} \centerline{\small\tt s.ryotaku@icc.ub.edu}

\vskip 1.cm
\centerline{\bf Abstract} \vskip 0.2cm \noindent

\noindent 
We explore the rich phase space of singly spinning
(both neutral and charged) black hole solutions in the large $D$ limit.
We find several 'bumpy' branches which are connected to multiple (concentric) black rings, and black Saturns. Additionally, we obtain stationary solutions without axisymmetry that are only stationary at $D\rightarrow \infty$, but correspond to long-lived black hole solutions at finite $D$. These multipolar solutions can appear as intermediate configurations in the decay of ultra-spinning  Myers-Perry black holes into stable black holes. Finally, we also construct stationary solutions corresponding to the instability of such a multipolar solution.

\end{titlepage}
\pagestyle{empty}
\small

\addtocontents{toc}{\protect\setcounter{tocdepth}{2}}
{
	\hypersetup{linkcolor=black,linktoc=all}
	\tableofcontents
}
\normalsize
\newpage
\pagestyle{plain}
\setcounter{page}{1}

%%%%%%%%%%%%%%%%%%%%%%%%%%%%%%%%%%%%%%%%%%%%%%%%
\section{Introduction}
\label{sec:introduction}
%%%%%%%%%%%%%%%%%%%%%%%%%%%%%%%%%%%%%%%%%%%%%%%%
Black hole solutions in higher dimensional gravity show a far richer behavior than their counterparts in four spacetime dimensions. In higher dimensions, the rotation plays a significant role to fertilize a variety of new solutions. Since in $D>5$, the (Newtonian) gravitational potential $\sim\frac{G M}{r^{D-3}}$ falls off more rapidly than the centrifugal barrier $\sim\frac{J^2}{M^2r^2}$, the horizon can be deformed to an extended shape at large angular momentum, and hence becomes vulnerable to a Gregory-Laflamme type instability~\cite{Gregory:1993vy,Gregory:1994bj}. This allows a family of non-uniform stationary solutions to branch off from the zero modes of the instabilities~\cite{Emparan:2008eg}.

The increased number of degrees of freedom in a higher dimensional theory, however, complicate the construction of black hole solutions and analysis of their dynamics. To tackle this problem, several approximation techniques have been developed. One such approximation is 
 the {\it blackfold} approach~\cite{Emparan:2007wm}, which has been successful in elucidating the black hole phases in the ultra-spinning regime: for example for black (multi-)rings/Saturns in which the horizon has highly elongated shape
, that allows to locally approximate them as loosely bent black strings/branes.

Another successful effective approach is the large spacetime dimension limit, or the {\it large $D$ limit}~\cite{Asnin:2007rw, Emparan:2013moa}, which has been proven to be  useful in various problems involving higher dimensional black holes~\cite{Asnin:2007rw,Emparan:2013moa,Emparan:2014cia, Emparan:2014aba,Emparan:2015rva,Emparan:2014jca,Suzuki:2015iha,Suzuki:2015axa,Emparan:2015hwa,
	Emparan:2015gva,Tanabe:2015hda,Tanabe:2015isb,Tanabe:2016opw,Emparan:2016sjk,Chen:2017wpf,
	Mandlik:2018wnw,Emparan:2018bmi,Iizuka:2018zgt,Andrade:2018zeb,Li:2019bqc,Casalderrey-Solana:2018uag,Guo:2019pte,Andrade:2018nsz,Andrade:2019edf,Andrade:2018rcx, Emparan:2019obu}.
%This limit allows black holes to have a common simple near horizon structure only depending on their charges, but independent of the presence of rotation and a cosmological constant~\cite{Emparan:2013xia}.
This limit allows black holes to have a simple near horizon structure decoupled from the asymptotic region%, with/without the rotation, cosmological constant or charges
~\cite{Emparan:2013xia}.
As a result, the Einstein's
equation reduces to an effective theory on the horizon surface expanded in $1/D$, namely the {\it large $D$ effective theory}~\cite{Emparan:2015hwa,Emparan:2015gva,Bhattacharyya:2015dva,Bhattacharyya:2015fdk}.
Different to the blackfold approach, the large $D$ limit is naturally endowed with the separation of scales between gradients along and orthogonal to the horizon: the gradient orthogonal to the horizon becomes large compared to gradients along the horizon in the limit of large $D$ as a result of the steepening of the gravitational potential. This enables us to formulate an effective theory without the requirement that the gradients along the horizon have to be infinitesimal, which makes the large $D$ expansion a powerful tool to study the non-uniform 'bumpy' phases of black holes.

In this paper, we explore the phase space of compact stationary solutions with a single spin in the large $D$ limit, specifically, we focus on the (non-)axisymmetric deformed families branching off from the Myers-Perry family. The instability of ultra-spinning MP black holes and the existence of nearby `rippled' solution was first conjectured in \cite{Emparan:2003sy} and later, after the proof of existence of the zero modes and the instability \cite{Dias:2009iu,Dias:2010eu,Dias:2010maa,Dias:2011jg,Hartnett:2013fba}, the rippled solutions were constructed numerically and identified as solutions that connect to black rings and black Saturns \cite{Emparan:2007wm,Emparan:2014pra,Dias:2014cia,Figueras:2017zwa}.

Because of the strong suppression of gravitational radiation at large $D$~\cite{Andrade:2019edf}, the effective large $D$ description also admits  stationary non-axisymmetric branches such as {\it black bars}~\cite{Andrade:2018nsz} and other multipolar solutions. 
Here we apply the {\it blob} approximation developed in~\cite{Andrade:2018nsz,Andrade:2018rcx},
where localized black hole solutions such as the Myers-Perry black hole are identified as stationary lumps (``blobs") on a membrane which share the same horizon topology as the black brane solution but nevertheless encode most of the physics pertaining to the localized solution.
 
\begin{figure}[h]
	\begin{center}
		\includegraphics[width=0.75\textwidth]{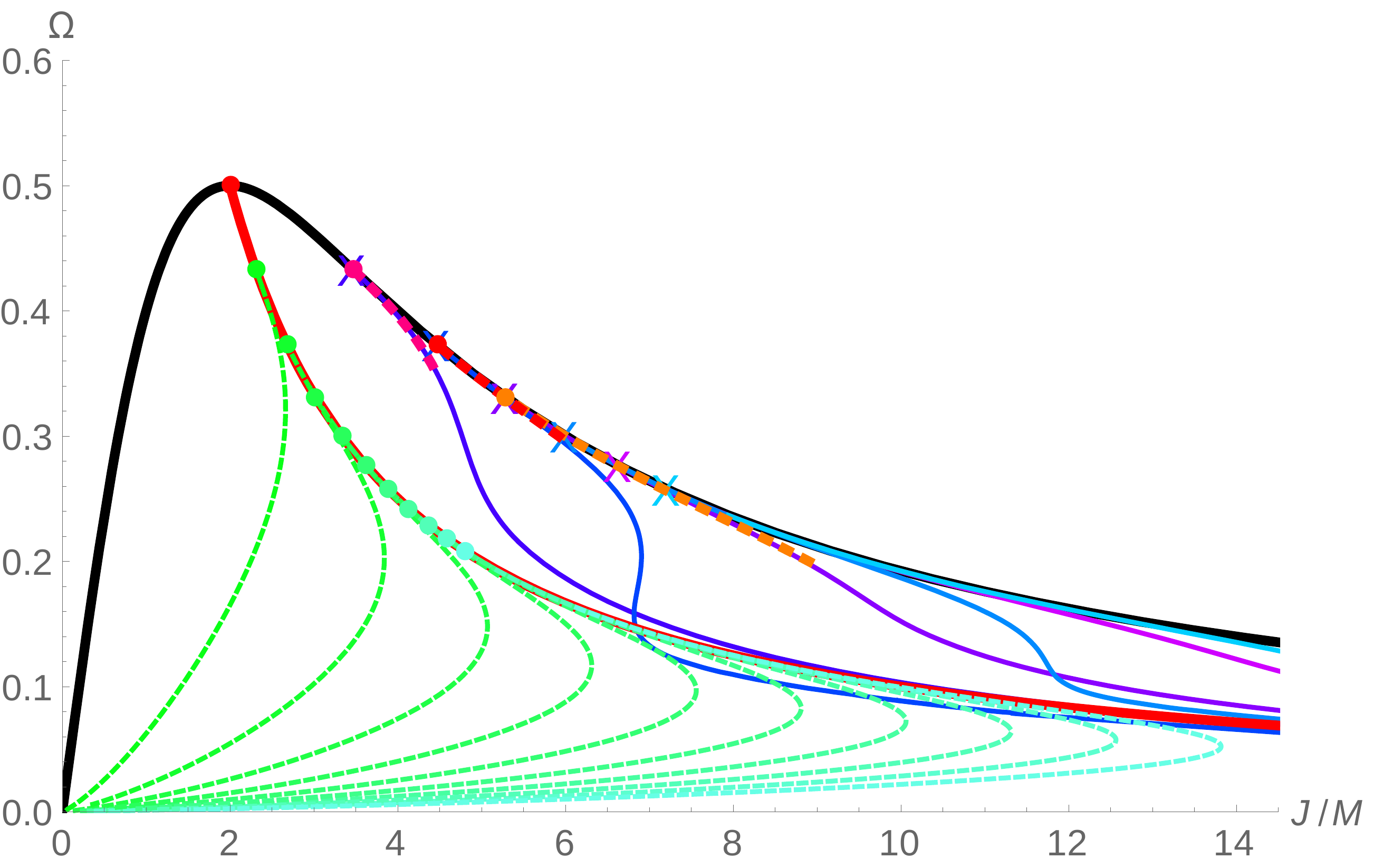}
		\includegraphics[width=0.23\textwidth]{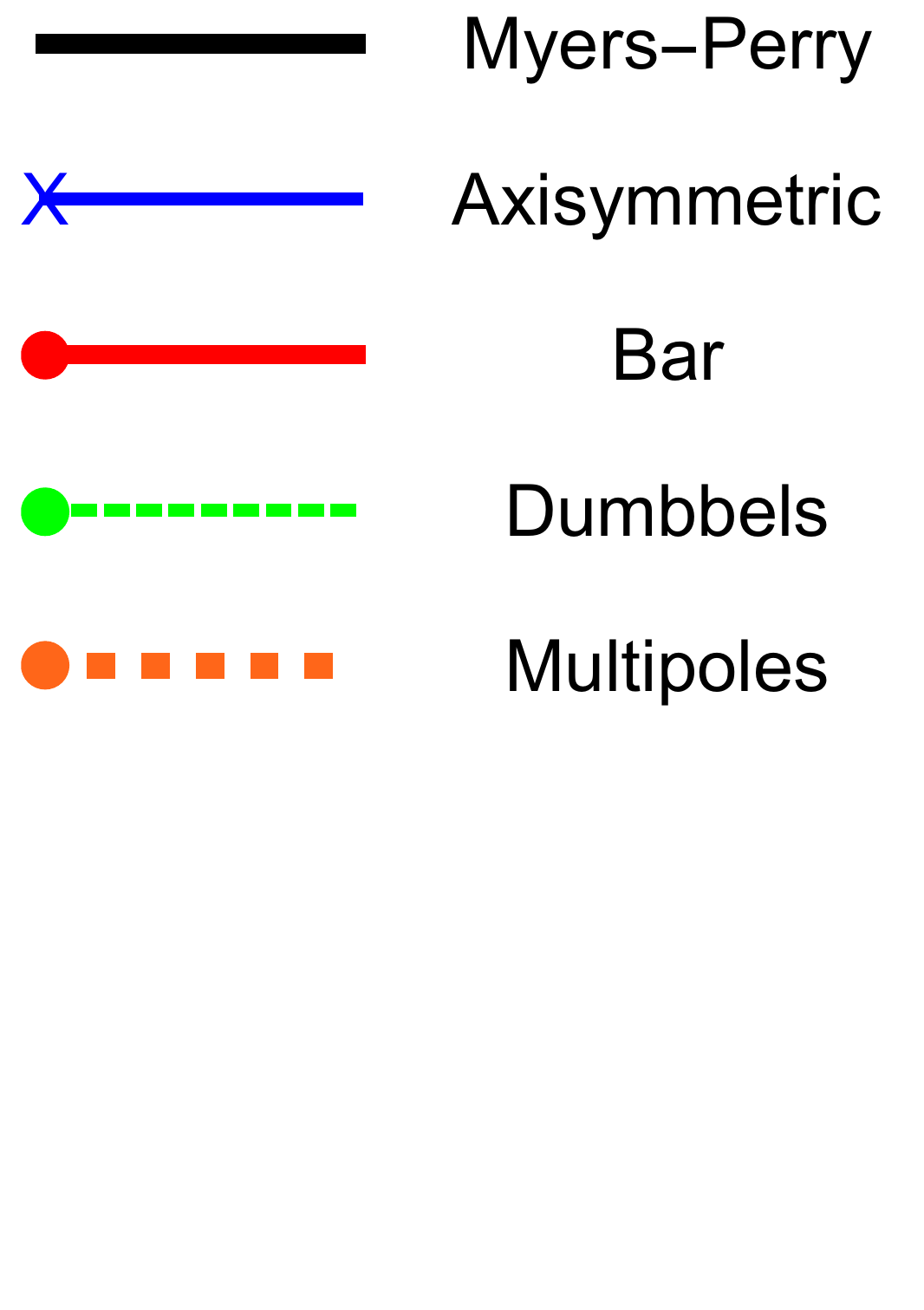}
		\caption{\small Phase space plot of the first appearing branches of solutions with a single angular momentum (per unit mass) $\cJ/\cM$ and angular velocity $\Omega$. In the ultra-spinning regime $\cJ/\cM>2$ the MP-BH develops instabilities and the corresponding zero modes appear at places marked with dots or crosses.
%				For the analytically known black bar we also study its non-uniform deformations. Branches are shown in different shadings of a color to make them more distinguishable. 
				For the analytically known black bar, we also study its non-uniform deformations ('dumbbells'), whose branches are shown in different shadings of a color to make them more distinguishable. 
			\label{fig:PhaseDiagram} }
	\end{center}
\end{figure}

\begin{figure}[H]
	\begin{center}
		\includegraphics[width=0.49\textwidth]{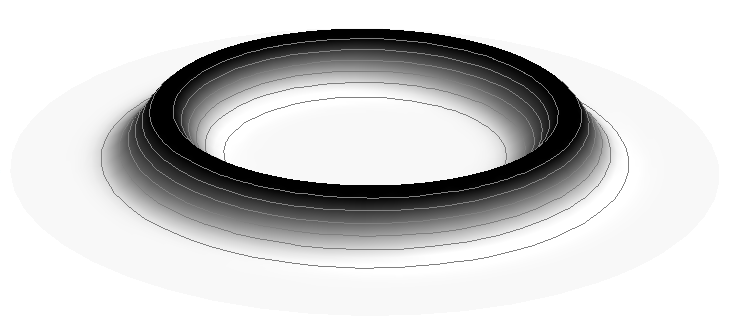}
		\includegraphics[width=0.49\textwidth]{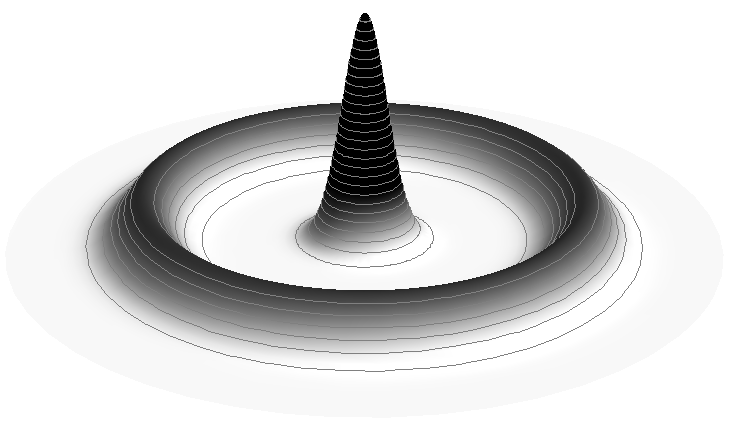}
	    \includegraphics[width=0.54\textwidth]{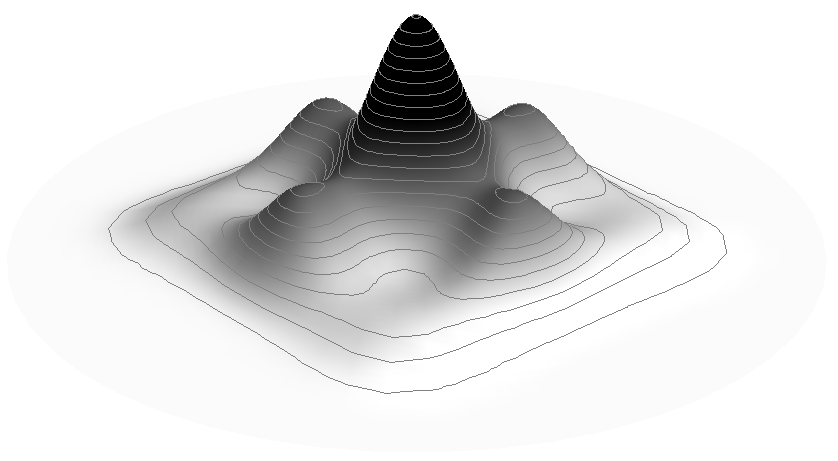}	
	    \includegraphics[width=0.45\textwidth]{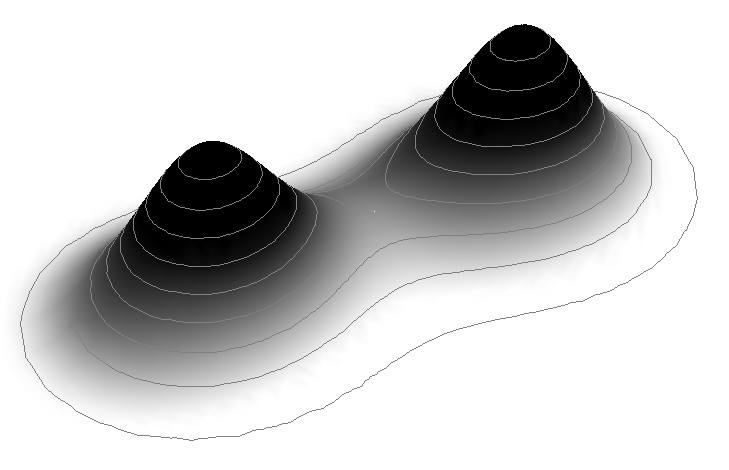}
		\caption{\small Four examples of bumpy solutions: Upper Left: Ring-like ripple. Upper Right: Saturn-like ripple Lower Left: Black flower with a quadrupolar deformation. Lower Right: Dumbbell. Plots show the mass density $m$. Coloring was chosen to highlight the important details of the solution, strictly speaking all solutions share the same horizon topology. \label{fig:ExampleProfiles}}
	\end{center}
\end{figure}

Figure \ref{fig:PhaseDiagram} shows the full phase space plot of solutions we obtain. The solutions correspond to Myers-Perry solutions and their axissymmetric `bumpy' deformations leading to black rings and black Saturns. We are also including stationary solutions without axisymmetry, which only can remain stationary at large $D$ since gravitational radiation decouples. These solutions have been shown to play an important role in dynamical evolutions of the ultra-spinning instability \cite{Andrade:2018yqu,Andrade:2019edf,Shibata:2009ad,Shibata:2010wz,Bantilan:2019bvf}. The first solution of this kind, a dipolar solution ``black bar" was found analytically in \cite{Andrade:2018nsz}. Here we study its stationary deformations and also find its multipolar generalizations ``black flowers". To illustrate features of the found solutions, we show plots of the mass density of four examples in figure \ref{fig:ExampleProfiles}.\footnote{The flower branches are hard to to construct far away from their branching points, so figure \ref{fig:PhaseDiagram} shows them only partially.} 

We observe that most of bumpy deformations remain tangential to their 'parent'-branch until the deformation becomes comparable to the original solution and new blobs start to form. At some point, these blobs barely have any overlap and the branches enter a new asymptotic behavior for small $\Omega$ becoming completely separated.
Some very short branches stick out non-tangentially above the parent-branch.

The paper is structured as follows: in section \ref{sec:SetupSection}, we outline the derivation of our large $D$ effective equations for black branes and describe how they also contain localized black hole solutions. In section \ref{sec:axisymmSol}, we construct perturbatively and numerically stationary `bumpy' deformations of the MP black hole that lead to (multiple) black rings and Saturns. 
In section \ref{sec:multipoleDef} and \ref{sec:blackBars}, we construct stationary non-axisymmetric solutions from multipolar deformations of MP black holes and deformations from black bars.
Section \ref{sec:effCharge} discusses effects of adding charge to obtain charged (but non-extremal) solutions. In the appendix we collect details of the perturbative calculation and describe our numerical procedure in greater detail.
%
%
%%%%%%%%%%%%%%%%%%%%%%%%%%%%%%%%%%%%%%%%%%%%%%%%
\section{Branes and localized black holes at large $D$}
\label{sec:SetupSection}
%%%%%%%%%%%%%%%%%%%%%%%%%%%%%%%%%%%%%%%%%%%%%%%%
%
%
%
%%%%%%%%%%%%%%%%%%%%%%%%%%%%%%%%%%%%%%%%%%%%%%%%
\subsection{Large $D$ effective equations}
\label{sec:largeD}
%%%%%%%%%%%%%%%%%%%%%%%%%%%%%%%%%%%%%%%%%%%%%%%%
%
We study possibly charged black holes in Einstein-Maxwell theory in higher dimensions
\begin{equation}
I=\int d^D x\sqrt{-g}\lp R-\frac14 F^2\rp \,,
\end{equation}
where
\begin{equation}
D=n+p+3\,,
\end{equation}
with $n$ large and $p$ a finite number. Ref. \cite{Emparan:2016sjk} developed an effective theory for fluctuations of $p$-branes along their extended directions $\sigma^i$ ($i=1,\dots,p$) ,
\begin{align}
\label{AFch}
ds^2=2dtdr-Adt^2-\frac{2}{n} C_i d\sigma^idt+\frac1{n}G_{ij}d\sigma^id\sigma^j +r^2d\Omega_{n+1}\,,
\end{align}
where $\sR=r^n$ and
\begin{align}
A=1-\frac{m(t,\sigma)}{\sR}+\frac{q(t,\sigma)^2}{2\sR^2}\,,\qquad 
C_i=\left(1-\frac{q(t,\sigma)^2}{2m(t,\sigma)\sR}\right) \frac{p_i(t,\sigma)}{\sR}\,,
\end{align}
\begin{align}
G_{ij}&=\delta_{ij}+\frac1{n}\left\{\left( 1-\frac{q(t,\sigma)^2 }{2m(t,\sigma)\sR}\right)\frac{p_i(t,\sigma) p_j(t,\sigma)}{m(t,\sigma)\sR} 
\right. \nonumber \\
& \hphantom{{} = \delta_{ij}+\frac1{n}\,}
\left.
-\ln\lp 1-\frac{m_-(t,\sigma)}{\sR}\rp
\left[ 2\delta_{ij}+  \nabla_i \frac{p_j(t,\sigma)}{m(t,\sigma)} +\nabla_j \frac{p_i(t,\sigma)}{m(t,\sigma)}\right] 
\right\} \,.
\end{align}
The electric potential is
\begin{equation}
A_t = -\frac{q(t,\sigma)}{\sR}\,.
\end{equation}

The degrees of freedom of the effective theory are the mass density $m(t,\sigma)$,  the charge density $q(t,\sigma)$ and the fields $p_i(t,\sigma)$. In the presence of charge it is convenient to introduce a new field $v_i(t,\sigma)$ defined by
\begin{equation}
p_i = m v_i+\nabla_i m\,,
\end{equation}
and the abbreviation
\begin{equation}
m_\pm =\frac{1}{2}\lp m\pm \sqrt{m^2-2q^2}\rp\,.
\end{equation}

The equations of motion of the effective theory are obtained by requiring that the Einstein-Maxwell equations are solve to leading order in a $1/D$-expansion and take the form of conservation equations
\begin{align}
&\partial_t m+\nabla_i (m v^i)=0\,,\\
&\partial_t (m v^i)+\nabla_j (m v^i v^j +\tau^{ij})=0\\
&\partial_t q+\nabla_i j^i=0
\end{align}
where
\begin{align}
\tau_{ij}&= -\lp m_+ -m_- \rp\delta_{ij}  -2m_+\nabla_{(i}v_{j)}-(m_+-m_-)\,\nabla_i\nabla_j \ln m \,,\\
j_i&=q v_i -m\nabla_i\lp\frac{q}{m}\rp\,.
\end{align}
These equations simplify further if we consider only stationary configurations, that satisfy
\begin{equation}
(\partial_t+v^i\partial_i) m=0\,,\qquad (\partial_t+v^i\partial_i) q=0\,,
\end{equation}
and $v^i$ is a time-independent killing vector \ie
\begin{equation}
\partial_t v^i=0, \qquad \nabla_{(i}v_{j)}=0\,.
\end{equation}
which implies the absence of dissipative effects. Absence of charge diffusion requires 
\begin{equation}
\nabla_i\lp\frac{q}{m}\rp=0\,,
\end{equation}
which states that the charge density is everywhere proportional to the mass density  via the proportionality constant
\begin{equation}
\label{gothq}
\fq \equiv \frac{q}{m}\,.
\end{equation} 
Under these assumptions the equations of motion are reduced to a single master equation that is most elegantly formulated in terms of the area-radius 
\begin{equation}
\cR=\ln m\,,
\end{equation}
and is given by 
\begin{align}
\label{eq:chargedMasterEq}
\nabla_i\lp \frac{v^2}{2}+\frac{m_+-m_-}{m}\lp\mathcal{R}+\nabla_j\nabla^j\mathcal{R}+\frac{1}{2}\nabla^j\mathcal{R}\nabla_j\mathcal{R}\rp\rp =0\,.
\end{align}
Using the scale invariance of the effective equations, which manifests itself in a shift symmetry of $\cR$,  the above equation can be formally mapped to the uncharged equation by defining the charge rescaled velocity field
\begin{align}
\label{eq:chargeRescalVel}
v^i_q=\sqrt{\frac{m}{m_+-m_-}}\,v^i=\frac{v^i}{\lp 1-2\fq^2\rp^{1/4}}\,,
\end{align}
and shifting $\cR$ to obtain the \emph{soap bubble equation}~\cite{Emparan:2016sjk}
\begin{equation}\label{mastercharge}
\frac{v_q^2}{2}+\mathcal{R}+\nabla_j\nabla^j\mathcal{R}+\frac{1}{2}\nabla^j\mathcal{R}\nabla_j\mathcal{R}=0\,.
\end{equation}
Which has the same form as the uncharged equation (\ie eq.~(\ref{eq:chargedMasterEq}) with $\fq=0$) but with the difference that the role of $v^2$ is now taken by the norm of the charged rescaled velocity field. Since the charged equation can be mapped to the uncharged one, solving eq.~(\ref{mastercharge}) for a given value of $v_q$ always gives a one parameter family of solutions, parameterized by the charge parameter $\fq$. In the case of non-vanishing charge, $v_q$ is not directly the physical velocity field and allows to study the effect of charging up the solution.

%
%%%%%%%%%%%%%%%%%%%%%%%%%%%%%%%%%%%%%%%%%%%%%%%%
\subsection{Black holes as Gaussian blobs on a membrane}
\label{sec:MP}
%%%%%%%%%%%%%%%%%%%%%%%%%%%%%%%%%%%%%%%%%%%%%%%%
%
Even though these equations were initially formulated to capture the dynamics of black branes. Ref. \cite{Andrade:2018nsz} found that this large $D$ effective theory also contains localized black hole solutions when solved with different boundary conditions. We recapitulate here the findings of \cite{Andrade:2018nsz,Andrade:2018rcx}.

To capture effects of a single spin we consider the case of $p=2$ and require the stationary solutions to have a purely rotational velocity field. Choosing angular coordinates for the spatial brane directions $\sigma^i=(r,\phi)$, the only non-vanishing component of the (charge rescaled) velocity field can be set to $v^\phi =\Omega_q$ and equation (\ref{mastercharge}) becomes
\begin{align}
\pd^2_r \cR+\frac{\pd_r \cR}{r}+\frac{\pd_\phi^2 \cR}{r^2}+\frac12 \lp \lp\pd_r \cR\rp^2+\frac{\lp \pd_\phi \cR\rp^2}{r^2}\rp +\cR +\frac{\Omega_q^2 r^2}{2}=0\,,
\label{eq:MasterEquationRotation}
\end{align}
where $\Omega_q$ is the charge rescaled angular velocity, according to eq.~(\ref{eq:chargeRescalVel}).

The Myers-Perry (MP) black hole solution (and its charged Kerr-Newman counterpart described in \cite{Andrade:2018rcx}) corresponds to the axisymmetric solution
\begin{equation}
\cR_{\text {KN}}(r)=\frac{2}{1+a_q^2}\lp 1-\frac{r^2}{4}\rp\,,
\label{eq:MPsol}
\end{equation}
with $a_q$ defined via
\begin{equation}
\Omega_q=\frac{a_q}{1+a_q^2}\,.
\end{equation}

Since this  corresponds to a Gaussian in the mass variable $m=\exp{\cR}$, this solution is strongly localized in the directions $\sigma^i$, but still shares the same horizon topology as the black brane (\ref{AFch}). This feature of the solution is due to the fact that the rescaling of the spatial directions $\sigma^i \rightarrow \sigma^i/\sqrt{n}$ assumed in eq.~(\ref{AFch}) leads for localized solutions to a magnification of the region around the center of one of its hemispheres. Since at large $D$ most of the surface of the black hole is concentrated in this region, a description of it can capture most of the physics connected to the localized black hole.

The aforementioned localization of the mass density motivates the following definition of a localized black hole: We call a solution of eq.~(\ref{eq:MasterEquationRotation}) a (stationary) localized black hole, if it has a finite mass $\cM$ according to  
\begin{equation}
\cM=\int_0^{2\pi}d\phi\int_0^\infty dr\, r \, m(r,\phi)\,.
\label{eq:normCond}
\end{equation}
And it has an angular momentum given by
\begin{eqnarray}
\cJ=\int_0^{2\pi}d\phi\int_0^\infty dr\, r \, p_\phi(r,\phi)=\int_0^{2\pi}d\phi\int_0^\infty dr\, \Omega\, r^3 \, m(r,\phi)\,. \label{eq:angularMomentum}
\end{eqnarray}
where we used $p_\phi = \partial_\phi m+ \Omega \, r^2 m$.
%%%%%%%%%%%%%%%%%%%%%%%%%%%%%%%%%%%%%%%%%%%%%%%%
\section{Axisymmetric sector: Black Ripples}
\label{sec:axisymmSol}
First, we consider the axisymmetric deformation of the Myers-Perry, which leads to an infinite number of 'bumpy'  solutions, or {\it black ripples}.
%%%%%%%%%%%%%%%%%%%%%%%%%%%%%%%%%%%%%%%%%%%%%%%%

%%%%%%%%%%%%%%%%%%%%%%%%%%%%%%%%%%%%%%%%%%%%%%%%
\subsection{Zero mode deformations}
\label{sec:axisymmSolPert}
%%%%%%%%%%%%%%%%%%%%%%%%%%%%%%%%%%%%%%%%%%%%%%%%
The MP-solution (\ref{eq:MPsol}) allows axisymmetric co-rotating zero mode deformations according to\footnote{For brevity of presentation we restrict to the case of zero charge for now and drop the subscript $q$. We will discuss the effects of non-zero charge in section \ref{sec:effCharge}.}
\begin{equation}
\cR(r) =  \cR_{\rm MP}(r) + \delta \cR(r).
\end{equation}
Plugging this into eq.~(\ref{eq:MasterEquationRotation}), we obtain
\begin{equation}
\delta \cR''(r) + \fr{r} \frac{1+a^2-r^2}{1+a^2} \delta \cR'(r)  + \delta \cR(r)   = - \frac{1}{2} \delta \cR'(r)^2.
\end{equation}
Introducing a new radial variable $z$ via
\begin{equation}
z := \frac{r^2}{2(1+a^2)},
\end{equation}
the deformation equation becomes a Laguerre equation with a quadratic source term
\begin{equation}
\cL_{(a^2+1)/2} \left[ \delta \cR \right] := z \delta \cR''(z) + (1-z)\delta \cR'(z) + \frac{a^2+1}{2} \delta \cR(z) =  -\frac{z}{2} \delta \cR'(z)^2\, , \label{eq:perturb-eq-axisym}
\end{equation}
where we introduced the Laguerre operator $\cL$.
We note that, in terms of the new variable, the MP-solution is now written as
\begin{equation}
\cR_{\rm MP}(z) = \frac{2}{a^2+1}- z.
\end{equation}
Perturbations of this solution should be normalizable in the sense of eq.~(\ref{eq:normCond}), which means for the perturbed profile $m=\exp(\cR_{\rm MP}+\delta \cR)$
\begin{equation}
\int_0^\infty dr\, r  \, m(r)  \sim \int_0^\infty dz e^{-z} \exp \lp\delta \cR(z) \rp <\infty, \label{eq:normCond-1},
\end{equation}
which is accomplished if the perturbation grows as a polynomial at each order, not showing exponential growth $\sim e^z$ or any divergences. 

At leading order, the regular and normalizable perturbations are given by Laguerre polynomials~\cite{Andrade:2018nsz},
\begin{equation}
\delta \cR(z) = \veps L_{N}(z) + \ord{\veps^2},\label{eq:axisym-linear-sol}
\end{equation}
only if $a^2+1 = 2N$, for integer $N$.
Non-trivial solutions have $N\geq2$. $N$ has the interpretation of a 'radial overtone' number, \ie it counts the number of oscillations along $r$. Since these zero modes correspond to 'bumpy black holes'~\cite{Emparan:2003sy,Emparan:2014pra,Dias:2014cia}, $N$ can also be interpreted as the number of bumps in the cross-section of the corresponding solution.

\subsection{Nonlinear perturbations}
In the following, we study how to include higher order perturbations for these zero-modes obtaining better control over the phase space of stationary solutions and also to support the later numerical analysis.

The general perturbative soution to eq.~(\ref{eq:perturb-eq-axisym}) is written as
\begin{equation}
\delta \cR(z) = \sum_{k=0}^\infty \veps^{k+1} f_k(z).
\end{equation}
and for a leading order solution with $a^2+1=2N \, , ( N=2,3,4,\dots)$,
the deformation equation~(\ref{eq:perturb-eq-axisym}) becomes
\begin{equation}
\cL_N \left[f_k(z)\right] =  \cS_k(z) \label{eq:perturb-eq-axisym-k}
\end{equation}
at each perturbation order $k$. 
As usual, the source term $\cS_k(z)$ is expressed by the solution up to $(k-1)$-th order.

A similar higher order perturbation analysis has been performed in~\cite{Suzuki:2015axa,Emparan:2018bmi} for perturbations (non-uniformities) of black strings. It was found there, that the length of the black string has to be renormalized to avoid secular terms that would break the periodic boundary condition. Here, for spinning localized solutions, it turns out that we have to renormalize the angular velocity $\Omega$ or the corresponding spin parameter $a$ which changes the blob size, to avoid secular behavior that would break the normalization condition~(\ref{eq:normCond-1}).

\subsubsection{Resonance and secular perturbation}
Secular behavior in perturbation theory is typically encountered when the dependence of some physical parameter on the perturbation parameter $\varepsilon$ is ignored. A common example for this is the slightly anharmonic oscillator
\begin{equation}
 \ddot{x}(t) + {\omega_0}^2 x(t) = -\varepsilon x(t)^3,
\end{equation}
Note that if we assume $x\ll 1$ the lowest order effect of the anharmonic term $\varepsilon x^3$ is to modify the frequency: $\omega_0 \rightarrow \omega_0+\varepsilon \omega_1$. The appropriate ansatz accordingly should be $x(t)=\sin((\omega_0+\varepsilon \omega_1)t)$, but naive perturbation theory $x(t)=x_0(t)+\varepsilon x_1(t)$ leads to the solution 
\begin{align}
x_0(t)&=\sin(\omega_0 t)\,,\label{eq:exampleSource}\\
x_1(t)&= t \cdot\sin(\omega_0t)+\dots\,, \label{eq:exampleSecTerm}
\end{align}
where the first correction grows unboundedly invalidating the perturbative ansatz and violating conservation of energy. Note here that the secular term (\ref{eq:exampleSecTerm}) results from a resonance phenomenon between the zeroth order solution (\ref{eq:exampleSource}) acting as a resonant source for the first order correction.

For our perturbative problem (\ref{eq:perturb-eq-axisym-k}), a similar resonant behavior occurs. Assuming $S_k(z)$ can be decomposed into a linear combination of Laguerre polynomials $L_M(z)$, we have to distinguish two cases in
\begin{equation}
 \cL_N f(z) = L_M(z). \label{eq:perturb-eq-axisym-LM}
\end{equation}
For $M \neq N$, the solution remains regular and normalizable, 
\begin{align}
f(z) = \frac{L_M(z)}{N-M}. \label{eq:int-laguerre-nm}
\end{align}
However, for $M=N$, which we are going to call the \emph{resonant} case, the solution is
\begin{align}
f(z) =  -L_N(z)\log z - \sum_{I=0}^{N-1} \frac{2}{N-I} L_I(z)+B \Psi(N,0,z)\, \label{eq:int-laguerre-nn} 
\end{align}
with $B$ an integration constant and $\Psi(N,0,z)$ a Laguerre function of the second kind (see~eq.(\ref{eq:laguerre-2-PM})).
Since $\Psi(N,0,z)$ has both a logarithmic divergence at $z=0$ and exponential growth for $z\to\infty$,
the solution can never be regular and normalizable at the same time.
This  corresponds to secular behavior because the resonant term can be eliminated by a infinitesimal shift of $a$ in eq.~(\ref{eq:perturb-eq-axisym}) since,
\begin{align}
\left. \partial_\alpha L_\alpha(z)\right|_{\alpha=N} = \Psi(N,0,z)+ L_N(z) \log z + ({\rm polynomial \ of} \ z).
\end{align}

\subsubsection{Recurrence formula}

The perturbative solution can be  obtained systematically by removing resonant terms in the sources order by order, which leads to an algebraic recurrence relation.
For this, we assume both $\delta \cR(z)$ and $a$ are expanded in $\veps$,
\begin{align}
 \delta \cR(z) = \sum_{k=0}^\infty \veps^k f_k(z),\quad a^2+1= 2N \left(1+\sum_{k=1}^\infty \veps^k \mu_{k} \right), \label{eq:axisym-expand-dR-a}
\end{align}
where we set
\begin{equation}
 f_0(z) = L_N(z).
\end{equation}
Plugging this into eq.~(\ref{eq:perturb-eq-axisym}) and expanding in $\veps$, we obtain the perturbation equation for each order in $\veps$,
\begin{align}
 \cL_N f_k(z) = -\fr{2} \sum_{\ell=0}^{k-1} z f'_\ell(z) f'_{k-1-\ell}(z) - N \sum_{\ell=1}^{k} \mu_\ell f_{k-\ell}(z) =: \cS_k(z) .\label{eq:perturb-src}
\end{align}
Assuming that $f_{\ell}(z)$ are polynomials for $\ell < k$, the source term also becomes a polynomial, and hence should be decomposed to the linear combination of the Laguerre polynomials,
\begin{align}
&\cS_k(z) :=  \sum_{K=0}^M \cC_K L_K(z) - N\mu_k L_N(z),
\end{align}
where $M$ is a finite positive integer. 
After eliminating $L_N(z)$ from the source by using $\mu_k$, $f_k(z)$ can be expressed as a polynomial as well.
And we can decompose the solution at each order into a finite linear combination of Laguerre polynomials
\begin{equation}
 f_k(z) = \sum_{I} {\cal C}_{k,I} L_I(z)\,.
 \label{eq:laguerre-exp}
\end{equation}
The coefficients of the resonant term $\cC_{k,N}$ correspond to the reparametrizations of $\veps$, and hence can be set to $0$.

So the problem reduces to determining the coefficients $\cC_{k,I}$ and $\mu_k$ at each order.
Substituting eq.~(\ref{eq:laguerre-exp}) into the source term~(\ref{eq:perturb-src}), we obtain
\begin{align}
& \cS_k(z) = \cL_N \left[ - \sum_{M\neq N} \left( \sum_{I,J}  \sum_{i=0}^{k-1} \cC_{i,I} \cC_{k-1-i,J}
 \frac{I+J-M}{4(N-M)} \cX_{I,J}^M \right)L_M(z) \right.\nonum
 &\left. \hspace{4cm}- \sum_{M\neq N} \sum_{i=1}^{k-1}\frac{N \mu_i C_{k-i,M}}{N-M} L_M(z)  
 \right] \nonum
 & \qquad  -\left[ N \mu_k +  \fr{4}  \sum_{I,J} \sum_{i=0}^{k-1}(I+J-N)\cC_{i,I} \cC_{k-1-i,J}\cX^N_{I,J}  +  \sum_{i=1}^{k-1} N\mu_i \cC_{k-i,N}\right] L_N(z),\label{eq:axisym-source-k}
\end{align}
where $\cX^K_{I,J}$ comes from the decomposition of the product of Laguerre polynomials~\cite{Watson1938},
\begin{align}
& L_I(z)L_J(z) = \sum_{K=|I-J|}^{I+J}\cX^K_{I,J} L_K(z),
\end{align}
which is written as
\begin{equation}
\cX^K_{I,J}  = \frac{(-2)^{I+J-K} K!}{(K-I)!(K-J)!(I+J-K)!} \hgfunc{3}{2}\left(\begin{array}{c}K+1,\fr{2}(K-I-J),\fr{2}(K-I-J+1)\\
 K-I+1,K-J+1\end{array};1\right).
\end{equation}
The last line in eq.~(\ref{eq:axisym-source-k}) is proportional to the resonant term, and hence should be removed by setting
\begin{subequations}
\begin{equation}
 \mu_k = - \fr{4N}  \sum_{I,J} \sum_{i=0}^{k-1}(I+J-N)\cC_{i,I} \cC_{k-1-i,J}\cX^N_{I,J}  -  \sum_{i=1}^{k-1} \mu_i \cC_{k-i,N}\,.
 \label{eq:recurrence-mu}
\end{equation}
For non-resonant terms, the $k$-th order coefficients are determined by
\begin{align}
&\cC_{k,M\neq N} = -   \sum_{I,J}  \sum_{i=0}^{k-1} \cC_{i,I} \cC_{k-1-i,J}
 \frac{I+J-M}{4(N-M)} \cX_{I,J}^M -  \sum_{i=1}^{k-1}\frac{N \mu_i C_{k-i,M}}{N-M}.
   \label{eq:recurrence-C-p}
\end{align}\label{eq:recurrence}
\end{subequations}
The coefficient of $L_N(z)$ is set to zero $\cC_{k,N}=0$ for $k\geq1$.
With these recurrence equations, the perturbation equation can be solved algebraically.

\subsubsection{Perturbation solution}
To solve the recurrence equation~(\ref{eq:recurrence}), we first set
\begin{equation}
 \cC_{0,M} = \delta_{N,M}.
\end{equation}
Then, we have the solution for $k=1$
\begin{equation}
 \mu_1 = -\frac{1}{4} \cX^N_{N,N},\quad \cC_{1,M\neq N} = - \frac{2N-M}{4(N-M)} \cX^M_{N,N}.\label{eq:bumpy-perturb-sol-1}
\end{equation}
Repeating the calculation, we get the result at $k=2$,
\begin{align}
\mu_2 
 = \sum_{I\neq N}\frac{(2N-I)I}{8N(N-I)}\cX^I_{N,N} \cX^N_{I,N},
\end{align}
and
\begin{align}
 & \cC_{2,M\neq N} 
  =  \sum_{I\neq N} \frac{(I+N-M)(2N-I)}{8(N-M)(N-I)} \cX^M_{N,I}\cX^I_{N,N} - \frac{N(2N-M)}{16(N-M)^2}
 \cX^N_{N,N} \cX^M_{N,N} .
\end{align}
Especially, the leading order shift in $a$ is given by
\begin{equation}
\mu_1 = -\frac{1}{4}\cX^N_{N,N} = -(-2)^{N-2}   \hgfunc{3}{2}\left[\begin{array}{c}N+1,-\frac{N}{2},-\frac{N-1}{2}\\
1,1\end{array};1\right].\label{eq:axisym-mu1}
\end{equation}
Here we note that $\mu_1$ alternates in sign with $N$.
For the first values of $N$, we obtain
\begin{equation}
\mu_1\bigr|_{N=2,3,4,5} = -\frac{5}{2}\ ,\ 14\ ,\ -\frac{173}{2}\ ,\  563.
\end{equation}
Using the relation to the Franel number (see Appendix.~\ref{sec:franel}), one can show the amplitude of $\mu$ grows very rapidly with $N$,
\begin{equation}
 \mu_1 \sim  (-1)^{N+1} \frac{2^{3N}}{N}.
\end{equation}

\subsubsection{Phase diagram}
Given the perturbative solution we can calculate the physical quantities $\cM,\, \cJ$ and the value at the origin $\cR_0=\cR(0)$ (which is used as an initial condition in the numerical analysis) perturbatively as follows.

\paragraph{Angular velocity and center thickness}
By definition, the angular velocity has the expansion
\begin{equation}
\Omega = \frac{a}{1+a^2} = \frac{\sqrt{2N-1}}{2N}\left(1-\frac{N-1}{2N-1}\mu_1 \veps+\ord{\veps^2}\right).\label{eq:axisym-perturb-Omega-res}
\end{equation}
The center thickness is given by
\begin{equation}
\cR_0 = \frac{2}{1+a^2}+ \veps+\ord{\veps^2} = \fr{N}\left(1+(N-\mu_1)\veps + \ord{\veps^2}\right).
\end{equation}
Which gives the gradient on the branching point is given by
\begin{equation}
\left.\frac{ \partial_\veps \log \Omega}{\partial_\veps \log \cR_0}\right|_{\veps=0} = \frac{N-1}{2N-1} \frac{\mu_1}{\mu_1-N}.\label{eq:gradient-omega-to-R0}
\end{equation}
Since $\mu$ grows much faster than $N$, the gradient rapidly approaches to that of the Myers-Perry branch for the larger value of $N$.
For the first few values of $N$, we obtain
\begin{equation}
\frac{ \partial_\veps \log \Omega}{\partial_\veps \log \cR_0} \bigr|_{N=2,3,4,5} = \frac{5}{27}\ , \quad \frac{28}{55}\ , \quad \frac{519}{1267}\ ,\quad \frac{1126}{2511}.
\end{equation}
At higher order, the center thickness is given by
\begin{equation}
 \cR_0 = \frac{a}{1+a^2} + \sum_{k=0} \veps^{k+1} \left(\sum_{I} \cC_{k,I}\right)
\end{equation}
where $\cC_{k,I}$ is the coefficients of the Laguerre expansion at each order in eq.~(\ref{eq:laguerre-exp}).
To compare with the numerical result~(figure~\ref{fig:OmegaR0}), we calculated the formula for $(\cR_0,\Omega)$-space up to $\veps^2$,
\begin{equation}
 \Omega = \frac{\sqrt{2N-1}}{2N} \left(1 + \omega_1 \bar{\veps} + \omega_2 \bar{\veps}^2\right),
\end{equation}
where
\begin{equation}
\bar{\veps} := N \cR_0 -1.
\end{equation}
$\omega_1$ coincides with eq.~(\ref{eq:gradient-omega-to-R0}).
Here we do not show the explicit formula for $\omega_2$, since it no longer reduces to the simple form.
The coefficients for several branches are
\begin{align}
& \omega_1\bigr|_{N=2,3,4,5} =\frac{5}{27},\, \frac{28}{55},\, \frac{519}{1267},\, \frac{1126}{2511},\\
& \omega_2\bigr|_{N=2,3,4,5} =\frac{118}{729},\, -\frac{172629}{66550},\, \frac{82075592}{290557309},\, 
-\frac{1528095425}{4691010024}.
\end{align}

\paragraph{Mass and angular momentum}
Provided that the perturbation is normalizable, the mass~(\ref{eq:normCond}) and angular momentum~(\ref{eq:angularMomentum}) are easily obtained by
\begin{align}
&\cM = \cM_{\rm MP}  \int_0^\infty e^{-z} \exp \lp \delta \cR(z) \rp dz,\label{eq:mass-integral}\\
%&\cJ = \cJ_{\rm MP}  \int_0^\infty e^{-z} z \exp \lp \delta \cR(z) \rp dz,\label{eq:ang-integral}
&\cJ = 2a\cM - 2a\cM_{\rm MP}  \int_0^\infty e^{-z} L_1(z) \exp \lp \delta \cR(z) \rp dz,\label{eq:ang-integral}
\end{align}
where $\cM_{\rm MP}$ is the mass of the Myers-Perry of the same $a$ and $z = L_0(z) - L_1(z)$ is used.
Since these integrations take the form of the inner product of the Laguerre polynomials,
it is convenient to use the expansion of the perturbative solution into the Laguerre polynomials,
\begin{equation}
\delta \cR (z) = \sum_{k=0}^\infty \sum_{M} \veps^{k+1} \cC_{k,M} L_M(z),
\end{equation}
where $\cC_{0,M} = \delta_{M,N}$ for the $N$-branch and $M$ runs over some finite at each perturbative order $k$.
Up to $\ord{\veps^2}$, one can expand as
\begin{equation}
\exp \lp \delta \cR(z) \rp = 1 + \veps L_N(z) - \veps^2  \sum_{M\neq N} \frac{M\cX_{N,N}^M}{4(N-M)}L_M(z)\, ,
\end{equation}
where we made use of the second order solution~(\ref{eq:bumpy-perturb-sol-1}).
Putting this into eqs.~(\ref{eq:mass-integral}) and (\ref{eq:ang-integral}), we obtain
\begin{equation}
\frac{\cJ}{\cM} = 2a\left[1- \frac{\cX^1_{N,N}}{4(N-1)}\veps^2\right],
\end{equation}
in which $a$ also should be expanded according to (\ref{eq:axisym-expand-dR-a}).
We see that the ratio of  angular momentum to mass only differs by $\ord{\veps^2}$ from the Myers-Perry branch.

%%%%%%%%%%%%%%%%%%%%%%%%%%%%%%%%%%%%%%%%%%%%%%%%
\subsection{Numerical construction}
%\subsection{Construction of bumpy phases}
\label{sec:axisymmSolNum}
%%%%%%%%%%%%%%%%%%%%%%%%%%%%%%%%%%%%%%%%%%%%%%%%
%
%
To construct fully non-linear solutions we have to solve numerically the axisymmetric version of the soap bubble equation (\ref{eq:MasterEquationRotation})

\begin{equation}
\RR'' + \frac{\RR'}{r} + \frac12 \RR'^2 + \RR + \frac{\Omega^2 r^2}{2} = 0\,,
\label{eqn:master}
\end{equation}
which is a second order nonlinear differential equation for $\RR(r)$. Since $r$ is a radial coordinate, any physical solution of eq.~(\ref{eqn:master}) must satisfy the regularity condition $\RR'(0) = 0$. This leaves the parameter $\RR_0 \equiv \RR(0)$ as the initial condition that is needed to integrate the differential equation outwards radially. However, not all values of $\RR_0$ will result in physical solutions. In general, as a consequence of the nonlinearity of eq.~(\ref{eqn:master}), $\RR(r)$ will become singular at a finite value of $r=r_s$ and only a discrete set of initial conditions will allow for solutions that that extend to $r \to \infty$. To find these branches our numerical procedure consists in maximizing the value $r_s$ where the singularity appears. Solutions appear as singularities/ peaks of $r_s$ as a function of the initial conditions. See Appendix \ref{app:method} for a more detailed description of the numerical method.

For fixed $\Omega \in [0, 1/2]$, the two branches (stable and unstable) of the MP black hole (\ref{eq:MPsol}) correspond to two such solutions. In terms of the parameter $a$, the MP solutions describe an ellipse in the $(\RR_0, \Omega)$ plane as

\begin{equation}
\RR_0 = \frac{2}{1+a^2}, \qquad \Omega = \frac{a}{1+a^2}.
\label{eqn:MPellipse}
\end{equation}

\begin{figure}[h]
	\begin{center}
		\includegraphics[width=210pt]{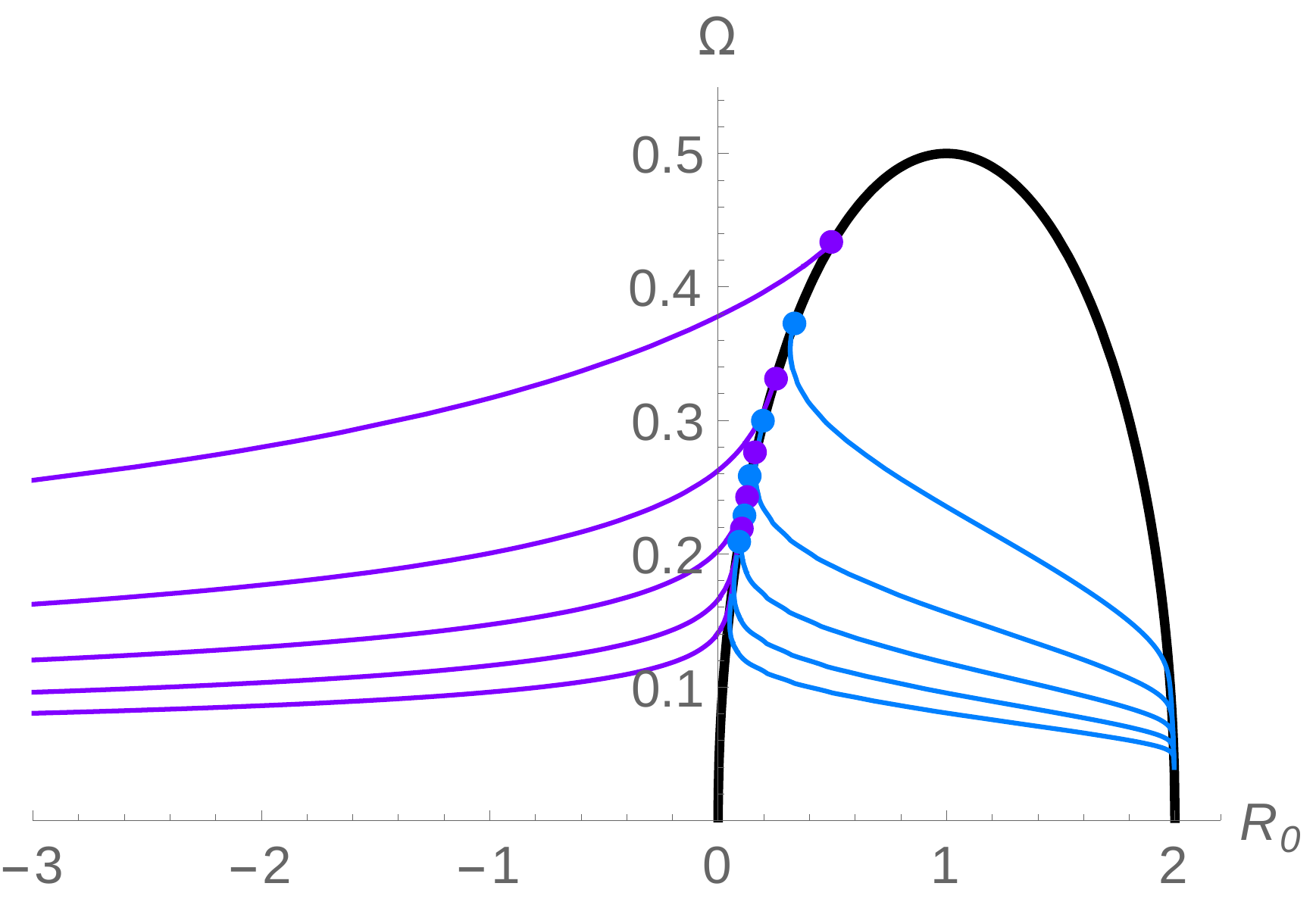}
		\includegraphics[width=210pt]{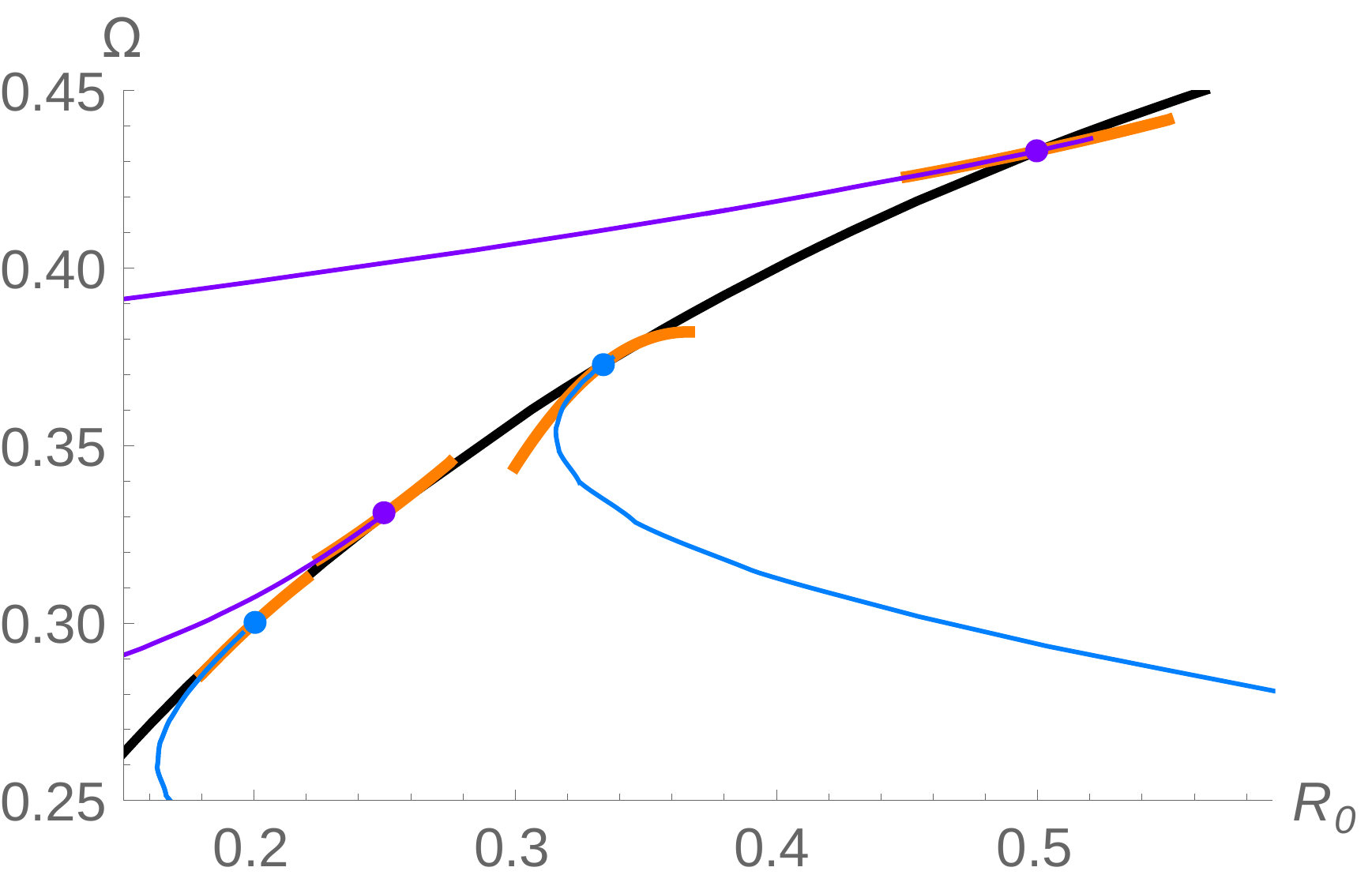}
		\caption{\small Branches of axisymmetric deformations (blue) of MP black hole (black) on the $(\RR_0,\Omega)$ plane. Branches moving towards negative $\RR_0$ connect to black rings. And have a decreasing mass density at the origin. While the branches moving towards positive $\RR_0$ connect to black Saturns and $\RR_0$ approaches a value of the stable MP black hole. The right plot is a close-up showing good agreement with the analytic expansions (orange). The right plot also shows the very short ($-$)-branches. \label{fig:OmegaR0}}
	\end{center}
\end{figure}

Apart from the MP black hole solutions, we find that multiple branches of bumpy solutions extend from every axisymmetric zero-mode. They can be represented in $(\RR_0, \Omega)$ plane as curves that extend from the Myers-Perry ellipse, as shown in figure \ref{fig:OmegaR0}.

\begin{figure}[!htb]
	\begin{center}
		\includegraphics[width=\textwidth]{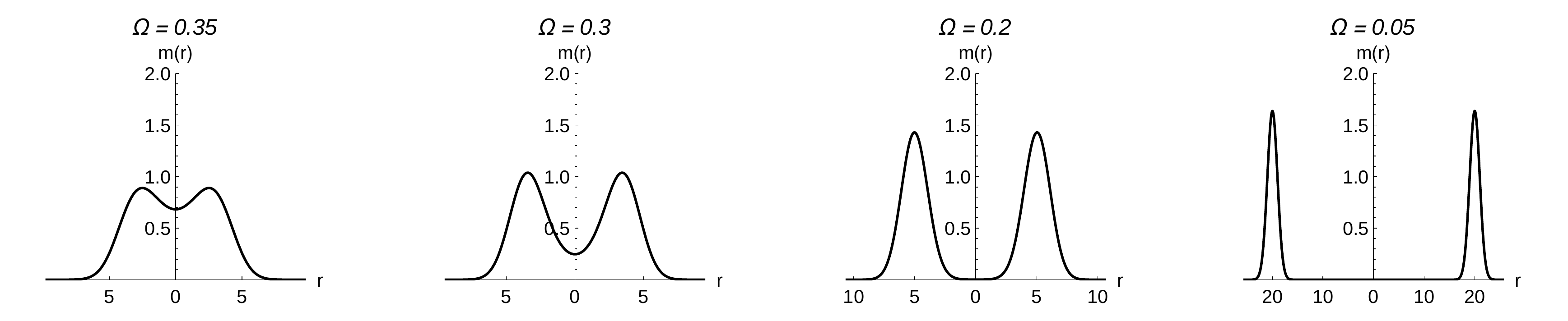}
		\includegraphics[width=\textwidth]{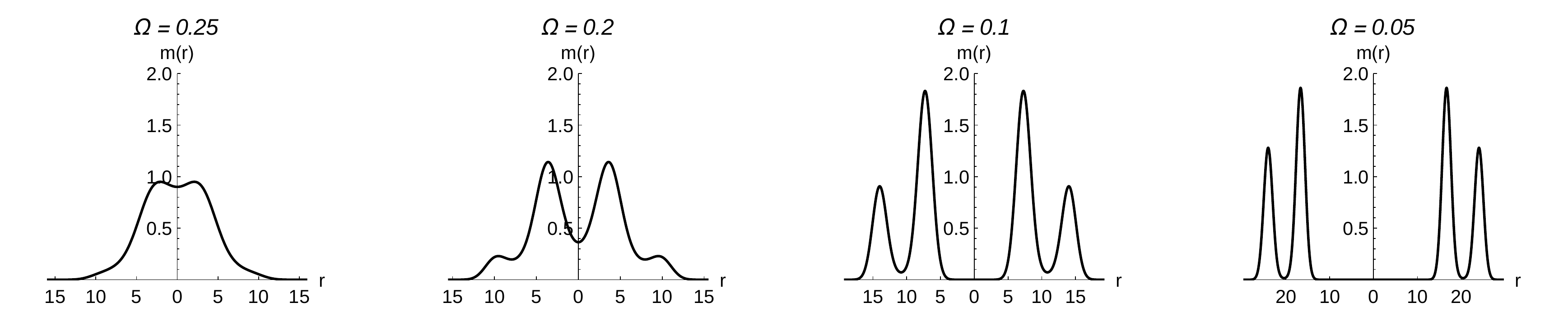}
		\caption{\small Cross-sections of the mass density $m$ for black ripples leading to black rings  corresponding to the zero modes $N = 2, 4$, at different values of $\Omega$. Close to the branching points the solutions develop bumpy deformations whereas far away from it the solutions closely resemble separated black rings. The (expected) pinching of the necks as we move away from the MP-branch follows a behavior described already in \cite{Emparan:2014pra}: For multiple rings the pinching starts at the interior necks and later on the outer ones.\label{fig:rings}}
	\end{center}
\end{figure}

\begin{figure}[!htb]
	\begin{center}
		\includegraphics[width=\textwidth]{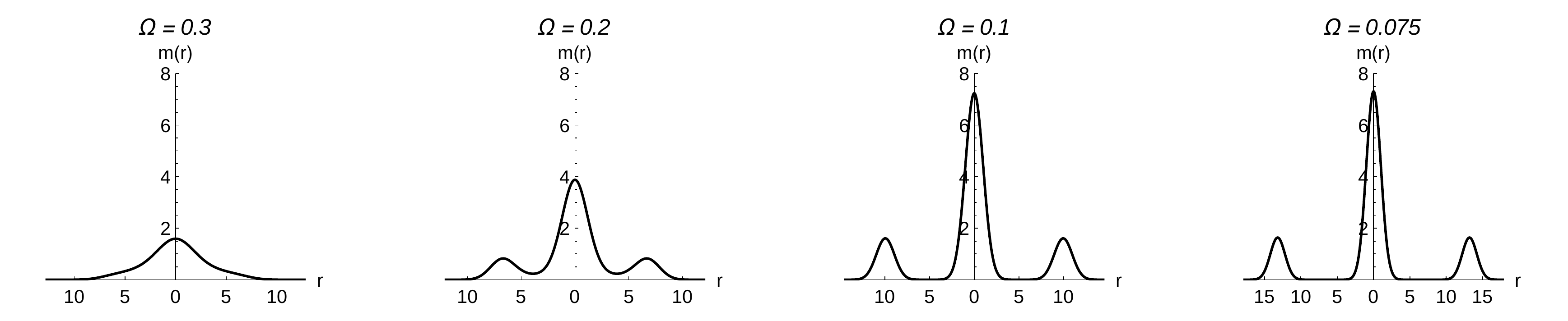}
		\includegraphics[width=\textwidth]{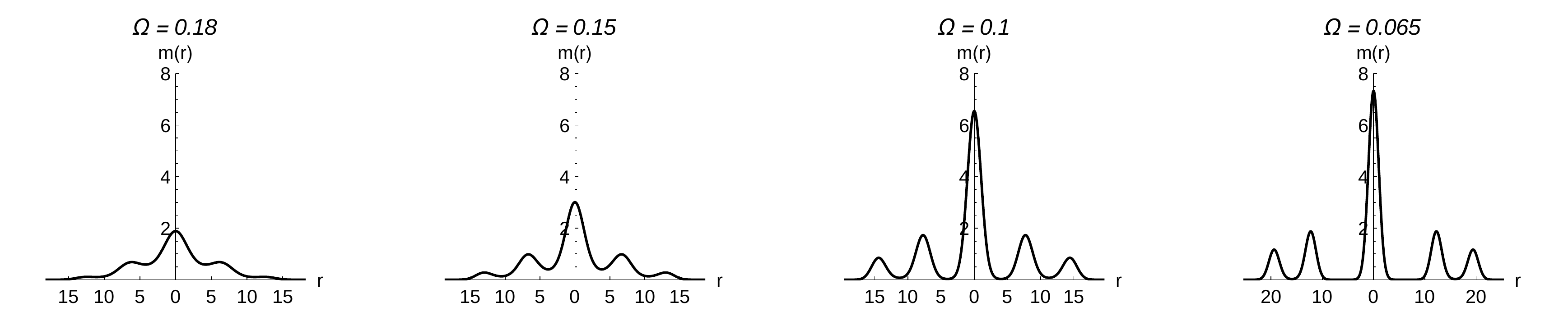}
		\caption{\small Cross-sections of the mass density $m$ for black ripples leading to black Saturns corresponding to the zero modes $N = 3, 5\,$, at different values of $\Omega$.\label{fig:saturns}}
	\end{center}
\end{figure}

We observe that the bumpy branches fall in two distinct categories. Those branches that arise from even $N$ zero modes, as defined in eq.~(\ref{eq:axisym-linear-sol}), tend to $\RR_0 \to -\infty$ as $\Omega \to 0$ (asymptotically like $\cR_{0}\propto -\frac{1}{\Omega^2}$). This is equivalent to a rapidly decreasing mass density at the rotation axis as one moves along the branch. These bumpy branches connect the MP-branch to families of $N-1$ concentric black rings. In figure \ref{fig:rings}, the mass density profiles $m = e^\RR$ are shown. On the other hand, for the zero modes with odd values of $N$, we have $\RR_0 \to 2$, which means that the mass density at the center will closely approach that of a stable MP black hole. These branches will resemble black Saturns with $N-2$ rings, as shown in figure \ref{fig:saturns}.

As discussed in \cite{Dias:2014cia, Emparan:2014pra}, every axisymmetric branch extends in both directions from the zero mode. This corresponds to the fact that linear perturbations of the Myers-Perry black hole can be added with either a positive or a negative amplitude. 
According to the convention in~\cite{Emparan:2014pra}, branches adding the amplitude of the sign $(-1)^{N+1}$ on the axis are called ($+$)-branches, which deform the MP-black hole towards the black rings or black Saturns,
and the opposites, ($-$)-branches, which develop a singularity on the equator of the horizon. It is so far unclear if this ($-$)-branch connects to some singly spinning black hole solution.

Agreeing with this, we find that  the negative side of the branches extends only for a very short interval, after which the allowed solutions cease to exist. This behavior is to some extent expected, since our approach can not resolve singular or conical solutions in phase space. Numerically the vanishing of a solution manifests itself as a vanishing pole in $r_s$. The ($-$)-branches are shown in the close-up plot of figure \ref{fig:OmegaR0}, as the very short blue lines extending into the opposite site of the ($+$)-branches. 
From the perturbative result~(\ref{eq:axisym-perturb-Omega-res}), one can also see that all ($-$)-branches increase $\Omega$, and vice versa at the linear level.

\begin{figure}[h]
	\begin{center}
    \includegraphics[width=250pt]{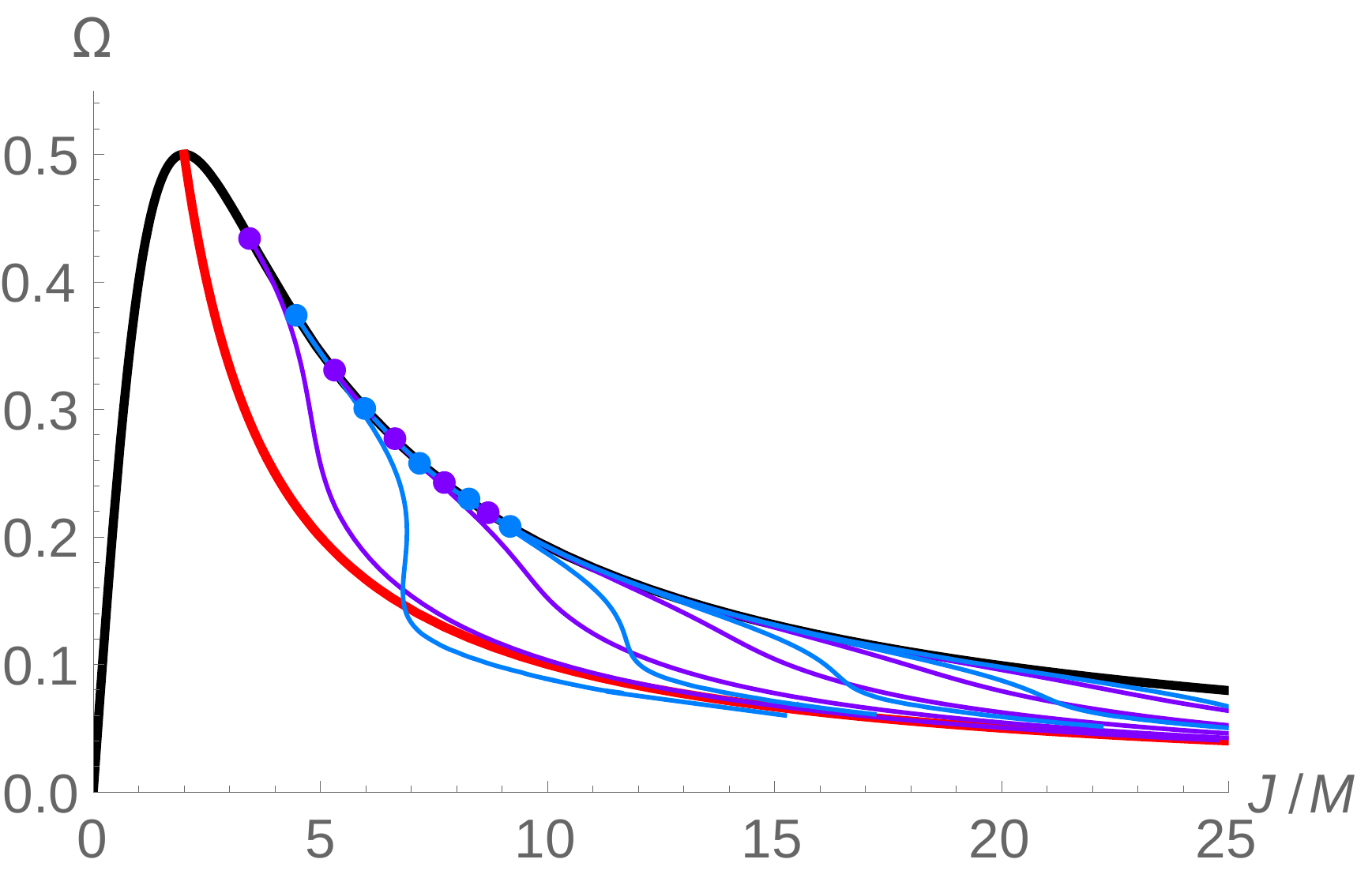}
	\caption{\small Phase diagram for axisymmetric solutions, we show the 10 first  appearing branches: Ring-branches are shown in purple, and Saturn branches in light-blue.
The Myers-Perry and black bar solutions are also plotted by the black and red curves. We do not expect the Saturn branches to terminate, but they become harder to construct for low $\Omega$. %The Myers-Perry solutions are represented by the thick black curve, and the black bars by the thick red curve.
	\label{fig:PhaseDiagramAxisymmetric}}
	\end{center}
\end{figure}

The angular momentum (per unit mass) is calculated numerically according to eq.~(\ref{eq:angularMomentum}). The bumpy branches can then be represented on the usual $(\cJ/\cM, \Omega)$ phase diagram, as depicted in figure \ref{fig:PhaseDiagramAxisymmetric}.

Figures \ref{fig:rings}, \ref{fig:saturns}, \ref{fig:PhaseDiagramAxisymmetric} show that the bumpy branches for black rings and black Saturns seem to extend to arbitrary angular momentum\footnote{Saturn type solutions become harder to construct numerically, since the different Saturn-type solutions pile up in initial condition space as can be seen in figure \ref{fig:OmegaR0}, but we see no evidence that the corresponding poles in $r_s$ vanish.} 
without encountering any conical singularities. 
For a sufficiently high angular momentum, the deformation ends up as multiple lumps/rings barely connected by exponentially thin necks.
Figure \ref{fig:PhaseDiagramAxisymmetric} also shows this in a change of behavior of the curves: All branches show three phases of qualitative behaviors:  In the first stage the branches are nearly tangential to the MP-branch. After that in an  intermediate stage new (ringlike) blobs start to form until they reach a new asymptotic phase. In this final phase the blobs are practically separated and do barely deform further but the distance between the blobs keeps increasing, the angular momentum behaves asymptotically like $\cJ/ \cM \propto 1/\Omega$.

For solutions with multiple ripples, we find that at low $\Omega$ the radii of ringlike blobs follow two different behaviors. The innermost ring has an approximate radius  growing like $\Omega^{-1}$, while the distance between the following outer rings increases slower than that and we estimate it to be 
$\sim\sqrt{|\log\Omega|}$.
The $\Omega^{-1}$-behavior agrees with the blackfold result for multi-rings if the separations of the rings are much shorter than the ring radius~\cite{Emparan:2010sx}.
These observations on the far extended branches lead us to the expectation that our ring/Saturn-like bumpy solutions will be connected via a topology changing transition to the single bumpy rings/Saturns, not directly to multi-rings or higher Saturns. This picture is consistent with the numerical result in $D=6$ bumpy Myers-Perrys~\cite{Emparan:2014pra}.

%
%
%%%%%%%%%%%%%%%%%%%%%%%%%%%%%%%%%%%%%%%%%%%%%%%%
%\section{Multipole Deformations: Black bars and black flowers}
\section{Multipole deformations: Black Flowers}
\label{sec:multipoleDef}
%%%%%%%%%%%%%%%%%%%%%%%%%%%%%%%%%%%%%%%%%%%%%%%%
%
In the large $D$ limit, the soap bubble equation (\ref{eq:MasterEquationRotation}) also admits non-axisymmetric stationary solutions, because gravitational waves are completely decoupled as a non-perturbative effect in $1/D$ and solutions with time-dependent multipoles do not radiate.

%%%%%%%%%%%%%%%%%%%%%%%%%%%%%%%%%%%%%%%%%%%%%%%%
\subsection{Multipolar zero modes}
\label{sec:multipoleDefPert}

We study again perturbations of the MP-black hole, but this time allow for angular dependence of the perturbations 
\begin{equation}
\cR(z,\phi) = \cR_{\rm MP}(z)+ \delta \cR(z,\phi)\,.
\end{equation}
Then, the deformation equation becomes
\begin{equation}
 \cL_{z,\phi} \delta \cR(z,\phi) =  \cS(z,\phi),\label{eq:deformedeq-nonaxisym}
\end{equation}
where we defined
\begin{align}
& \cL_{z,\phi} := z \partial_z^2+(1-z) \partial_z   + \fr{4z}\partial_\phi^2 +\frac{a^2+1}{2}\,, \\
& \cS(z,\phi)   := -\frac{1}{2} z (\partial_z \delta \cR(z,\phi))^2-\frac{1}{8z}(\partial_\phi \delta \cR(z,\phi))^2.
\end{align}
It is convenient to expand the angular dependence as a Fourier series
\begin{equation}
\delta \cR(z,\phi) = \sum_{k=0}^\infty z^\frac{k}{2} f^{(k)}(z) \cos k \phi,
\end{equation}
where each radial function is expanded in $\veps$,
\begin{equation}
 f^{(k)}(z) = \sum_{p=0}^\infty \veps^{p+1} f^{(k)}_p(z).\label{eq:fk-perturb-ex}
\end{equation}
With the Fourier decomposition, the linear part reduces to the generalized Laguerre equation
\begin{equation}
 \cL_{z,\phi} \delta \cR(z,\phi) =  \sum_{k=0}^\infty z^\frac{k}{2} \cL_{(a^2+1-k)/2}^{(k)} f^{(k)}(z) \cos(k\phi),
\end{equation}
which admits normalizable solutions for $k=m$ when
\begin{equation}
 a^2+1 -m = 2N \ (N=0,1,2,\dots).
\end{equation}
We also decompose the source terms into Fourier modes
\begin{equation}
 \cS(z,\phi) = \sum_{k=0} z^\frac{k}{2} \cS^{(k)}(z)\cos k\phi,
\end{equation}
where
\begin{align}
& \cS^{(k)}(z) = - \fr{4} \sum_{\ell = 0}^\infty z^{\ell-1} \left( \ell(\ell+k) f^{(\ell)}(z) f^{(\ell+k)}(z) + (\ell+k) z f^{(\ell)}{}'(z) f^{(\ell+k)}(z) \right.\nonum
&\left.\hspace{5cm}+ \ell z f^{(\ell)}(z) f^{(\ell+k)}{}'(z) + 2 z^2 f^{(\ell)}{}'(z) f^{(\ell+k)}{}'(z) \right)\nonum
 & \quad- \fr{8} \sum_{ \ell =0}^k  \left( (k-\ell) f^{(\ell)}{}'(z)f^{(k-\ell)}(z) + k f^{(\ell)} (z)f^{(k-\ell)}{}' (z)+ 2  z f^{(\ell)}{}' (z)f^{(k-\ell)}{}'(z)\right).\label{eq:nonaxisym-source-k}
\end{align}
Here the last line exists only for $k>0$.

\subsection{Nonlinear perturbation}
For  higher order perturbations, we proceed in almost the same manner as for the axisymmetric sector.
The generalized Laguerre operators ${\cal L}_N^{(m)}$ also show resonant behavior if they are sourced by  the corresponding resonant term $L_N^{(m)}(z)$, provided $N$ is a non-negative integer.
Therefore, for the solution to be regular and normalizable, the resonant term has to be removed
from the source for every mode by renormalizing the angular velocity as
\begin{equation}
   a^2 + 1 = \left(N+\frac{m}{2}\right)\left(1+\sum_{p=1}^\infty \mu_p \veps^p\right).\label{eq:omega-renom-nonaxisym}
\end{equation}
A new phenomenon we observe, is that some modes can not independently excited at linear order, otherwise the renormalization of the angular velocity becomes impossible.
To show this, let us assume to the contrary that we start at linear order only with the zero mode corresponding to $a^2+1 -m= 2N$,
\begin{equation}
 f_0^{(m)}(z) =  L^{(m)}_N(z).  \label{eq:nonaxim-ex-lin-0}
\end{equation}
Then, this mode acts as a source for the neighboring perturbations $f^{(0)}_1$ and $f^{(2m)}_1$ at next-to-leading order,
\begin{align}
& \cL^{(0)}_{N+m/2} f^{(0)}_1(z) = \cS^{(0)}(z)\,, \label{eq:sourcedNLOneighbor1}\\
& \cL^{(2m)}_{N-m/2} f^{(2m)}_1(z) = \cS^{(2m)}(z)\,.\label{eq:sourcedNLOneighbor2}
\end{align}
If $m$ is a even, eqs.~(\ref{eq:sourcedNLOneighbor1}) and (\ref{eq:sourcedNLOneighbor2}) will contain resonant sources.\footnote{For odd $m$, the neighboring modes would have half integer parameters, so resonant behavior only can appear starting at third order.}
However, since we did not include the corresponding linear order term at leading order, the parameter renormalization cannot absorb the resonant terms. This implies that we are forced to include also the neighboring overtone modes at leading order
\begin{equation}
 f_0^{(0)}(z) = \alpha_0 L^{(0)}_{N+m/2},\quad f^{(m)}_0(z) = \alpha_1 L^{(m)}_N(z) ,\quad f^{(2m)}_0(z) =  \alpha_2 L^{(2m)}_{N-m/2}(z).  \label{eq:nonaxim-ex-lin-1}
\end{equation}
Repeating the same argument for the new linear solution, one might be concerned that now we need an infinite tower of overtone modes to regularize the secular behavior. However, if $N-(i-1) m/2 <0$ for the $i$-th overtone, the equation
\begin{equation}
 \cL_{N-(i-1)m/2}^{(i m)} f_1^{(im)}(z) = S^{(im)}(z)
\end{equation}
ceases to produce secular behavior as long as the source term is a polynomial.
Therefore, given $m$ and $N$, the linear order solution should be a linear combination of its overtone modes whose overtone number does not exceed $2N/m+1$.\footnote{This limit is the same in the case of odd $m$, taking into account that only odd overtone modes are involved.}

\subsubsection{Recurrence formula}
Using the expansion of the spin parameter~(\ref{eq:omega-renom-nonaxisym}) we can derive a recurrence formula for all orders in perturbation theory. Eq.~(\ref{eq:deformedeq-nonaxisym}) can be rewritten as
\begin{equation}
 \cL_{N+(m-k)/2}^{(k)} f^{(k)}(z) = \bar{\cS}{}^{(k)}(z)\,, \label{eq:deformedeq-nonaxisym-k-bar1}
\end{equation}
where
\begin{equation}
\bar{\cS}{}^{(k)}(z) = \cS^{(k)}(z) - \left(N+\frac{m}{2}\right)\sum_{p=1}^\infty \mu_p \veps^p f^{(k)}(z)\,,
\label{eq:nonaxisym-source-k-bar2}
\end{equation}
and $\cS^{(k)}(z)$ given through eq.~(\ref{eq:nonaxisym-source-k}).
Under the perturbative expansion~(\ref{eq:fk-perturb-ex}), we also expand the source term by
\begin{equation}
 \bar{\cS}{}^{(k)}(z) = \sum_{p=1}^\infty \veps^p \bar{\cS}{}_p^{(k)}(z).
\end{equation}
Using an inductive argument, the regular normalizable perturbations are shown to be polynomials to all orders of the perturbation.
Therefore, we expand the radial functions at each order
by the associated Laguerre polynomials,
\begin{equation}
 f_p^{(k)}(z) = \sum_{I}  \cC^{(k)}_{p,I} L_I^{(k)}(z).
\end{equation}
As discussed in the previous section, the linear order solution should include all the overtone modes with $N-im/2>0$,
\begin{equation}
 \cC_{0,N+m/2}^{(0)}:=\alpha_0,\quad \cC_{0,N}^{(m)}:=\alpha_1,\quad \cC_{0,N-m/2}^{(2m)}:=\alpha_2,\ \dots\ ,\cC_{0,N-(\eta-1) m/2}^{(\eta m)}:=\alpha_\eta,
\end{equation}
where $\eta:=\lfloor 2N/m \rfloor+1$. If $m$ is odd, the even overtones are turned off.
Using the reparametrization of $\veps$, we set
\begin{equation}
\cC_{p,N}^{(m)}=0 \quad (\text{if }p>0).
\end{equation}
Substituting this expansion into eq.~(\ref{eq:nonaxisym-source-k-bar2})
, the source term can be decomposed into a resonant part and a normalizable part
\begin{equation}
 \bar{\cS}^{(k)}_p(z) = {\cal T}^{(k)}_p L^{(k)}_{N+(m-k)/2}(z) + \cL^{(k)}_{N+(m-k)/2}\bigr[({\rm polynomial\ of \ }z)\bigl]
 \label{eq:nonaxisym-source-k-bar-result}
\end{equation}
where ${\cal T}^{(k)}_p = 0$ gives the normalization condition%
\footnote{If $N+(m-k)/2$ is not a non-negative integer, ${\cal T}^{(k)}_p$ becomes trivially zero.}.
To extract the resonant term from the source, the following decomposition formula of the product of the associated Laguerre polynomials is used
\begin{equation}
 z^\frac{i+j-k}{2} L_I^{(i)}(z) L_J^{(j)}(z) = \sum_{K=0} {\cal Y}^{(i,j,k)}_{I,J,K} L^{(k)}_K(z),
\end{equation}
where the coefficients are written by the integral of the triple product of the associated Laguerre polynomials
\begin{align}
&{\cal Y}^{(i,j,k)}_{I,J,K} = \frac{K!}{(K+k)!}\cI^{(i,j,k)}_{I,J,K}\,
\end{align}
with
\begin{align}
\cI^{(i,j,k)}_{I,J,K} := \int_0^\infty dz e^{-z} z^\frac{i+j+k}{2} L_I^{(i)}(z) L_J^{(j)}(z) L_K^{(k)}(z).
\end{align}
This integration can be expressed through Lauricella's generalized hypergeometric functions (see Appendix.~\ref{sec:associated-laguerre-lauricella})~\cite{Erdelyi1936}. \footnote{An English reference is found, for example, in~\cite{Lee2000}.}

Since the LO-perturbation only contains the fundamental mode $m$ and its overtones, 
also at NLO only $m$ and its overtones are excited. To eliminate the resonant part in 
(\ref{eq:nonaxisym-source-k-bar-result}), we require for $i=0,\dots,\eta$ (again, only odd $i$ if $m$ is odd)
\begin{align}
&\left(N+\frac{m}{2}\right)\sum_{q=1}^{p} \mu_{q} \cC_{p-q,N+(1-i)m/2}^{(im)}  \nonum
&= - \fr{4}\sum_{j=0}^\infty \sum_{q=0}^{p-1} \sum_{I,J}   \cC^{(jm)}_{q,I} \cC^{((i+j)m)}_{p-1-q,J} (I+J-N+(i+2j-1)m/2) {\cal Y}^{(jm,(i+j)m,im)}_{I,J,N+(1-i)m/2} \nonum
&\quad -\fr{8}\sum_{j=0}^i \sum_{q=0}^{p-1} \sum_{I,J}  \cC^{(jm)}_{q,I} \cC^{((i-j)m)}_{p-1-q,J} (I+J-N+(i-1)m/2)  {\cal Y}^{(jm,(i-j)m,im)}_{I,J,N+(1-i)m/2}, \label{eq:nonaxisym-recurrence-normalizable-con}
\end{align}
where the last line only exists for $i>0$.
Other than the resonant terms, we also obtain the coefficients
\begin{subequations}\label{eq:nonaxisym-recurrence}
\begin{align}
& \cC_{p,K}^{(im)} = -\sum_{q=1}^{p-1} \frac{N+m/2}{N+(1-i)m/2-K}  \mu_q \cC_{p-q,K}^{(im)}   \nonum
&\qquad - \sum_{j=0}^\infty\sum_{q=0}^{p-1} \sum_{I,J}   \cC^{(jm)}_{q,I} \cC^{((i+j)m)}_{p-1-q,J} \frac{I+J+jm-K}{4(N+(1-i)m/2-K)} {\cal Y}^{(jm,(i+j)m,im)}_{I,J,K} \nonum
& \qquad - \sum_{j=0}^i \sum_{q=0}^{p-1} \sum_{I,J}   \cC^{(jm)}_{q,I} \cC^{((i-j)m)}_{p-1-q,J} \frac{I+J-K}{8(N+(1-i)m/2-K)}  {\cal Y}^{(jm,(i-j)m,im)}_{I,J,K}.
\label{eq:nonaxisym-recurrence-coeff}
\end{align}
\end{subequations}
Again, we do not have the last line for $i=0$.

\subsubsection{Comparison to the numerical results}
For later comparison with the numerical result, we derive an expression for the center value of each angular Fourier mode. As in the axisymmetric sector, the center thickness is defined by
\begin{subequations}\label{eq:nonaxisym-thickness}
\begin{equation}
\cR_0 = \frac{2}{1+a^2} + \sum_{i=0}^\infty\veps^{i+1} \sum_{I}\cC_{i,I}^{(0)},\label{eq:nonaxisym-thickness-0}
\end{equation}
and for the multipoles, we define\footnote{Which will serve as initial conditions in the numerical setup~(\ref{eq:numAnsatzFourMode}).}
\begin{equation}
\cR_{k} = \sum_{i=0}^\infty \veps^{i+1} \sum_{I}\frac{ (I+k)!\cC_{i,I}^{(k)}}{(2(1+a^2))^{k/2}I!k!}.
\label{eq:nonaxisym-thickness-k}
\end{equation}
\end{subequations}

\subsubsection{Even multipoles}
The analysis for different fundamental modes $(N,m)$ differs in important aspects, so we are going to distinguish several cases in the following. Let us begin with the case $m$ even. 
As opposed to the axisymmetric modes, the normalization condition~(\ref{eq:nonaxisym-recurrence-normalizable-con}) already gives the coupled equation that determines the linear coefficients and the parameter renormalization,
\begin{subequations}
\begin{align}
 & \mu_{1} \alpha_0  = - \fr{4} \sum_{j=0}^\eta  {\cal A}_{0,j}  \alpha_j^2,\label{eq:nonaxisym-normalize-leading-0}\\
& \mu_{1} \alpha_i  = - \fr{4} \sum_{j=0}^{\eta-i}  {\cal A}_{i,j}  \alpha_j \alpha_{i+j}  -\fr{8}\sum_{j=0}^i   {\cal B}_{i,j} \alpha_j \alpha_{i-j} \quad (i>0),\label{eq:nonaxisym-normalize-leading-i}
\end{align}\label{eq:nonaxisym-normalize-leading}
\end{subequations}
where
\begin{align}
&{\cal A}_{i,j}  =   {\cal Y}^{(jm,(i+j)m,im)}_{N+(1-j)m/2,N+(1-i-j)m/2,N+(1-i)m/2}\,, \\
& {\cal B}_{i,j} =  {\cal Y}^{(jm,(i-j)m,im)}_{N+(1-j)m/2,N+(1-i+j)m/2,N+(1-i)m/2}\,.
\end{align}
The nonlinear eq.~(\ref{eq:nonaxisym-normalize-leading}) is hard to solve in general and we will further distinguish different cases.

\begin{figure}[t]
\begin{center}
\includegraphics[width=220pt]{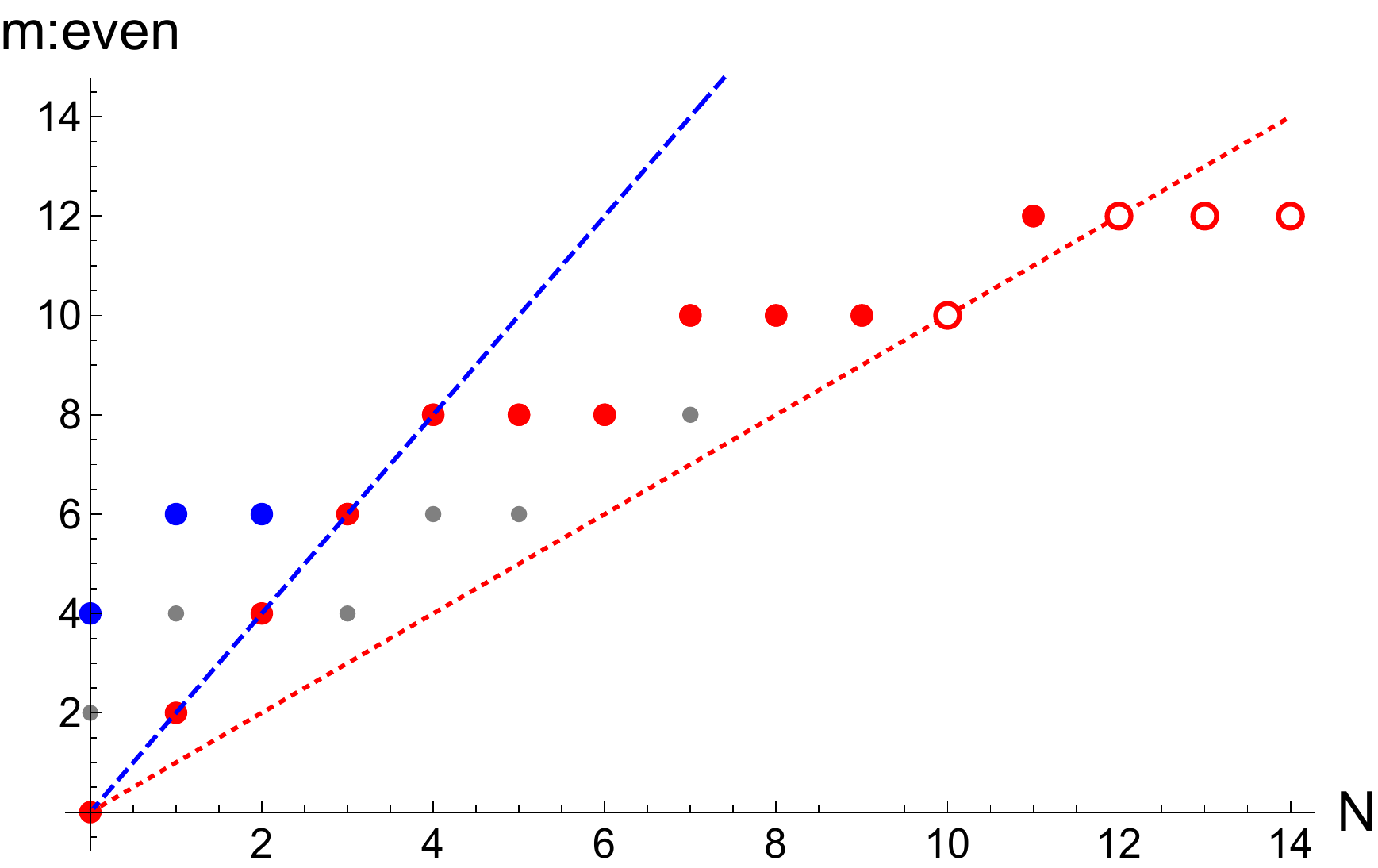}

\caption{\small The maximum values of $m$ in the $2N<m$ sector (blue circles), defined by the constraint~(\ref{eq:cond-nonaxim-1}), and in the $N<m\leq 2N$ sector (red and red empty circles), defined by the positivity of eq.~(\ref{eq:nonaxim-even-ellipse-rad}). The blue dashed and red dotted curves denote $m=2N$ and $m=N$, respectively.
Branches in each sector should be above each curve. The maximum values below $m=N$ (which can not be realized physically) are shown by red empty circles. Gray dots denote possible branches below the maxima. \label{fig:nonaxim-even-m-bound}}

\end{center}
\end{figure}

\paragraph{Even multipoles with $2N<m$}
Here the leading order solution consists of only two modes
\begin{equation}
f^{(0)}_0(z) = \alpha_0 L_{N+m/2}^{(0)}(z),\quad  f_0^{(m)} (z) = \alpha_1 L_{N}^{(m)}(z)\,.
\end{equation}
The normalization condition~(\ref{eq:nonaxisym-normalize-leading}) becomes
\begin{align}
& \mu_1 \alpha_0 = -\frac{\cI_0}{4} \, \alpha_0^2  -\, \frac{(N+m)!}{4N!}  \cI_1 \alpha_1^2\,, \\
&\mu_1 \alpha_1 = - \fr{2} \cI_1  \, \alpha_0\alpha_1 \,,
\end{align}
where
\begin{equation}
\cI_0 =\cX^{N+m/2}_{N+m/2,N+m/2}\,,\quad \cI_1 = {\cal Y}_{N+m/2,N,N}^{(0,m,m)}\,.
\end{equation}
Setting $\alpha_1=0$ immediately reproduces the axisymmetric result~(\ref{eq:axisym-mu1}).
Therefore assuming $\alpha_1\neq 0$, we obtain
\begin{equation}
 \mu_1 = - \fr{2}\cI_1   \alpha_0\,,
\end{equation}
and
\begin{equation}
\left(  2\cI_1-\cI_0 \right)  \, \alpha_0^2 = \frac{(N+m)!}{N!}\cI_1 \alpha_1^2.\label{eq:cond-nonaxim-2}
\end{equation}
Which has real solutions only if 
\begin{equation}
 % 2\geq \frac{\cI_0}{\cI_1} 
\frac{\cI_0}{\cI_1} \leq 2  \label{eq:cond-nonaxim-1}\,.
\end{equation}
This leads to an upper bound for $m$ (see figure~\ref{fig:nonaxim-even-m-bound}).
Since the sign of $\alpha_1$ does not matter, we obtain
\begin{equation}
\alpha_1/\alpha_0 =  \sqrt{\frac{N!}{(N+m)!}} \sqrt{2-\frac{\cI_0}{\cI_1} }\,.
\end{equation}
The only branches satisfying $2N<m$ and the constraint (\ref{eq:cond-nonaxim-1}) are
\begin{subequations}
\begin{align}
&(N,m)=(0,2): \quad \mu_1 = 1, \quad \alpha_1 = \frac{1}{\sqrt{2}} \quad ({\rm black \ bar}),\\
&(N,m)=(0,4): \quad \mu_1=-3,\quad \alpha_1 = \frac{1}{6\sqrt{2}},\\
&(N,m) = (1,4): \quad\mu_1 = 20, \quad\alpha_1 = \fr{10 \sqrt{2}},\\
&(N,m) = (1,6): \quad \mu_1 = -\frac{175}{2}, \quad\alpha_1 = \fr{210 \sqrt{5}},\\
&(N,m) = (2,6): \quad\mu_1 =658, \quad \alpha_1 = \fr{168}\sqrt{\frac{19}{47}},
\end{align}
\end{subequations}
where we set $\alpha_0=1$.

The right hand side in eq.~(\ref{eq:cond-nonaxim-1}) monotonically grows in $N$, and for $N\geq3$, the bound~(\ref{eq:cond-nonaxim-1}) finally starts to exclude all of $m>2N$. 
We will see that a similar bound appears also in the sector $N<m\leq 2N$ for $N \geq 3$.
This upper bound does not mean the absence of the higher multipole deformation,
but rather implies such deformation should be coupled with the lower companions even in the linear order.
For example, $(N,m)=(0,6)$ can be coupled with $(N,m)=(2,2)$ (together with $(3,0)$ and $(1,4)$), which is in $\frac{2}{3}N<m\leq N$ sector. 

Lastly, we evaluate the center values and angular velocity in eq.~(\ref{eq:nonaxisym-thickness}) up to $\ord{\veps}$,
\begin{equation}
\cR_0 = \frac{2}{1+a^2} + \alpha_0 \veps = \frac{1}{N+m/2} \left(1-(\mu_1-(N+m/2)\alpha_0) \veps \right),
\end{equation}
and
\begin{equation}
\cR_m =  \frac{(N+m)!\alpha_1}{(4n+2m)^{m/2}N!m!} \veps.
\end{equation}
By defining $\bar{\veps}:=(N+m/2)\cR_0-1$ we obtain
\begin{equation}
 \Omega = \frac{\sqrt{2N+m-1}}{2N+m} \left(1+\omega_1 \bar\veps\right),\quad  \cR_m = r_1\bar\veps.\label{eq:nonaxisym-even-1-omega-r0-grad}
\end{equation}
with the expansion coefficients
\begin{subequations}
\begin{align}
&(N,m) = (0,4):  \quad \omega_1 = \fr{5}, \quad r_1 = \fr{1920 \sqrt{2}},\\
&(N,m) = (1,4):  \quad \omega_1 = \frac{8}{17}, \quad r_1 = -\fr{4896 \sqrt{2}},\\
&(N,m) = (1,6):  \quad \omega_1 = \frac{25}{61}, \quad r_1 = \fr{11243520 \sqrt{5}},\\
&(N,m) = (2,6):  \quad \omega_1 = \frac{2632}{5877}, \quad r_1 = -\fr{31344000}\sqrt{\frac{19}{47}}.
\end{align}
\end{subequations}
Some of these results are compared with the numerical analysis in figure \ref{fig:OmegaR0Stars}.

\paragraph{Even multipoles with $N < m \leq 2N$ }
Here three modes have to be excited at leading order
\begin{equation}
f^{(0)}_0(z) = \alpha_0 L_{N+m/2}^{(0)}(z),\quad  f_0^{(m)} (z) = \alpha_1 L_{N}^{(m)}(z) ,\quad 
f^{(2m)}_0(z) = \alpha_2 L_{N-m/2}^{(2m)}(z).
\end{equation}
The normalization condition~(\ref{eq:nonaxisym-normalize-leading}) leads to a quadratic constraint for the relative amplitudes
\begin{subequations}\label{eq:nonaxisym-even-2-eq}
\begin{align}
& \mu_1 \alpha_0 = -\fr{4}\cI_0  \alpha_0^2  - \fr{4}\cI_1' \alpha_1^2 - \fr{4}\cI_2' \alpha_2^2,\label{eq:nonaxisym-even-2-eq1}\\
&\mu_1 \alpha_1 = - \fr{2} \cI_1   \alpha_0\alpha_1 -\fr{4}\cI_3 \alpha_2 \alpha_1,\label{eq:nonaxisym-even-2-eq2}\\
&\mu_1 \alpha_2 = - \fr{2} \cI_2   \alpha_0\alpha_2 - \fr{8}  \cI_3' \alpha_1^2,\label{eq:nonaxisym-even-2-eq3}
\end{align}
\end{subequations}
where the coefficients are given by
\begin{align}
&\cI_0 =\cX^{N+m/2}_{N+m/2,N+m/2}\,,&& \cI_1 = {\cal Y}_{N+m/2,N,N}^{(0,m,m)}\,,\\
&\cI_2 = {\cal Y}^{(2m,2m,0)}_{N-m/2,N-m/2,N+m/2}\,,&& \cI_3 = {\cal Y}^{(m,2m,m)}_{N,N-m/2,N}\,,
\end{align}
and
\begin{align}
\cI_1' = \frac{(N+m)!}{N!}  \cI_1,\quad \cI_2' = \frac{(N+3m/2)!}{(N-m/2)!} \cI_2,\quad
\cI_3' = \frac{(N-m/2)!}{(N+3m/2)!} \frac{(N+m)!}{N!}\cI_3.
\end{align}
Setting $\alpha_1=0$ immediately reproduces the previous analysis in which $m$ is replaced by $2m$. Therefore, we consider $\alpha_1\neq 0$ and (\ref{eq:nonaxisym-even-2-eq2}) reduces to
\begin{equation}
\mu_1 = - \fr{2} \cI_1  \alpha_0 -\fr{4}\cI_3 \alpha_2.\label{eq:nonaxisym-even-2-mu1}
\end{equation}
Substituting this to the rest of eq.~(\ref{eq:nonaxisym-even-2-eq}), we obtain two quadratic equations
\begin{align}
&   (2 \cI_1-\cI_0)  \alpha_0^2 + \cI_3 \alpha_2\alpha_0 - \cI_2' \alpha_2^2 =   \cI_1' \alpha_1^2 ,\label{eq:nonaxisym-even-con-a}\\
& 4(\cI_1-\cI_2)  \alpha_0\alpha_2 +2\cI_3 \alpha_2^2  =  \cI_3' \alpha_1^2.
\label{eq:nonaxisym-even-con-b}
\end{align}
$\cI_1$ and $\cI_2$ (and accordingly $\cI_1'$ and $\cI_2'$) have to have the same sign for fixed $N$ and $m$. Thus eq.~(\ref{eq:nonaxisym-even-con-a}) and eq.~(\ref{eq:nonaxisym-even-con-b}) describe an ellipse and a hyperbola in the $(\alpha_1/\alpha_0,\alpha_2/\alpha_0)$ plane. The
curves always have two (or no) intersections, which are shown to be identical by a constant shift in the angular coordinate $\phi \to \phi+\pi/m$.
Therefore, we have at most one branch for each $(N,m)$ with $N<m\leq 2N$. 

The radii of the ellipse from eq.~(\ref{eq:nonaxisym-even-con-a}) are proportional to
\begin{align}
2-\frac{\cI_0}{\cI_1} + \frac{\cI_3^2}{4\cI_0\cI_2'}\,. \label{eq:nonaxim-even-ellipse-rad}
\end{align}
The positivity of this value is the necessary condition for the existence of the branch, which gives the upper bound for $m$ (figure~\ref{fig:nonaxim-even-m-bound}). 
Since the last term in eq.~(\ref{eq:nonaxim-even-ellipse-rad}) decays very quickly in $N$, the upper bound coincides with that from eq.~(\ref{eq:cond-nonaxim-1}) for $N\geq 3$.
And for $N>11$ the upper and the lower bound can not be satisfied at the same time. Accordingly this sector only contains a finite finite number of branches, like the $m>2N$ sector.

We show the result for the lower branches
\begin{subequations}
\begin{align}
&(N,m)=(1,2) :\quad \mu_1=-4.48,\quad \alpha_1 = 0.382,\quad \alpha_2 = 0.00243\,,\\
&(N,m) = (2,4) :\quad\mu_1= -132.5,\quad \alpha_1 = 0.0439,\quad \alpha_2 = -3.84 \times 10^{-8}\,,\\
&(N,m) = (3,4) :\quad \mu_1 =  903.0,\quad \alpha_1 = 0.0299,\quad \alpha_2 = -1.20\times 10^{-9}\,,\\
&(N,m) = (3,6) :\quad\mu_1 = -4851.0,\quad \alpha_1 = 0.00268,\quad \alpha_2 = -2.87\times 10^{-13}\,,
\end{align}
where we set $\alpha_0=1$.
\end{subequations}
One can observe that the amplitude of the overtone mode will be strongly suppressed for larger $N$ and $m$. The gradient of the angular velocity and the center values~(\ref{eq:nonaxisym-even-1-omega-r0-grad}) are also evaluated for the same branches as
\begin{subequations}
\begin{align}
&(N,m)=(1,2) :\quad \omega_1 = 0.230, \quad r_1 = 0.0221, \quad r_2 = -4.89\times10^{-7}\,,\\
&(N,m) = (2,4) :\quad \omega_1 = 0.416, \quad r_1 = 0.0000189, \quad r_2 = -2.56\times 10^{-18}\,,\\
&(N,m) = (3,4) :\quad \omega_1 = 0.447, \quad
r_1 = -2.92\times 10^{-6}, \quad
r_2 = 2.49\times 10^{-20}\,,\\
&(N,m) = (3,6) :\quad \omega_1 = 0.454, \quad
r_1 = 3.36\times 10^{-9}, \quad r_2 =-4.64\times 10^{-32}\,,
\end{align}
\end{subequations}
where we also evaluated the amplitude of the overtone $r_2$ defined via
\begin{align}
 \cR_{2m} = \frac{(N+m/2)!\alpha_2}{(4n+2m)^m (N-m/2)!(2m)!}\veps =: r_2 \bar{\veps}\,.
\end{align}

\subsubsection{Odd multipoles with $2N<m$}
For odd $m$ the leading order modes do not produce secular behavior at second order, but starting from third order it will also appear in this case.
Here the LO-solution consists of a single mode,
\begin{equation}
  f^{(m)}_{0} (z)= L_N^{(m)}(z).
\end{equation}
At second order the even $m$ modes have to be excited
\begin{align}
& \cC_{1,K}^{(0)} = -   \frac{2N+m-K}{4(N+m/2-K)} {\cal Y}^{(m,m,0)}_{N,N,K},\\
& \cC_{1,K}^{(2m)} =  -   \frac{2N-K}{8(N-m/2-K)}  {\cal Y}^{(m,m,2m)}_{N,N,K},
\end{align}
without any renormalization,
\begin{equation}
 \mu_1 = 0.
\end{equation}
Iterating eq.~(\ref{eq:nonaxisym-recurrence}) reveals that there are only even $m$ modes for every odd order in $\varepsilon$, and vice versa. Which results in  $\mu_k=0$ for odd $k$.
At third order, the normalization condition~(\ref{eq:nonaxisym-recurrence-normalizable-con}) becomes
\begin{align}
&\mu_{2}  
=-  \sum_K\left[  \cC^{(0)}_{1,K}  \frac{K}{2N+m} {\cal Y}^{(0,m,m)}_{K,N,N}+ \cC^{(2m)}_{1,K}\frac{K+m}{2(2N+m)} {\cal Y}^{(2m,m,m)}_{K,N,N}\right] \nonum
&=\frac{N!}{(N+m)!}\left[ \sum_{K=0}^{2N+m}   \frac{K(2N+m-K)}{4(2N+m)(N+m/2-K)} \left( \cI^{(0,m,m)}_{K,N,N} \right)^2\right. \nonum
&\left. \hspace{4cm}+ \sum_{K=0}^{2N} \frac{(K+m)(2N-K)}{16(2N+m)(N-m/2-K)} \frac{K!}{(K+2m)!}\left(\cI^{(2m,m,m)}_{K,N,N}\right)^2\right].\label{eq:nonaxisym-odd-mu2}
\end{align}
Different from the even cases, the normalization condition for the simplest odd multipoles does not lead to a bound for $m$. For the lower sector $m\leq 2N$, we will have multiple overtones at linear order, which leads to coupled equations at third order as in the even modes. This may bound $m$ as in the even modes.

In contrast to the case of $m$ even, $\Omega$ and $\cR_0$ only have even powers of $\veps$ appearing in their expansion
\begin{align}
& \Omega = \frac{\sqrt{2N+m-1}}{2N+m}\left(1-\frac{N+m/2-1}{2N+m-1}\mu_2 \veps^2\right)\,,\\
&  \cR_0 = \fr{N+m/2} \left[1 + \veps^2\left( (N+m/2)\sum_{K=0}^{2N+m} \cC_{1,K}^{(0)}
 - \mu_2 \right)\right]\,,
\end{align}
while $\cR_m$ is odd in $\veps$,
\begin{align}
\cR_m = \frac{(N+m)!}{(4N+2m)^{m/2}N!m!}\veps.
\end{align}
This means that odd branches go out from the Myers-Perry branch only in one direction.\footnote{Changing the sign of $\veps$ in $\cR_m$ is equivalent to the constant rotation $\phi \to \phi+\pi/m$, and hence does not lead to another branch.}
The leading order corrections can be written as
\begin{equation}
 \Omega = \frac{\sqrt{2N+m-1}}{2N+m}\left(1+\omega_2 \veps^2\right),\quad 
 \cR_0 = \fr{N+m/2} \left(1+\rho_0 \veps^2\right),\quad \cR_m = \rho_m \veps.
\end{equation}
And the first few branches satisfy,
\begin{subequations}
\begin{align}
& (N,m)=(0,3): \, \mu_2 = 0,\,\, \omega_2 = 0,\quad \rho_0 = 36, \quad \rho_m = \fr{6\sqrt{6}},\\
& (N,m)=(0,5): \, \mu_2 = 0,\,\, \omega_2=0, \quad \rho_0 = -6400, \quad \rho_m = \fr{100\sqrt{10}},\\
& (N,m)=(1,3): \, \mu_2 = -6592,\,\, \omega_2 = 2472,\quad \rho_0 = 4352, \quad \rho_m = \fr{5}\sqrt{\frac{2}{5}}\, .
\end{align}
\end{subequations}
For $N=0$ branches, eq.~(\ref{eq:nonaxisym-odd-mu2}) gives $\mu_2=0$ for any odd $m$,
\begin{equation}
\left. \Omega\right|_{N=0} = \frac{\sqrt{m-1}}{m}\left(1+\ord{\veps^4}\right).
\end{equation}
For $N>0$, for example, we have
\begin{equation}
\left.\frac{d \ln \Omega}{d \ln \cR_0}\right|_{(N,m)=(1,3)} = \frac{309}{544}.
\end{equation}

%%%%%%%%%%%%%%%%%%%%%%%%%%%%%%%%%%%%%%%%%%%%%%%%

%%%%%%%%%%%%%%%%%%%%%%%%%%%%%%%%%%%%%%%%%%%%%%%%
\subsection{Numerical construction}
%\subsection{Non-axisymmetric solutions}
\label{sec:multipoleNum}
%%%%%%%%%%%%%%%%%%%%%%%%%%%%%%%%%%%%%%%%%%%%%%%%
To obtain the fully non-linear multipole solutions numerically, we use a Fourier decomposition corresponding to overtones of a fundamental mode $m$
\begin{equation}
\cR_m(r,\phi)=\sum_{n=0}^\infty \cR^{(n m)}(r) \, r^{n m} \cos(n m \phi)\,.
\label{eq:numAnsatzFourMode}
\end{equation}
Plugging this into the stationary master equation (\ref{eq:MasterEquationRotation}), we obtain a countable set of coupled equations for the fundamental Fourier mode $\cR^{(m)}(r)$ and its overtones $\cR^{(n \cdot m)}(r)$ ($n=2,3,\dots$). From the perturbative analysis, we know that close to the MP-branch higher overtones will only be weakly excited. So we truncate the Fourier series for some $n_{\text{max}}$ to obtain a finite dimensional problem. The resulting coupled ODEs can be now solved using the shooting method described in appendix \ref{app:method}, where now the space of initial conditions is spanned by the amplitudes of the Fourier modes $\cR^{(nm)}(r)$ close to the origin, which we will denote as $\cR_{0},\cR_{m},\cR_{2m}, \dots, \cR_{n_{\text{max}} m}$.

\begin{figure}[H]
	\begin{center}
		\includegraphics[width=210pt]{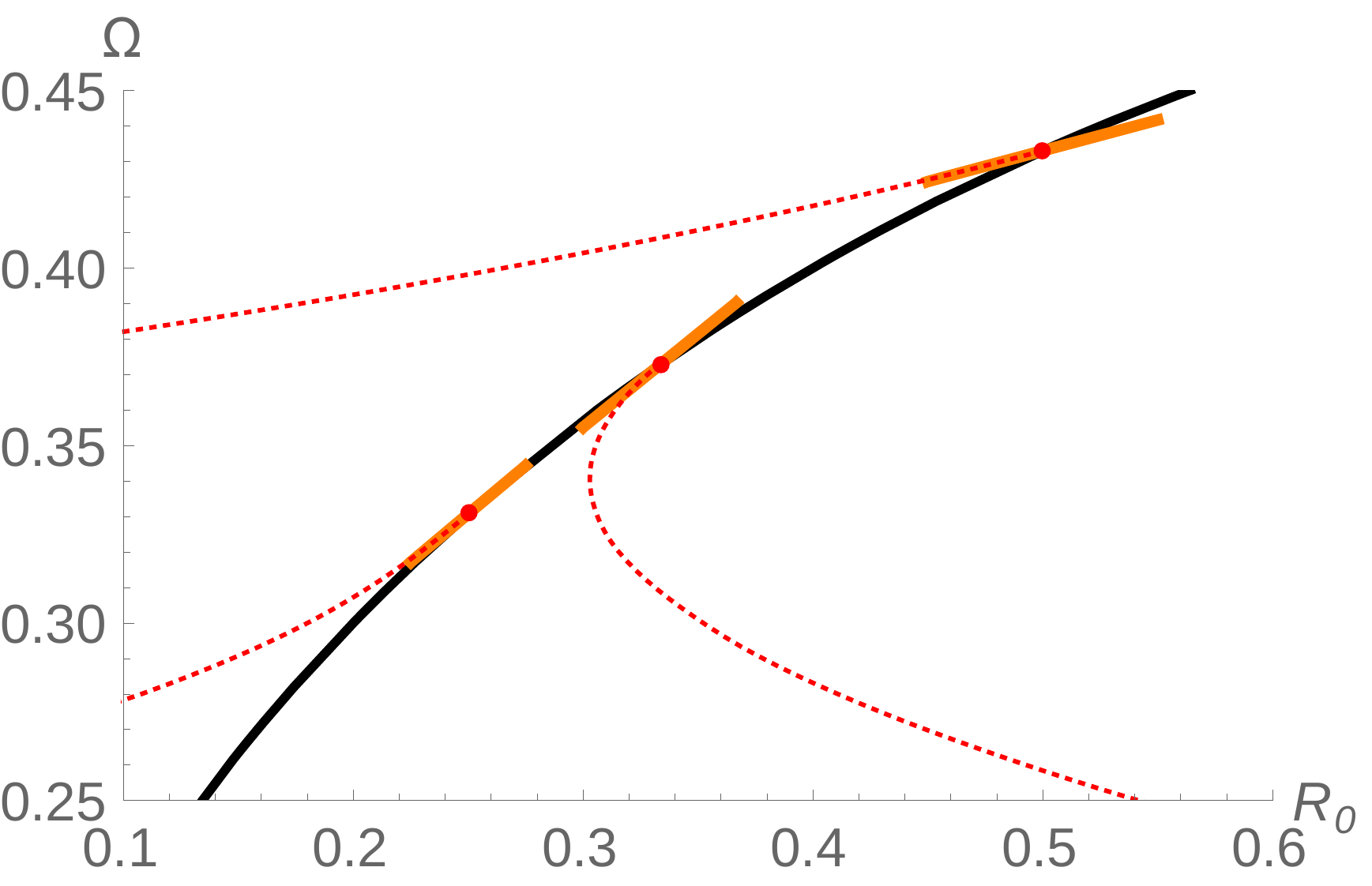}
		\caption{\small Beginning of the branches for $(N,m) = (0,4)$, $(1,4)$ and $(1,6)$ on the $(\RR_0,\Omega)$ plane.\label{fig:OmegaR0Stars}}
	\end{center}
\end{figure}

In figure \ref{fig:OmegaR0Stars}, we show examples of branches extracted numerically with only the fundamental Fourier mode, \ie $n_{max}=1$, and compare them to the perturbative result. We checked that the truncation $n_{max}=1$ is consistent for the beginning of the branch we show by comparing the results to a higher truncation with $n_{max}=2$ and finding good agreement of the results. To extend the branches further overtones should be included.

The inclusion of overtones however makes our numerical procedure much less efficient (see appendix \ref{sec:AppNumMet} for details), s.t. at this point we do not find conclusive results for odd multipole branches and even multipole branches corresponding to the opposite sign of the perturbation.

\begin{figure}[H]
	\begin{center}
		\includegraphics[width=0.35\textwidth]{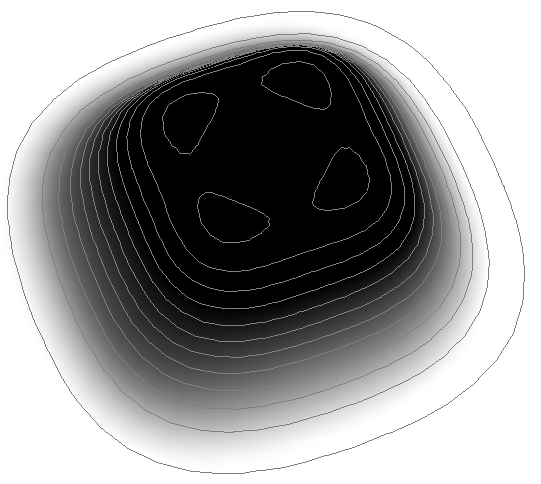}
		\qquad\,\,
		\includegraphics[width=0.35\textwidth]{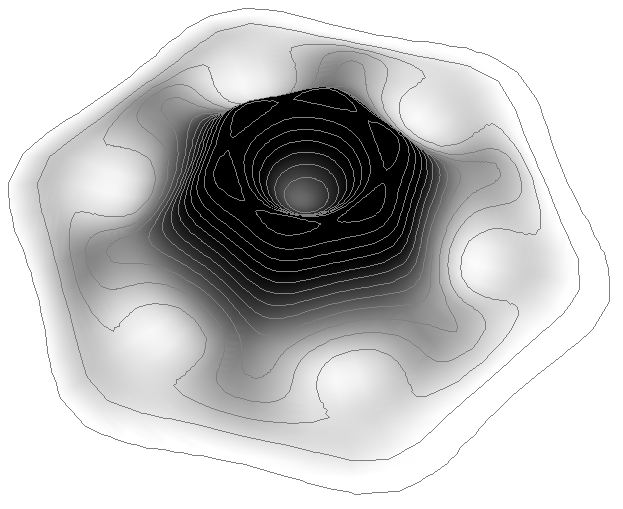}
		\caption{\small Mass profiles for branches with $(N,m) = (0,4)$ (left) and $(N,m) = (1,6)$ (right).\label{fig:flowerSamples}}
	\end{center}
\end{figure}

\begin{figure}[H]
	\begin{center}
		\includegraphics[width=210pt]{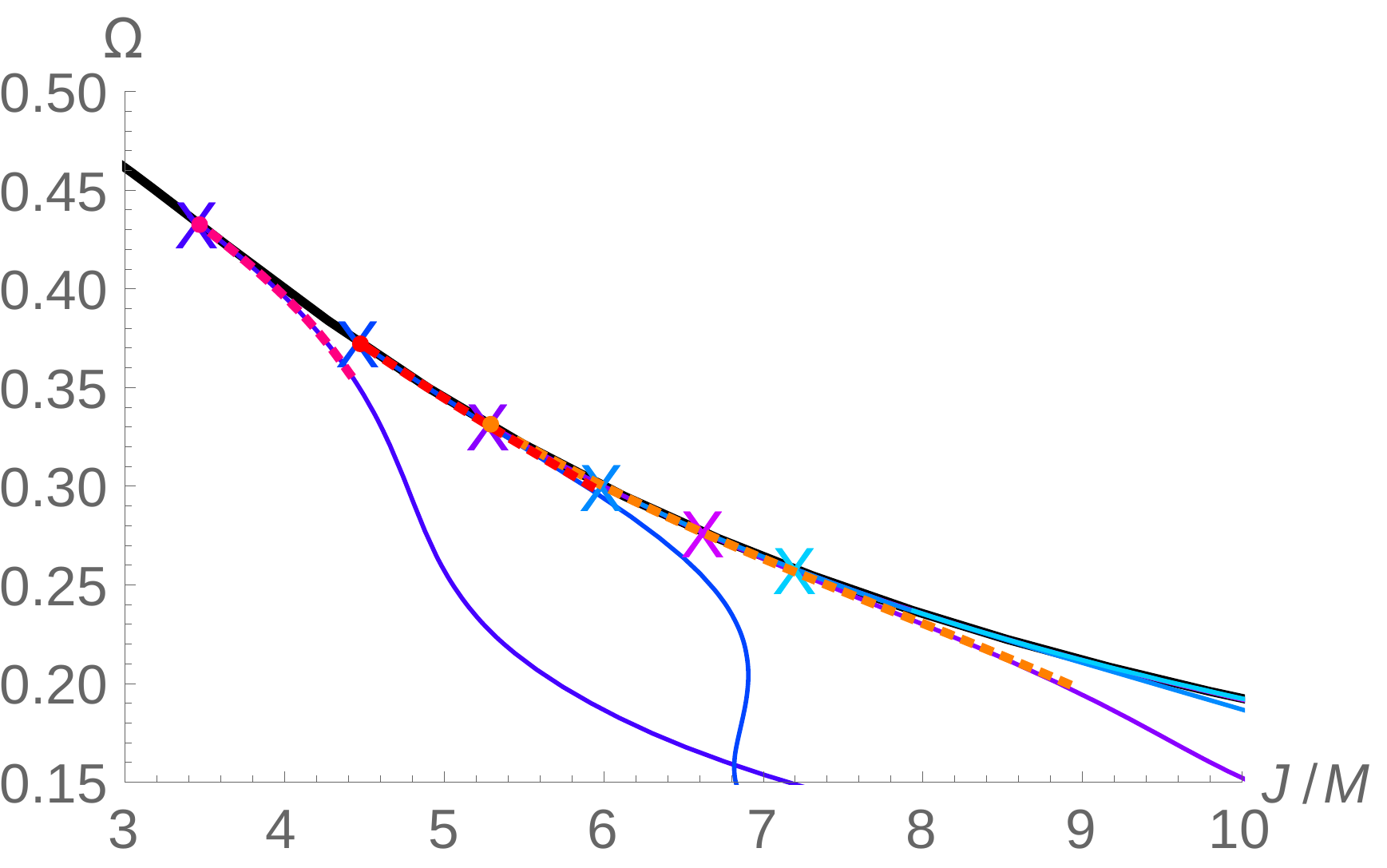}
		\includegraphics[width=210pt]{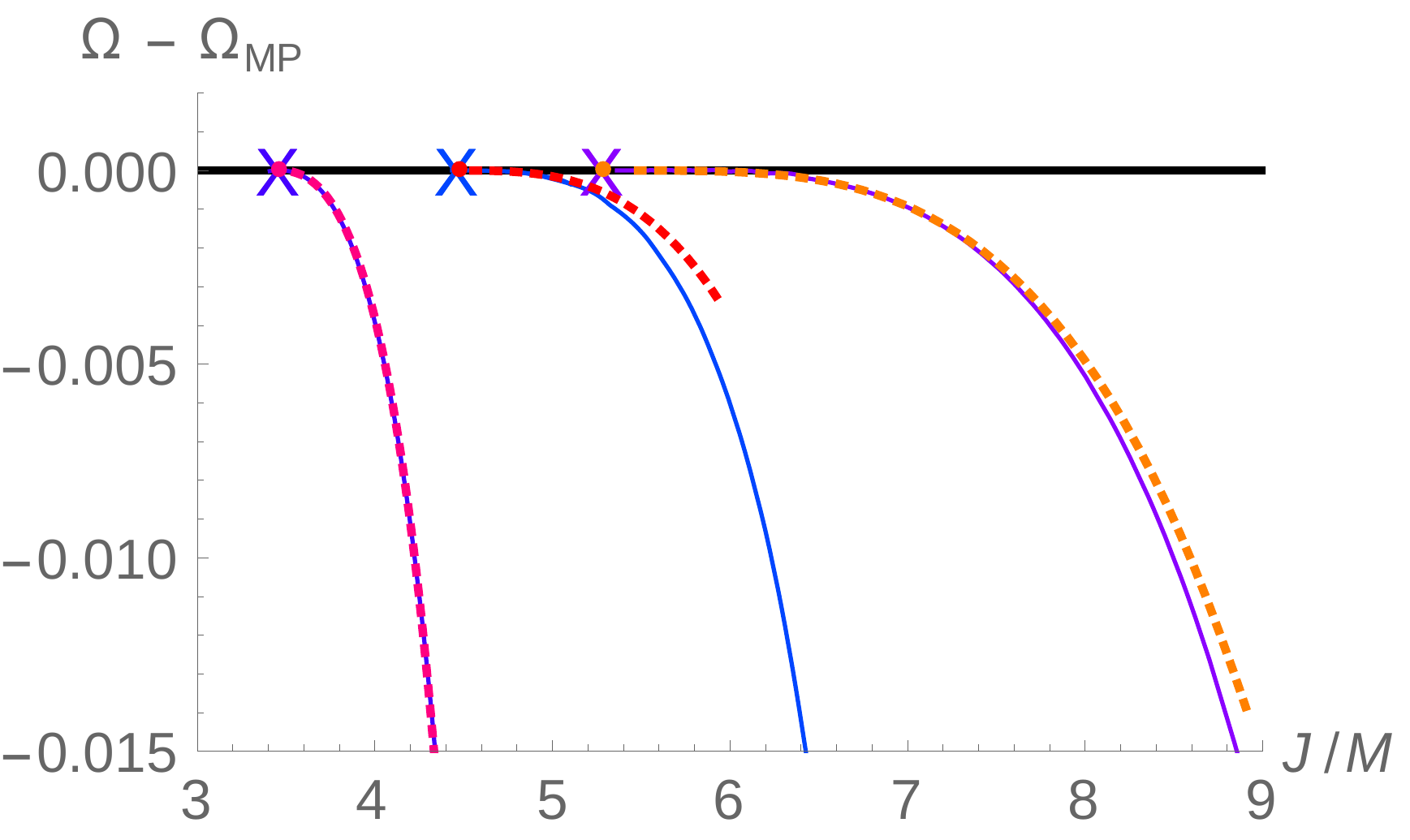}
		\caption{\small Dashed lines: Branches for $(N,m) =(0,4)$, $(1,4)$ and $(1,6)$ in the $(\cJ/\cM,\Omega)$ plane. Solid lines: Branches of axisymmetric solutions. It can be observed that black flower curves take a similar path to the ripple branches originating from the same zero modes. \label{fig:PhaseDiagramStars}}
	\end{center}
\end{figure}

In figure \ref{fig:flowerSamples}, we show representative plots of mass densities for some of the branches. The profiles for even multipoles show a behavior that can be related to the perturbative result that modes of different $N$ and $m$ couple to each other: The black flower branches show mass profiles, which when averaged over the angular direction resemble the corresponding axisymmetric branch that starts at the same branching point, which results in a similar $(\cJ/\cM,\Omega)$-curve see figure \ref{fig:PhaseDiagramStars}.

%%%%%%%%%%%%%%%%%%%%%%%%%%%%%%%%%%%%%%%%%%%%%%%%
%\section{Deformed Black bars}
\section{Deformed black bars: Dumbbells and Spindles}
\label{sec:blackBars}
As already studied in the previous section the large $D$ effective equations allow for stationary solutions without axisymmetry around the rotation axis, the first (and so far only) analytically known solution is the dipolar black bar~\cite{Andrade:2018nsz}. Like the other multipolar solutions, the black bar plays an important role in the decay of the ultra-spinning instability of MP-black holes \cite{Andrade:2018yqu,Andrade:2019edf,Bantilan:2019bvf}. At high enough angular momentum the bar gets very elongated and develops a Gregory-Laflamme type instability. In this section, we are going to study the zero mode configurations corresponding to this instability.

The black bar is best studied in Cartesian coordinates in the co-rotating frame
\begin{equation}
 x =  r \cos(\phi-\Omega t),\quad y = r \sin(\phi-\Omega t),
\end{equation}
where it can be written as
\begin{equation}
 \cR_{\rm bar}(x,y) = 1-\frac{x^2}{2\ell_\perp^2}-\frac{y^2}{2\ell_\parallel^2}
\end{equation}
where
\begin{equation}
 \ell_\perp^2 = \frac{2}{1+\sqrt{1-4\Omega^2}},\quad \ell_\parallel^2 = \frac{2}{1-\sqrt{1-4\Omega^2}}.
\label{eq:ellDef}
\end{equation}
Note that for small $\Omega$ the bar becomes very elongated and in the limit $\Omega \rightarrow 0$ the solution connects to a non-rotating black string along the $y$-direction.

%%%%%%%%%%%%%%%%%%%%%%%%%%%%%%%%%%%%%%%%%%%%%%%%

%%%%%%%%%%%%%%%%%%%%%%%%%%%%%%%%%%%%%%%%%%%%%%%%
\subsection{Co-rotating zero modes}
\label{sec:blackBarsPert}
We deform the bar perturbatively via $\cR = \cR_{\rm bar}(x,y) +  \delta \cR (x,y)$, where the deformation $\delta \cR(x,y)$ satisfies
\begin{equation}
\left[\partial_x^2 - \frac{x}{\ell_\perp^2} \partial_x + \partial_y^2 - \frac{y}{\ell_\parallel^2} \partial_y + 1\right] 
\delta \cR = - \frac{1}{2} ((\partial_x \delta \cR)^2 +(\partial_y \delta \cR)^2)
\end{equation}
At linear order, the regular solutions are given by Hermite polynomials
\begin{equation}
 \delta \cR(x,y) = \veps H_{n_x}\left(\frac{x}{\sqrt{2}\ell_\perp}\right) H_{n_y}\left(\frac{y}{\sqrt{2}\ell_\parallel}\right)+\ord{\veps^2},
\end{equation}
where $n_x,n_y$ are non-negative integers with
\begin{equation}
 \frac{n_x}{\ell_\perp^2} + \frac{n_y}{\ell_\parallel^2} = 1.
\end{equation}
Together with the constraint $\ell_\perp^{-2} + \ell_\parallel^{-2} = 1$, the regular and non-trivial perturbations are
available only for
\begin{equation}
n_x=0, \quad n_y = \ell_\parallel^2  \geq 2.
\end{equation}

\subsection{Nonlinear perturbations}
Considering the linear result, we can assume only $y$-dependence even in the non-linear regime.
Then, by rescaling
\begin{equation}
 z = \frac{y}{\sqrt{2}\ell_\parallel},
\end{equation}
the deformation equation reduces to
\begin{equation}
\cH_{\ell_\parallel^2} \delta \cR(z) = - \frac{1}{2} \delta \cR '(z)^2,
\end{equation}
where $\cH_N$ is the Hermite operator defined by
\begin{equation}
 \cH_N := \frac{d^2}{dz^2} - 2z \frac{d}{dz}+2N. \label{eq:bar-deform-0}
\end{equation}
Given the value of $\ell_\parallel$, $\Omega$ and $\ell_\perp$ is written by
\begin{equation}
 \Omega = \frac{\sqrt{\ell_\parallel^2-1}}{\ell_\parallel^2},\quad \ell_\perp = \frac{\ell_\parallel}{\sqrt{\ell_\parallel^2-1}}
 = \fr{\ell_\parallel\Omega}.
\end{equation}
The corrections beyond the linear order can be derived in the same manner as the bumpy deformation of the Myers-Perry.
First, we expand the deformation function by $\veps$
\begin{equation}
 \delta \cR (z) = \sum_{k=0}^\infty \veps^{k+1} f_k(z).
\end{equation}
If we consider  a branch bifurcating from the zero mode $\ell_\parallel^2 = N$ on the black bar branch, 
one can set
\begin{equation}
 f_0(z) = H_N(z).
\end{equation}
The length of the bar $\ell_\parallel$ for the deformed branch should be expanded by $\veps$,
\begin{equation}
  \ell_\parallel^2 = N \left(1+\sum_{k=1}^\infty \mu_k \veps^k \right),
\end{equation}
where the running coefficient $\mu_k$ is determined so that $f_k(z)$ remains to be normalizable at each order.
Expanding eq.~(\ref{eq:bar-deform-0}) by $\veps$, we obtain
\begin{equation}
 \cH_N f_k(z) = -\fr{2}\sum_{i=0}^{k-1} f_i'(z) f_{k-1-i}'(z) - 2N\sum_{i=0}^{k-1} \mu_{k-i} f_i(z) =: \cS_k(z).
\end{equation}
Similar to the bumpy solutions, the higher order corrections can be solved algebraically.
Assuming $f_k(z)$ is a polynomial, each order solution can be expanded by the Hermite polynomials,
\begin{equation}
 f_k(z) = \sum_{M=0} \cC_{k,M}H_M(z),
\end{equation}
where the linear order solution is supposed to be $\cC_{0,M} = \delta_{M,N}$.
Substituting this, the source term of each order becomes
\begin{equation}
 \cS_k(z) = - \fr{2} \sum_{i=0}^{k-1} \sum_{I,J} \cC_{i,I} \cC_{k-1-i,J} H_I'(z)H_J'(z) 
 - 2 N \sum_{i=0}^{k-1} \sum_{I} \mu_{k-i} \cC_{i,I} H_I(z).
\end{equation}
Using the properties of the Hermite polynomials, the source term can be decomposed to the resonant and non-resonant terms,
\begin{align}
&S_k(z) = \cH_N \left[-\fr{4} \sum_{K \neq N}   \sum_{I,J} \sum_{i=0}^{k-1} \cC_{i,I} \cC_{k-1-i,J} \frac{I+J-K}{N-K} {\cal Q}_{I,J}^K  H_K(z)-\sum_{K\neq N}  \sum_{i=1}^{k-1} \frac{N\mu_{k-i} \cC_{i,K}}{N-K} H_K(z)  \right]
\nonum
&\hspace{3cm} -\left[ \fr{2}\sum_{I,J} \sum_{i=0}^{k-1} (I+J-N)\cC_{i,I} \cC_{k-1-i,J} {\cal Q}_{I,J}^N+  2N\sum_{i=0}^{k-1} \mu_{k-i} \cC_{i,N}\right]  H_N(z) \,,
\end{align}
where $Q^K_{I,J}$ is given by eq.~(\ref{eq:hermite-product-cf}).
Using $\cC_{0,M} = \delta_{M,N}$, the regularizing condition is given by
\begin{subequations}
\begin{equation}
 \mu_k =-  \sum_{I,J} \sum_{i=0}^{k-1}\frac{I+J-N}{4N}\cC_{i,I} \cC_{k-1-i,J} {\cal Q}_{I,J}^N-  \sum_{i=1}^{k-1} \mu_{k-i} \cC_{i,N}\,,
\end{equation}
and the non-resonant coefficients,
\begin{align}
&\cC_{k, M \neq N} =-\fr{4} \sum_{I,J}  \sum_{i=0}^{k-1} \cC_{i,I} \cC_{k-1-i,J} \frac{I+J-M}{N-M} {\cal Q}_{I,J}^M  - \sum_{i=1}^{k-1}  \frac{N\mu_{k-i} \cC_{i,M}}{N-M}.
\end{align}
\end{subequations}
For the resonant term, we simply set
\begin{align}
\cC_{k,N} =0 \quad (k>0). 
\end{align}
Using induction one can show for odd branches that $f_k(z)$ has only odd (even) power for the even (odd) order, and $\mu_k$ vanishes for every odd order. Similarly, for even $N$, it can be shown that at each order only even powers appear.

\subsubsection{Perturbation solution}
By solving the recurrence equation with $\cC_{0,M} = \delta_{M,N}$, one can obtain the solution to arbitrary order. 
The result for $\ord{\veps^2}$ is
\begin{align}
 \mu_1 = - \fr{4} {\cal Q}^N_{N,N},\quad \cC_{1,M\neq N} = -  \frac{2N-M}{4(N-M)}{\cal Q}^M_{N,N}, \label{eq:dumbbell-perturb-res-1}
\end{align}
and for $\ord{\veps^3}$,
\begin{align}
&\mu_2 =\fr{8} \sum_{I} \frac{I(2N-I)}{N(N-M)} {\cal Q}^I_{N,N} {\cal Q}_{N,I}^N, \label{eq:dumbbell-perturb-res-2-mu}\\
& \cC_{2,M\neq N} = \fr{8} \sum_{I\neq N} \frac{(N+I-M)(2N-I)}{(N-M)(N-I)}{\cal Q}^M_{I,N} {\cal Q}^I_{N,N} 
-\frac{N(2N-M)}{16(N-M)^2}{\cal Q}^N_{N,N} {\cal Q}^M_{N,N},
 \label{eq:dumbbell-perturb-res-2-C}
\end{align}
where ${\cal Q}^N_{N,N}=0$ for the odd $N$, giving $\mu_1=0$ for the odd dumbbells.

\subsubsection{Physical quantities}
Once, given the deformation $\delta \cR(z)$ as
\begin{equation}
 \delta \cR(z) = \sum_{i=0}^\infty \sum_{I}\veps^{i+1} \cC_{i,I} H_I(z),
\end{equation}
the physical quantities are calculated using properties of the Hermite polynomials.

\paragraph{Value at the origin}
Here we evaluate the center values $\cR_0 = \cR(0)$ and $\bar{\cR_0} = \cR'(0)$, which are also used as the boundary condition for the numerical analysis. Due to the mirror symmetry in the even case, $\bar{\cR_0}$ only exists for the odd branches.
The center thickness $\cR_0$ of the deformed bar is given by
\begin{equation}
\cR_0 = 1 + \sum_{i=0}^\infty \sum_{I} \veps^{i+1}\cC_{i,I} H_I(0), \label{eq:R0-dumbbell-perturb}
\end{equation}
where
\begin{equation}
 H_M(0) = \left\{ \begin{array}{cl} (-2)^{M/2} (M-1)!! & \quad (M: {\rm even}) \\ 0 &  \quad (M: {\rm odd})  \end{array}
 \right. .
\end{equation}
For the odd branch, only odd Hermite polynomials  appear at every odd order in $\veps$, so $\cR_0$ becomes the function of $\veps^2$.
Using $H_{I}'(0) = - H_{I+1}(0)$, $\bar{\cR_0}$ is similarly evaluated to
\begin{equation}
\bar{\cR_0} =  - \sum_{i=0}^\infty \sum_{I} \veps^{i+1}\cC_{i,I} H_{I+1}(0). \label{eq:R0bar-dumbbell-perturb}
\end{equation}
With eq.~(\ref{eq:dumbbell-perturb-res-1}), we obtain
\begin{align}
&\cR_0 = 1 + \veps H_N(0) - \veps^2\sum_{I\neq N} \frac{4N-I}{2(N-I)} {\cal Q}_{N,N}^I H_I(0) + \ord{\veps^3},\\
&\bar{\cR_0} = - \veps H_{N+1}(0) + \ord{\veps^3},
\end{align}
where $\bar{\cR_0}$ does not have $\ord{\veps^2}$ term, because ${\cal Q}^I_{N,N}$ vanishes for odd $I$.
For comparison with the numerical analysis~(figure~\ref{fig:OmegaR0Bars}), we obtain,
\begin{equation}
 \Omega = \frac{\sqrt{N-1}}{N}\left(1+\omega_1 \bar{\veps} + \omega_2 \bar{\veps}^2\right),\quad \bar{R_0} = \bar{\rho_0} \bar{\veps}
\end{equation}
where 
\begin{equation}
\bar{\veps} := \left\{ \begin{array}{cl}
\cR_0-1& \quad ( \rm even)\\
\sqrt{|\cR_0-1|} &\quad (\rm odd)
\end{array}\right.
\end{equation}
For  odd branches with $N=2n+3$, $\cR_0$ is given by $\cR_0 = 1 + (-1)^n \bar{\veps}^2$.
The even branches have
\begin{align}
 &\left. \omega_1 \right|_{N=4,6,8,10} = 2,\quad -16, \quad 129,\quad -896\\
 &\left. \omega_2 \right|_{N=4,6,8,10} = 52,\quad 8088,\quad \frac{4178816}{5}, \quad \frac{529505120}{7},
\end{align}
and the odd branches have $\omega_1=0$ and
\begin{align}
 &\left. \omega_2 \right|_{N=3,5,7,9} =\frac{12}{19},\quad \frac{19200}{1969},\quad \frac{5480160}{53939}, \quad\frac{23886707712}{24551641},\\
 &\left. \bar{\rho_0} \right|_{N=3,5,7,9} =-2 \sqrt{\frac{3}{19}},\quad 6 \sqrt{\frac{5}{1969}}, \quad-10 \sqrt{\frac{7}{53939}},\quad \frac{210}{\sqrt{24551641}} .
\end{align}
This shows that one always need to spin up the black hole for the transition to an odd branch.

\paragraph{Mass and angular momentum}
The mass~(\ref{eq:normCond}) and angular momentum~(\ref{eq:angularMomentum}) can be calculated by
\begin{equation}
 \cM = \cM_{\rm bar} \int_{-\infty}^\infty \frac{dz}{\sqrt{\pi}} e^{-z^2} \exp(\delta \cR(z)),
\end{equation}
and
\begin{equation}
 \cJ = \frac{\cM}{\Omega} + 4 \cM_{\rm bar} \ell_\parallel^2 \Omega  \int_{-\infty}^\infty
 \frac{dz}{8\sqrt{\pi}}e^{-z^2} H_2(z) \exp(\delta \cR(z)),
\end{equation}
where $\cM_{\rm bar} = 2\pi e/\Omega$ is the mass of the bar solution for the given $\Omega$.
Due to the orthogonal property of the Hermite polynomials, the integrals in $\cM$ and $\cJ$ pick up $H_0(z)$ and $H_2(z)$ components in $\exp(\delta \cR(z))$, respectively.

Using the result in the previous section, the ratio of the angular momentum to the mass is given by
\begin{equation}
 \frac{\cJ}{\cM} = \fr{\Omega}\left(1-\frac{2(N-1)}{N(N-2)}{\cal Q}^2_{N,N} \veps^2+\ord{\veps^3}\right),
\end{equation}
where we note that $\Omega$ should also varies in $\veps$.
For the odd branch, both $\cJ/\cM$ and $\Omega$ become a function of $\veps^2$.

%%%%%%%%%%%%%%%%%%%%%%%%%%%%%%%%%%%%%%%%%%%%%%%%

%%%%%%%%%%%%%%%%%%%%%%%%%%%%%%%%%%%%%%%%%%%%%%%%
\subsection{Numerical construction}
%\subsection{Dumbbells and spindles}
\label{sec:blackBarsNum}
%%%%%%%%%%%%%%%%%%%%%%%%%%%%%%%%%%%%%%%%%%%%%%%%

In order to find fully nonlinear deformations of the black bar, we begin by considering equation (\ref{mastercharge}) with the ansatz

\begin{equation}
\RR(x,y) = - \frac{x^2}{2 \ell^2_\perp} + \RR(y)\, ,
\label{eq:barAnsatz}
\end{equation}

where we imply that $\RR(y) \equiv \RR(0, y)$, and $\ell^2_\perp$ is defined by eq.~(\ref{eq:ellDef}). With this substitution, we are left with 

\begin{equation}
\RR'' + \frac12 \RR'^2 + \RR + \frac{\Omega^2 y^2}{2} = \ell^{-2}_\perp\, .
\label{eqn:masterDumbbells}
\end{equation}

Since $y$ is no longer a radial coordinate, the condition $\RR'(0) = 0$ is no longer required. We can define $\RR'(0) \equiv \bar \RR_0$ instead. Allowed solutions must extend regularly both to $y \to -\infty$ and $y \to \infty$ simultaneously. If we start the integration from $y=0$, the initial conditions are given by $\RR_0 \equiv \RR(0)$ and $\bar \RR_0 \equiv \RR'(0)$, which have to be tuned in order to get allowed solutions. 

The branches arising from even $N$ zero modes have a $y \to -y$ symmetry, so $\bar \RR_0 = 0$. These bars only require $\RR_0$ to be tuned, so they can be found in the same way as the axisymmetric solutions. Nonzero values of $\bar \RR_0$ give rise to the branches originating in odd $N$ zero modes. This requires a slightly more involved numerical algorithm, which is described in Appendix \ref{app:method}.

\begin{figure}[h!]
	\begin{center}
		\includegraphics[width=0.49\textwidth]{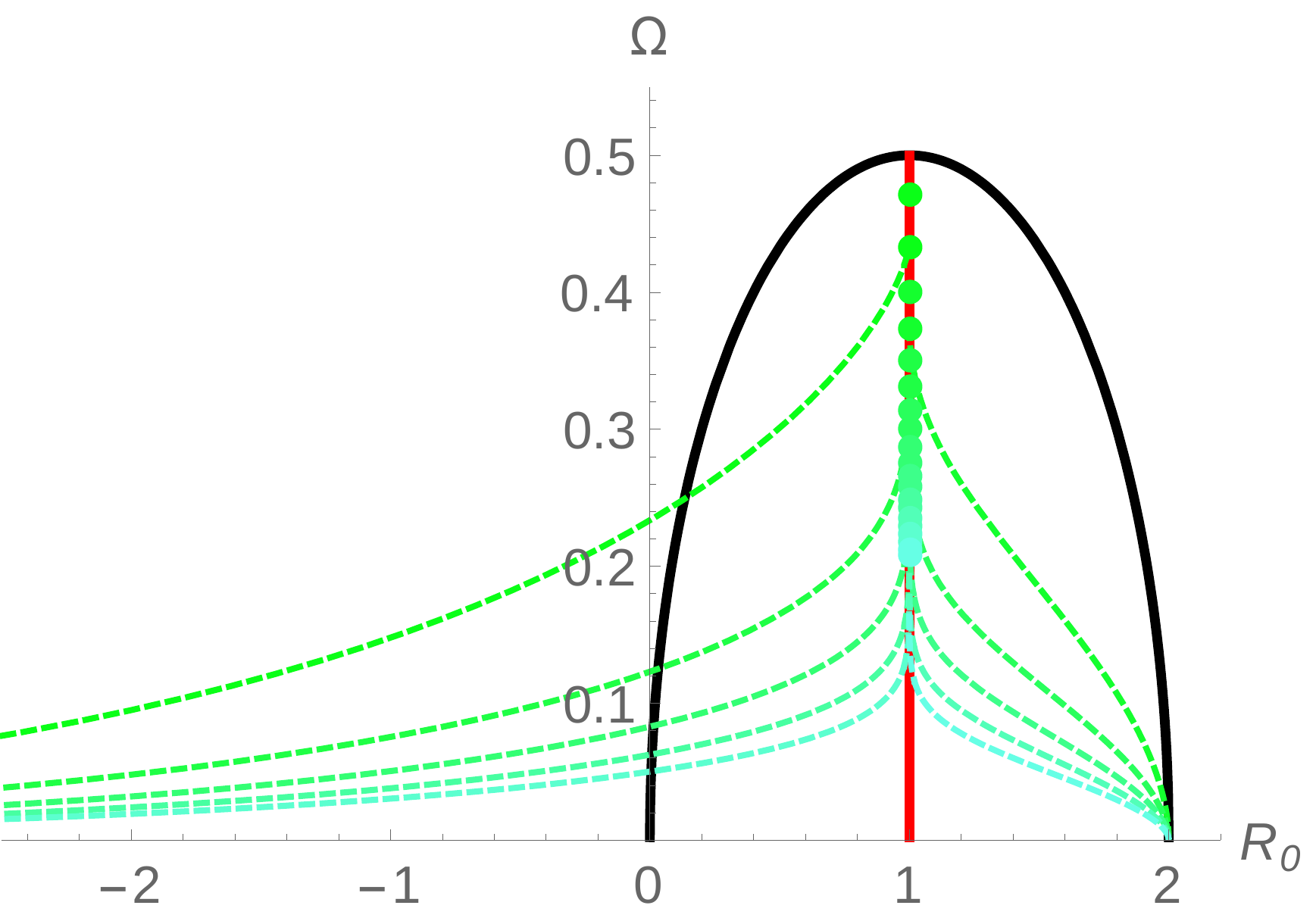}
		\includegraphics[width=0.49\textwidth]{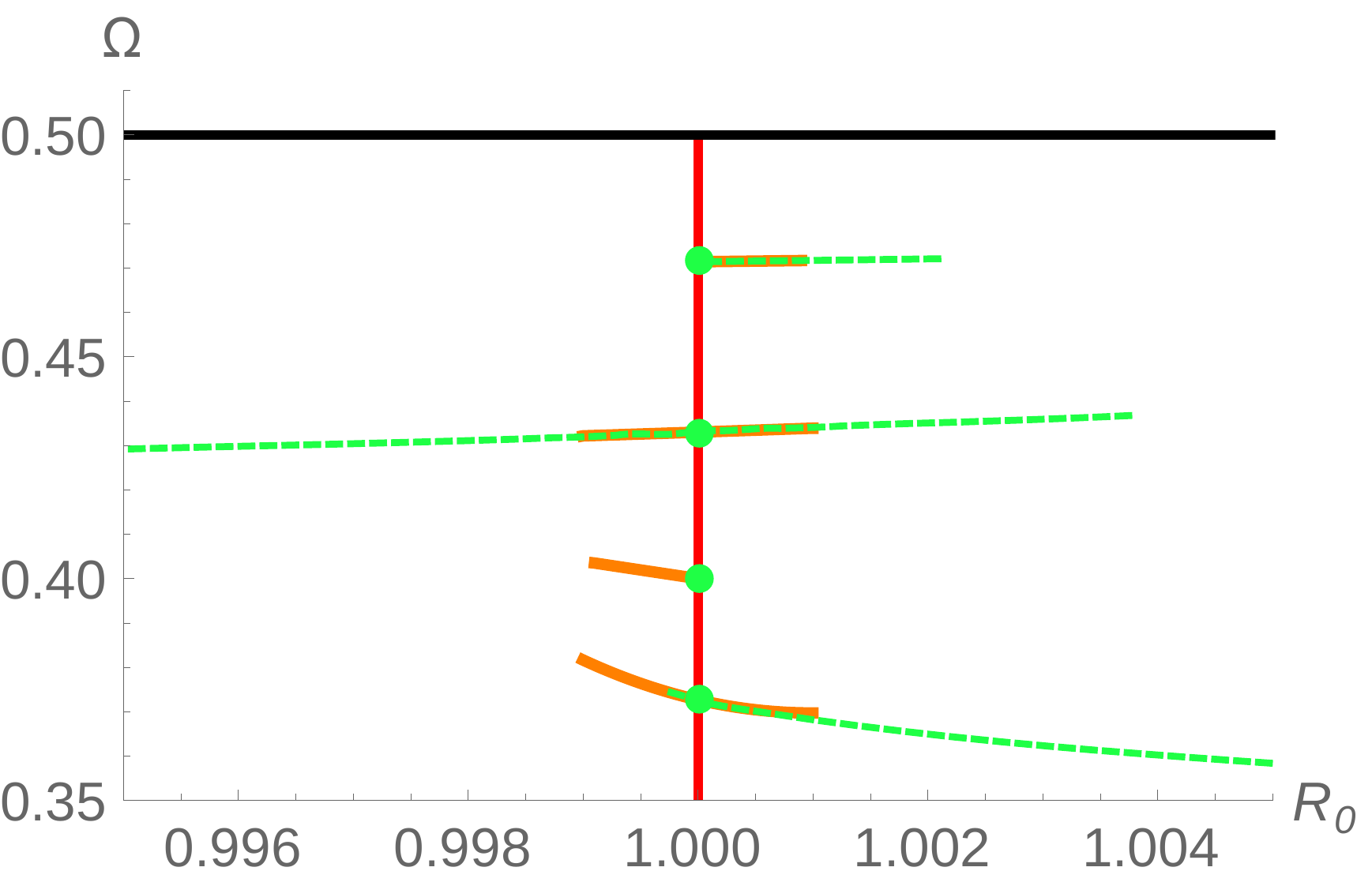}
		\caption{\small Branches of black bar deformations on the $(\RR_0,\Omega)$ plane. The right plot is a close-up showing good agreement with the analytic expansions (orange) and also zooms in on the short branches. Different tones of green are being used for different branches for the sake of clarity.\label{fig:OmegaR0Bars}}
	\end{center}
\end{figure}

In figure \ref{fig:OmegaR0Bars}, the first branches of deformed black bars are shown in the $(\RR_0,\Omega)$ plane. In this case, there is a strong qualitative difference between even and odd $N$. Odd branches extend only in one direction. This is to be expected, since in this case, reversing the sign of linear perturbations is equivalent to the gauge change $\phi \to \phi + \pi$. Surprisingly, for odd $N$ branches, $\Omega$ increases as we move away from the zero modes, and these branches are also very short.

\begin{figure}[h!]
	\begin{center}
    \includegraphics[width=\textwidth]{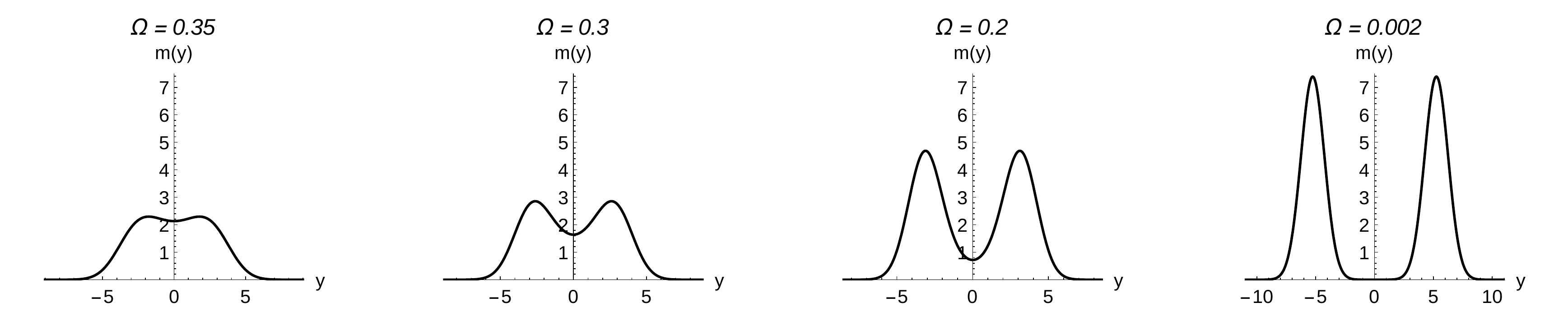}
    \includegraphics[width=\textwidth]{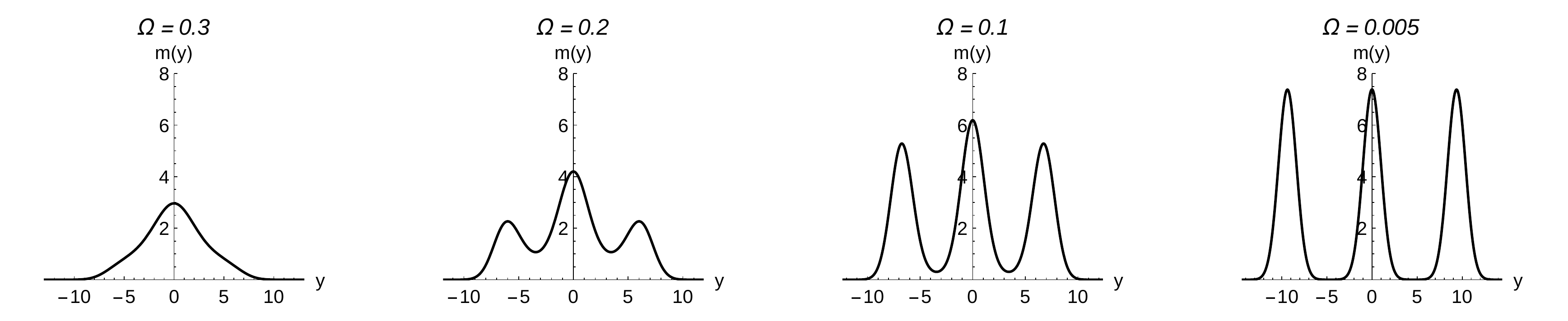}
	\caption{\small{Deformed black bars corresponding to $N = 4,6$ (dumbbells) for different values of $\Omega$. The deformation only shows $y$-dependence and the dumbbells remain Gaussian in $x$-direction. \label{fig:dumbbells}}}
	\end{center}
\end{figure}

Even $N$ branches result in the bar breaking apart in $N/2$ separated blobs. In $(\RR_0, \Omega)$ plane, they behave in a way that is qualitatively similar to the axisymmetric case, and can therefore be classified in two types. If $N$ is a multiple of 4, $\RR_0 \to 0$ and the mass density approaches zero at the origin. If $N$ is even but not a multiple of 4, then one of the blobs stays at the origin, with $\RR_0 \to 2$. The profiles of the first two symmetric bars ($N = 4,6$) are depicted in figure \ref{fig:dumbbells}.

Similar to the axisymmetric branches, even $N$ branches can be extended far away from the black bar to the arbitrarily small $\Omega$, in which the mass profile approaches to the multiple blobs located in the almost equal interval. Again, we observe these intervals grow very slowly at the same logarithmic rate
as that of ring-like blobs in the axisymmetric branches. Therefore, one can expect these branches finally would pinch off to the array of binary black holes.

The angular momentum per unit mass is calculated using eqs.~(\ref{eq:normCond}) and (\ref{eq:angularMomentum})

\begin{equation}
\frac{\mathcal{J}}{\mathcal{M}} = \frac{\int dx\,dy \, p_\phi}{\int dx\,dy \, m}\, ,
\end{equation}

with 

\begin{equation}
m(x, y) = \exp \left( \RR(y) - \frac{x^2}{2 \ell^2_\perp}\right), 
\end{equation}

\begin{equation}
p_\phi(x, y) =\left[(x^2 + y^2)\,\Omega + \frac{xy}{\ell^2_\perp} + x\RR'(y)\right]m(x, y)
\end{equation}

\begin{figure}[h!]
	\begin{center}
    \includegraphics[width=250pt]{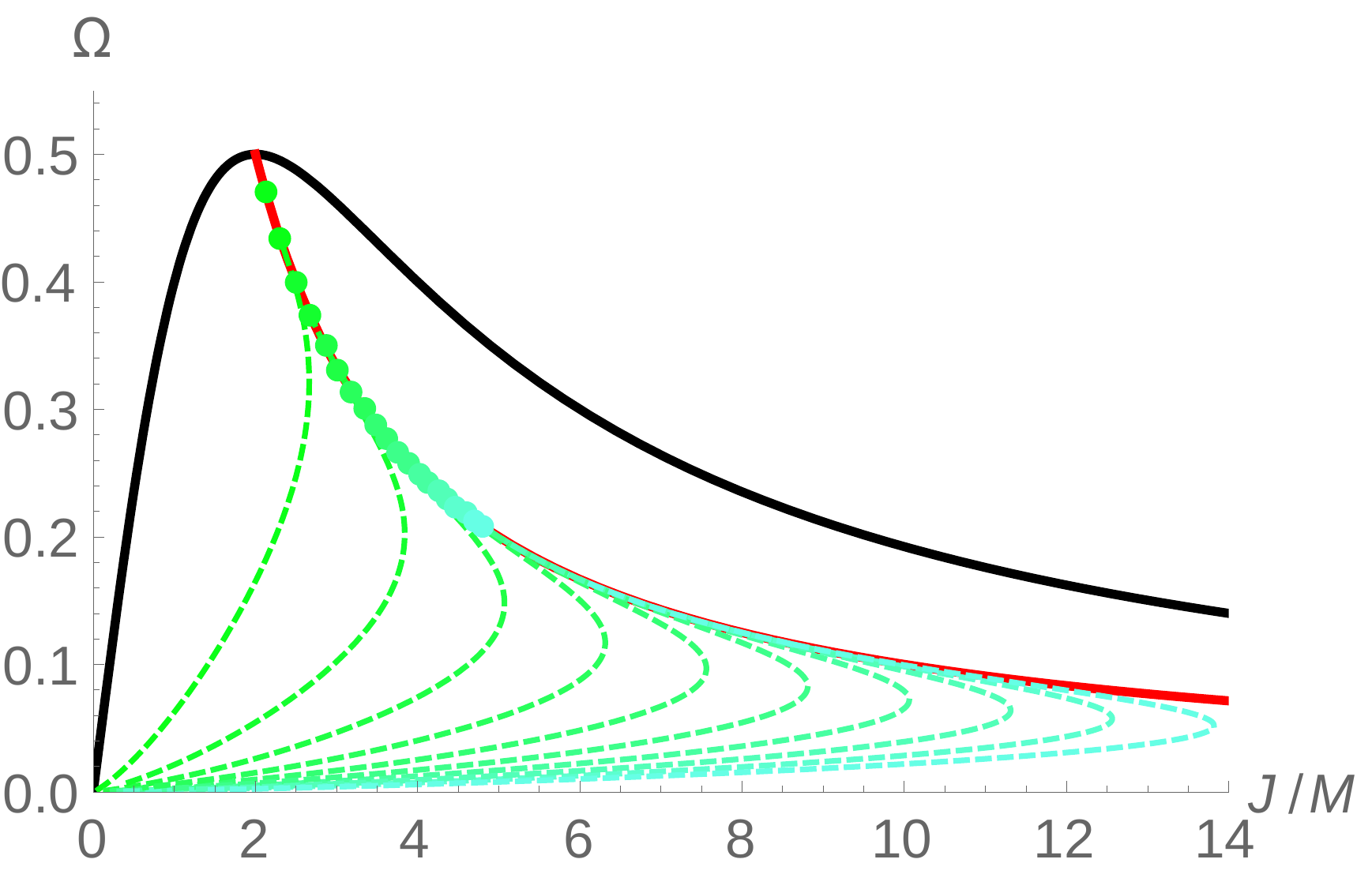}
	\caption{\small The 10 first dumbbell branches, we also plot the branching points of the odd bar perturbations marked by points that only give rise to short `spindle' branches.The Myers-Perry solutions are represented by the thick black curve, and the (non-deformed) black bars by the thick red curve. Different tones of green are being used for different branches for the sake of clarity.\label{fig:PhaseDiagramDumbbells}}
	\end{center}
\end{figure}

The phase diagram for the deformed bars is shown in figure \ref{fig:PhaseDiagramDumbbells}.

%%%%%%%%%%%%%%%%%%%%%%%%%%%%%%%%%%%%%%%%%%%%%%%%
\section{Effects of adding charge}
\label{sec:effCharge}
%%%%%%%%%%%%%%%%%%%%%%%%%%%%%%%%%%%%%%%%%%%%%%%%
%
Following the approach of \cite{Andrade:2018rcx} and as already described in section \ref{sec:largeD} we can easily construct the (non-extremal) charged solution corresponding to every uncharged solution. According to eq.~(\ref{eq:chargeRescalVel}) for a given charge parameter $\fq=\frac{Q}{M}$ and given $\Omega$, the charged solution has the profile of an uncharged solution with rotation parameter 
\begin{align}
\Omega_q=\frac{\Omega}{\lp 1-2\fq^2\rp^{1/4}}\,.
\end{align}

The $(\cJ/\cM,\Omega)$ phase diagrams for $|Q|>0$ are thus the same diagrams as in the uncharged case with a rescaling of the $\Omega$-axis by the factor $\lp 1-2\fq^2\rp^{-1/4}$. Accordingly the bumpy branches will appear at the same $\cJ/\cM$ but at a lower $\Omega$. As shown in the previous sections lower values of $\Omega$ correspond to more elongated/ further separated blobs, \ie adding charge to the black holes leads to stronger deformations. This intuitively can be understood as charge repulsion deforming the horizon.

%
%%%%%%%%%%%%%%%%%%%%%%%%%%%%%%%%%%%%%%%%%%%%%%%%%%%%%%%%%%%%%%%%%%%%%%%%%%
\section{Discussion}
\label{sec:discussion}
%%%%%%%%%%%%%%%%%%%%%%%%%%%%%%%%%%%%%%%%%%%%%%%%%%%%%%%%%%%%%%%%%%%%%%%%%%

In this paper we have demonstrated that the hydro-elastic equations \cite{Emparan:2016sjk} contain a whole new class of `rippled' stationary solutions, besides the already known black branes, their non-uniform deformations \cite{Emparan:2018bmi} and the non-deformed spinning localized black holes \cite{Andrade:2018nsz}.

We have constructed solutions that branch off from the singly spinning Myers-Perry solution directly or indirectly via the black bar branch, which has been already identified in \cite{Andrade:2018nsz}. We found both axisymmetric and non-axisymmetric solutions, and only the former ones can remain stationary at finite $D$, since  non-axisymmetric solutions will radiate gravitational waves. However,  with increasing number of dimension the emission of gravitational waves becomes weaker, which will allow the non-axisymmetric solutions to be long-lived.

The axisymmetric solutions described in this paper, we have identified as {\it ring-like} and {\it Saturn-like} bumpy black holes, or {\it black ripples} in short. They bifurcate from the axisymmetric zero modes of Myers-Perry in the ultra-spinning regime. As in the numerical studies in finite dimensions~\cite{Dias:2014cia,Emparan:2014pra}, we found that all branches extend in two directions: either with a positive or a negative amplitude of the deformation. The direction that increases the angular velocity leads to a very short branch, the other direction extends indefinitely at large $D$. This implies that the former directions lead to singular solutions, as observed in previous numerical constructions~\cite{Dias:2014cia,Emparan:2014pra}.

Multipolar deformations can not be stationary in a fixed number of dimensions, but are indicative of ultraspinning instabilities of the Myers-Perry black hole. In high enough dimension they correspond to long-lived transient objects. We generically call them {\it black flowers}, the simplest case among them is the black bar and it has an analytic solution. 

The black bar also has an infinite number of co-rotating zero modes, from which deformed branches develop: the {\it dumbbells} and the {\it spindles}. We classify the deformed bars by the parity of their zero mode as odd and even. Similarly to the ripples, the even branches go out in two directions. In the spin-down direction, the deformation grows a dumbbell-like profile with a distinct number of blobs for each branch, and hence we call them {\it dumbbells}. In the opposite direction, we could find only very short branches which we call {\it spindles}. Odd branches turned out very short as well. Odd branches and spindles correspond to solutions with increased angular velocity. One might expect that both the spindles and the odd branches end up forming a singularity.

It is very suggestive that the spindle branches correspond to the solutions that develop sharp pointy endings, as observed dynamically in~\cite{Bantilan:2019bvf,Andrade:2019edf}. These sharp endings of the deformed bar would be possibly affected by the Gregory-Laflamme instability, presenting a large number of zero modes close to the end of the short branch. The sharpened tips could, in principle, pinch off producing detached small black holes.

This process of a black hole developing long arms that end up pinching off has indeed been observed in~\cite{Bantilan:2019bvf,Andrade:2019edf}, not only for the spindles but also for higher multipole deformations. We find it likely that these dynamical solutions would correspond to the short branches described above, \ie  those resulting from exciting the zero modes in the direction with increasing $\Omega$. This would apply both to the spindle solutions and to multipolar deformations leading to multiple arms. This conjecture is supported by the fact that short branches go in the direction of decreasing $\cJ/\cM$, which should be favored in finite $D$ simulations since gravitational radiation tends to decrease the angular momentum to mass ratio of the evolving object.

The method used to identify axisymmetric solutions should be exhaustive, and thus we do not expect the ripple branches to have their own secondary axisymmetric zero modes. We expect, on the other hand, that the axisymmetric solutions will become unstable to multipolar deformations. An indication of a ring-like ripple breaking apart into four black holes via an $m=4$ deformation was already found at large $D$ in \cite{Andrade:2019edf}. 
Interestingly, black rings share the same kind of instabilities and subsequent pinch-offs~\cite{Arcioni:2004ww,Elvang:2006dd,Santos:2015iua,Figueras:2015hkb}.
Such instabilities would begin at zero modes along the branches of ripples. This fact leaves open the possibility of the `long' multipolar branches actually merging with the ripple branches at these zero modes. No conclusive results have been obtained about this intriguing possibility in this paper.

We have found no evidence that the long multipolar branches have bifurcations. This possibility could be analyzed in future work, possibly with an improved numerical setup. The dumbbell branches end as an array of separated black holes and thus seem unlikely to have further zero modes.

The nature of the boundary conditions that are imposed in the blob formalism, together with the nonlinearity of the large $D$ effective equations, leads to a remarkably challenging numerical problem. Ordinary relaxation and spectral techniques have not been shown to give reliable results so far. This fact is probably due to the requirement of imposing boundary conditions at spatial infinity, together with the equations of motion being numerically bad-behaved as $r \to \infty$. Additionally, the equations are nonlinear, which rules out direct eigenvalue-finding standard algorithms. Fortunately, the shooting approach used in this paper, which consists in identifying sharp peaks in the radius where the numerical solution becomes singular, seems to be enough to find the right solutions. It is remarkable that this technique works even though the numerical method is usually able to integrate only to a finite value of $r$. Axisymmetric solutions are easily found this way. For the case of multipolar deformations, one encounters a multidimensional shooting problem with a scalar-valued output function ($r_s$), which becomes increasingly difficult as one increases the number of overtones. For this reason, an alternative method, possibly based on relaxation techniques, would be desirable in the future.

In the formalism employed here, the effect of the charge is simply incorporated in the effective angular velocity $\Omega_q = \Omega/(1-2\fq^2)^{1/4}$ as in \cite{Andrade:2018rcx}. Therefore, with a given value of charge and $\Omega$, the corresponding charged solution is immediately obtained from the uncharged one. Due to the factor $\lp 1-2\fq^2\rp^{-1/4}$, the charged deformed branches will appear for the same $\cJ/\cM$ but for a lower $\Omega$, which corresponds to more elongated/further separated blobs. This can be interpreted as the effect of the charge repulsion.
Since all the analysis is written in terms of $\Omega_q$, one can take the extremal limit $\fq^2 \to 1/2$ of all branches, keeping $\Omega_q$ finite, resulting in a smooth limit, that leads to rather strange deformed `extremal' branches, both with and without rotation. The proper large $D$ limit of extremal horizons is however yet unclear, and a more careful analysis seems appropriate.

\paragraph{Fate of far extended branches}
All `long' branches (corresponding to bulging deformations) extend far away from the original bifurcating points in the phase space, where they develop broad thin regions. Currently, very little is known about how to interpret these nearly zero thickness regions in the large $D$ effective theory. In the case of spherical black holes the thickness falls off towards infinity as a Gaussian profile, which might be interpreted as the round tip of the black hole. Therefore, if the deformation develops a thin neck between blobs, and its size grows infinitely large, one can expect such deformation to end up as a pinch off of the horizon at finite $D$. This would correspond to a topology-changing transition.

We found that the ripple branches develop such long thin necks connecting Gaussian-shaped ring blobs (with a central blob in the case of Saturns) at their final stages of deformation. Particularly, we observed that the separation process involves two distinct length scales. From the numerical solutions, we could easily estimate that the radii of ring blobs grow like $\Omega^{-1}$ as $\Omega \to 0$. The same behavior has been derived in the blackfold approach~\cite{Emparan:2007wm,Emparan:2010sx}, which might imply that the blackfold approximation becomes already accurate in the pinch off phase, due to the localization of gravity at large $D$. Another scaling is that of the intervals between ring blobs, which are estimated as $\sim\sqrt{|\log\Omega|}$.
Due to the hierarchy in these two scales, we expect the first pinch off to occur always on the axis, indicating a first topology change to a bumpy black ring/Saturn, before transitioning to the multi-rings/Saturns, as observed in the $(+)_3$-branch of $D=6$ bumpy black holes~\cite{Emparan:2014pra}.

Dumbbell branches also extend far away from the black bar to arbitrarily small $\Omega$, where the mass profile approaches that of multiple evenly spaced blobs. As opposed to the ripples, dumbbells show only a single scaling, which has the same logarithmic growth as the intervals between the ring blobs in the case of ripples. Therefore, one can expect that these branches would finally pinch off 
to multiple black holes\footnote{Or one might say 'rotating black hole array'.}.

\paragraph{Finite $D$ effects}
The blob coordinate is supposed to be identified as the small patch of the $\sqrt{D}$-amplified entire coordinate.\footnote{This is only an estimate from the Myers-Perry solution, in which the exact coordinate match is known.}
Therefore, the blob approximation will break down if the length of the thin neck reaches $\sim\sqrt{D}$, when the $1/D$ corrections are included.
This breakdown will give some information on the transition in phase space.
For example, the pinch off from the ripples to black rings or Saturns will take place at $\Omega \sim 1/\sqrt{D}$.
Actually, black rings are already constructed by using the large D effective theory approach in the same scaling~\cite{Tanabe:2015hda,Tanabe:2016opw}. This implies that one can use the effective theory result as the global setup to solve the local topology-change.
For other logarithmic scalings $\sim \sqrt{|\log \Omega|}$, the break down will occur at much smaller spin $\Omega \sim e^{-D}$.
In the black string analysis, a similar type of breakdown is already seen after including $1/D$ corrections~\cite{Emparan:2018bmi}.
The black hole entropy is another important quantity to evaluate the stability of the solutions. Since the mass and entropy become degenerate at $D\to\infty$, we would need to know the next-to-leading order terms in $1/D$ expansion to calculate the entropy difference for a given mass.

\paragraph{Blob-Blob interactions}
For the ripples and dumbbells, we observed a universal scaling of the blob distance as $\sqrt{|\log \Omega|}$ at $\Omega \to 0$, implying an effective interaction between the blobs (or ring-like blobs). This indicates the possibility to reconstruct the large $D$ effective theory as a particle-like (or soliton-like) effective description of blobs weakly interacting via very thin necks. This possibility will be pursued elsewhere.

The origin of this logarithmic dependence, though very naively, might be understood as a force balance between the centrifugal force and the attraction between the blobs at large $D$. Assuming a black hole of radius $r_{\rm H}$ and an orbiting particle, the gravitational force is approximated as $(r_{\rm H}/r)^{D}$ and the centrifugal force as $\Omega^2 r$. The equilibrium is accomplished by $r/r_{\rm H}\sim 1-2D^{-1}\log \Omega$. Therefore, the particle orbit exists very close to the horizon $\sim|\log\Omega|/D$. This introduces the $|\log \Omega|$ scaling in the near horizon region.
Curiously, if we assume two adjacent black holes with the same mass, the equilibrium condition would be modified to $r/r_H \sim 2-2D^{-1} \log (e^{D/2} \Omega)$ with $e^{D/2}\Omega = \ord{1}$ or $|\log\Omega| \sim D$. 
This coincides with the value at which the neck length between blobs reaches $\sqrt{|\log \Omega|}\sim \sqrt{D}$ and the blob approximation breaks down.

\paragraph{Towards the topology change}
The topology-changing transition at large $D$ is described by the conifold metric which solves the Ricci flow equation~\cite{Emparan:2019obu}. Especially, the black string/black hole transition is completely solved by the King-Rosenau (KR) solution for the $2D$ Ricci flow. 
Some of the topology-changing transitions (Saturn-like ripples, dumbbells) can be reduced to the $2D$ Ricci flow problem in the co-rotating frame, since the transition occurs in a very narrow region. Hence, they should also be solved by the KR solution, due to the rigidity in $2D$ compact ancient flow~\cite{daskalopoulos2009classification}\footnote{A solution of the Ricci flow is called {\it ancient}, if it can be extended to the infinite past of the flow time (corresponding to the asymptotic infinity in the large $D$ conifold metric). }. For the transition between ring-like ripples and black rings, we need a better understanding of the $3D$ Ricci flow.

Here we should note that, in the case of the black string/black hole transition, one just has to give the global configuration (such as the black hole (blob) radius and the compactification scale) as boundary conditions for the conifold metric, without considering the force balance condition. Now, for example, if we consider the transition between a dumbbell and binary black hole, we also have the rotation $\Omega$, which will not appear in the large $D$ conifold analysis after switching to the co-rotating frame. To relate $\Omega$ with the mass and separation, one needs to find the proper force balance condition at large $D$, as roughly estimated in the previous paragraph.

In the current formalism, we could only follow the $(-)$-ripple branches for a very short range.
These $(-)$-branches are shown to develop a single-sided conical horizon on the equator when they approach the end of their branch~\cite{Emparan:2014pra}. Therefore, it should also be possible to study the ending phase of $(-)$-branches using the large $D$ conifold metric and Ricci flow methods. Different from the usual pinch off problem, one may have to find the non-compact Ricci flow solution, in which only one side is the horizon.

%%%%%%%%%%%%%%%%%%%%%%%%%%%%%%%%%%%%%%%%%%%%%%%%%%%%%%%%%%%%%%%%%%%%%%%%%%
\section*{Acknowledgments}
We would like to thank Roberto Emparan for comments on an earlier draft of this paper.
Work supported by ERC Advanced Grant GravBHs-692951 and MEC grant FPA2016-76005-C2-2-P. 
RL is supported by the Spanish Ministerio de Ciencia, Innovaci\'on y Universidades Grant FPU15/01414.
RS is supported by JSPS KAKENHI Grant Number JP18K13541 and partly by Osaka City University Advanced Mathematical Institute (MEXT Joint Usage/Research Center on Mathematics and Theoretical Physics).

\appendix

%%%%%%%%%%%%%%%%%%%%%%%%%%%%%%%%%%%%%%%%%%%%%%%%
\section{Numerical methods}
\label{app:method}
%%%%%%%%%%%%%%%%%%%%%%%%%%%%%%%%%%%%%%%%%%%%%%%%

\subsection{Axisymmetric sector}

\begin{figure}[h]
\begin{center}
\includegraphics[width=250pt]{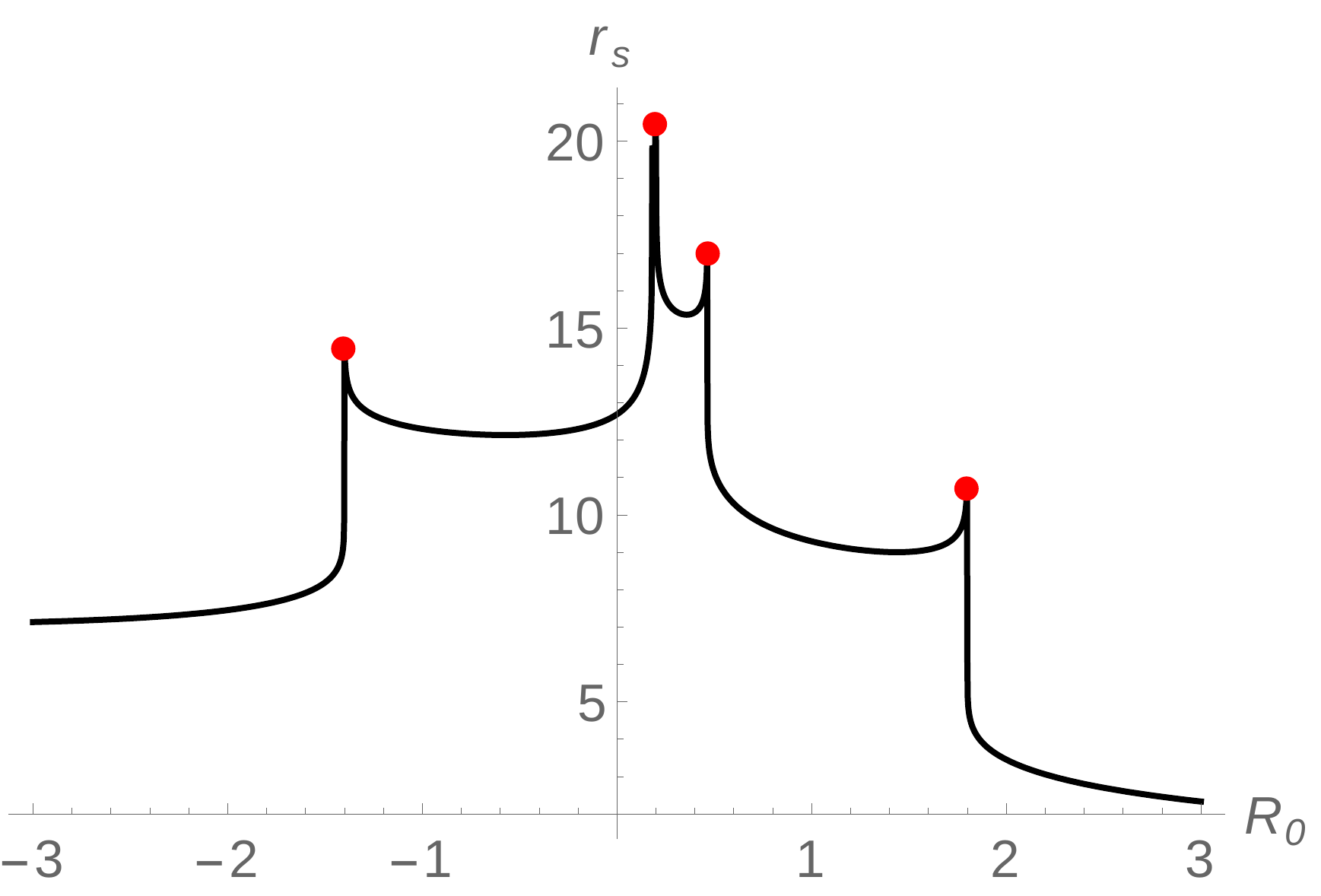}
\caption{\small Values of $r_s$ (radius where the solution becomes singular) for $\Omega = 0.3$. The solutions that have to be free of such singularities and extend to infinity appear as sharp peaks, which we marked here with red dots. \label{fig:singularRadius}}
\end{center}
\end{figure}

Stationary axisymmetric black holes are regular solutions of eq.~(\ref{eqn:master}) that extend from 0 to $r \to \infty$. Due to singular point at $r=\infty$ from the rotation term it is particularly difficult to use of spectral and relaxation methods. For this reason, the approach used in this paper is essentially a shooting method. By regularity at the origin the ODE can be generally integrated radially outwards with the initial conditions $\RR(0) = \RR_0$ and $\RR'(0) = 0$. The numerical solution will generally become singular at some finite $r = r_s$. In figure \ref{fig:singularRadius}, the values of $r_s$ are shown as a function of the initial condition parameter $\RR_0$, interestingly the appearance of singularities is (semi-) continuous in the space of initial conditions which makes it possible to look for singularities/ peaks where the solution extends to infinity. These peaks correspond to (approximate) locations of the allowed solutions. 

For this purpose, the $(\RR_0, \Omega)$ plane is not a very suitable representation. This is so because the branches of solutions become very closely packed at low $\Omega$, while the ring-like branches reach very large negative values. A numerical algorithm intended to find all these peaks with a high precision needs therefore an extremely high dynamic range of detection in $\RR_0$, so it can both find widely spread peaks and resolve extremely close packed ones. This is solved by introducing the coordinates $(\alpha,\beta)$ as 

\begin{equation}
\Omega = \frac{e^\beta}{2} \text{sech} \,\alpha \, , \qquad \RR_0 = 2-e^{\alpha + \beta} \text{sech} \,\alpha\,.
\end{equation}

These coordinates both range from $-\infty$ to $\infty$, and cover the $(-\infty, 2)\times(0, \infty)$ region in $(\RR_0, \Omega)$ plane. They are analytically invertible as

\begin{equation}
\alpha = \log \left( \frac{2-\RR_0}{2\Omega} \right), \qquad \beta = \log \left( \frac{(2-\RR_0)^2 + 4\Omega^2}{2(2-\RR_0)} \right)
\end{equation}

\begin{figure}[h!]
\begin{center}
\includegraphics[width=200pt]{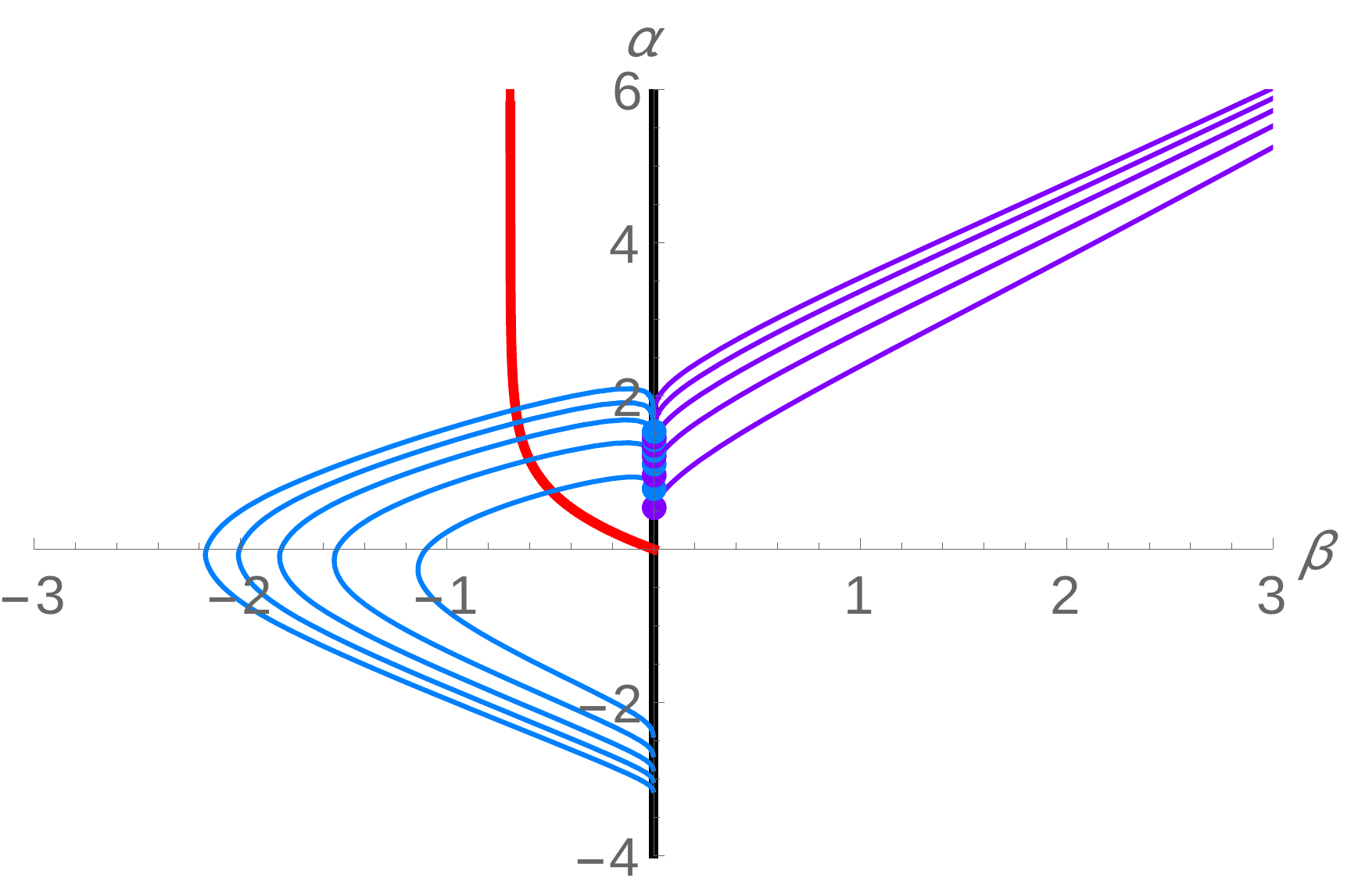}
\caption{\small Location of the first 10 branches of black ripples (5 ring-like and 5 Saturn-like) in the $(\alpha, \beta)$ plane.\label{fig:alphabetaPlane}}
\end{center}
\end{figure}

In these new coordinates, the Myers-Perry black holes lay on the vertical axis ($\beta = 0$), with the Schwarzschild black hole corresponding to $\beta = 0, \; \alpha \to -\infty$ (see figure \ref{fig:alphabetaPlane}). The ripple solutions become now much more suitable to be found numerically. In particular, the ring-like branches can be parametrized by $\beta$, and the Saturn-like by a polar angle $\theta$ such that $\alpha = \rho \cos \theta$ and $\beta = - \rho \sin \theta$.

When a branch ends, as for the negative amplitude modes, the peak that represents it becomes a local maximum, with no divergence whatsoever. This requires us to define a criterion for a local maximum to be considered a proper peak, or a {\it vanishing peak}. The criterion that has been taken for a peak to be valid is    

\begin{equation}
\max \left\{ r_s(\alpha, \beta) - r_s(\alpha + \delta\alpha, \beta), r_s(\alpha, \beta) - r_s(\alpha - \delta\alpha, \beta)\right\} > \Delta\, ,
\end{equation}

where $\delta\alpha = 0.01$ and $\Delta = 3$. When extracting the angular momenta of the solutions, it is also important to take into account that numerical error may result in extra (unphysical) oscillations of the $\cR (r)$ profiles. These oscillations appear as additional bumps, or {\it fake rings}. These have to be properly removed before the angular momentum integration, since they could add an erroneous contribution to the integration result.

\subsection{Black bar deformations}

Deformations with even values of $N$ are found in a way which is completely analogous to the axisymmetric case. In this case it is convenient to reparameterize the $(\RR_0, \Omega)$  by the coordinates $(\gamma, \delta)$,

\begin{equation}
\gamma = -\log (2\Omega)\, , \qquad \delta = -\log (2-\RR_0)
\end{equation} 

\begin{figure}[h!]
\begin{center}
\includegraphics[width=200pt]{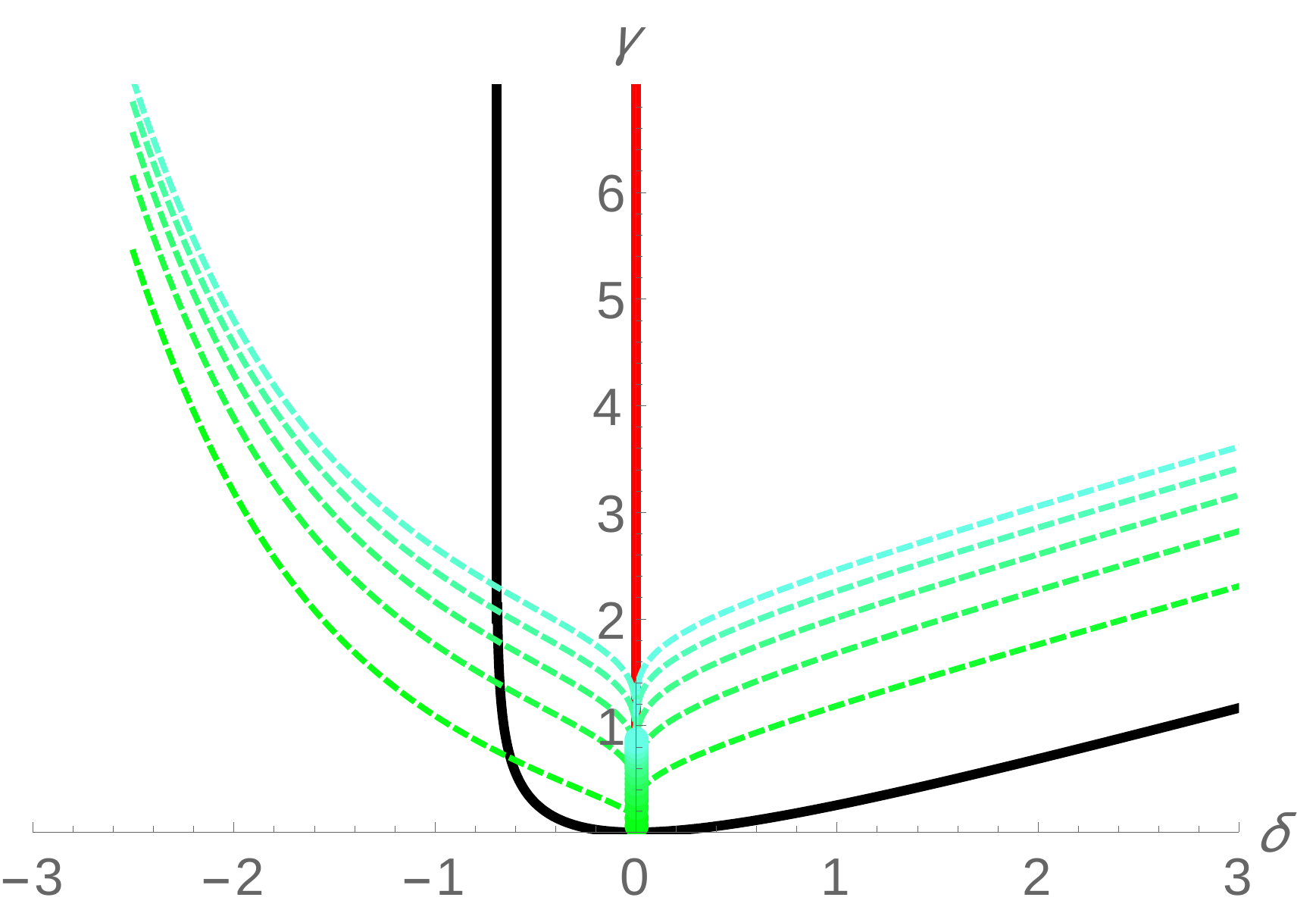}
\caption{\small Location of the first 10 branches of deformed bars in the $(\gamma, \delta)$ plane.\label{fig:gammadeltaPlane}}
\end{center}
\end{figure}

Odd deformations of bars are described by solutions of eq.~(\ref{eqn:masterDumbbells}) that have a nonzero value of $\bar \RR_0 = \RR'(0)$. This increases the complexity of the problem, since it now requires to tune both $\RR_0$ and $\bar \RR_0$ in order to get a solution that extends to infinity both for the negative and positive sides of the $y$ axis. This complication can be partially circumvented by noticing that, for the deformed black bars, the change $y \to - y$ is equivalent to $\bar \RR_0 \to - \bar \RR_0$. This means that, if $(\Omega, \RR_0, \bar \RR_0)$ gives an allowed solution, then so does $(\Omega, \RR_0, - \bar \RR_0)$. This fact allows the right values of $\RR_0$ to be found by requiring the peaks in $r_s(\Omega, \RR_0, \bar \RR_0)$ to be located at opposite values of $\bar \RR_0$. This is done by the secant root-finding method in a few iterations. Again, {\it vanishing peaks} and {\it fake blobs} are discarded in a similar way as in the axisymmetric case.

\subsection{Multipole deformations} \label{sec:AppNumMet}

By using the ansatz~(\ref{eq:numAnsatzFourMode}) truncated at some Fourier mode $\cos(n_{\text{max}} m \phi)$, we obtain a set of $n_{\text{max}} + 1$ coupled equations for the functions $\cR^{(n m)}(r)$. These equations, by imposing the regularity condition $\cR^{(n m)'}(0) = 0 \; \forall n$, can be solved by specifying the values of the radial functions at the origin. The problem reduces then to finding peaks in the singular radius $r_s(\Omega, \cR_{0},\cR_{m},\cR_{2m}, \dots, \cR_{n_{\text{max}} m})$.

Identifying peaks on a function with more than one variable is in general not an easy task, especially if there is no straightforward way of reducing the problem to one variable (as in the case of odd deformations of the black bar). For this reason, in this article we restrict ourselves to the fundamental Fourier mode, \ie  we maximize $r_s(\Omega, \cR_{0},\cR_{m})$. We use the {\it Mathematica} function {\bf NMaximize} to identify the peak by incrementing $\Omega$ in small steps, and constraining the search in a small region around the result of the previous step.

Even with this method, the values of the $\cR_{0},\cR_{m}$ still are affected by small fluctuations (which are likely due to numerical error) around the branch. We correct this by subsampling the data points.

\section{Matching to the entire hemisphere}
In general, blob solutions are thought to be identified as a polar cap of the compact black holes, in which the polar angle is stretched by $\sqrt{D}$ to match with the radial coordinate in the blob~\cite{Andrade:2018nsz},
\begin{equation}
 r = \sqrt{D} \theta. \label{eq:blob-hole-rtheta}
\end{equation}
In ref.~\cite{Andrade:2018nsz}, the linear order deformation of the blob and the perturbation in the Myers-Perry~\cite{Suzuki:2015iha} confirmed to be matched for $1 \ll r \ll \sqrt{D}$,
\begin{equation}
 \delta \cR \propto L_N\left( \frac{r^2}{2(1+a^2)} \right)\sim r^{2N} \sim \sin^{2N}\theta.\label{eq:blob-hole-match-lin}
\end{equation}
Here we show that this match is also consistent beyond the linear level, despite the increase in the degree of the polynomials in the higher perturbation order.
The degree of each perturbation solution can be estimated from the recurrence formula~(\ref{eq:recurrence-C-p}) as
\begin{equation}
{\rm deg} [f_{k}(z)] = \underset{i} {\rm max}\left({\rm deg} [f_{i}(z)] + {\rm deg} [f_{k-1-i}(z)]\right) -1,
\end{equation}
where the last $-1$ comes from $I+J-K$ factor in eq.~(\ref{eq:recurrence-C-p}).
Starting from $f_0(z) = L_N(z)$, the induction easily follows
\begin{equation}
{\rm deg} [f_{k}(z)] = (k+1) N -k.
\end{equation}
Since the coordinate match~(\ref{eq:blob-hole-rtheta}) leads to
\begin{equation}
 z \sim r^2 \sim D \sin^2 \theta,
\end{equation}
the perturbation at each order gives the match at $1 \ll z \ll D$,
\begin{equation}
 \veps^{k+1} f_k(z) \sim \veps^{k+1} z^{(k+1)N-k}\sim \bar{\veps}\, {}^{k+1} D^{-2k} (\sin\theta)^{2(k+1)N-2k}.
\end{equation}
where we rescaled the perturbation parameter by $\bar{\veps} = D^{2N}\veps$, so that the linear order remains finite at $D\to \infty$.
Therefore, the linear order match~(\ref{eq:blob-hole-match-lin}) turns out to be correct even up to the nonlinear order, and all the nonlinear perturbation will be matched with the subleading correction in $1/D$,
\begin{equation}
 \delta \cR \sim \bar{\veps} \sin^{2N}\theta+\ord{D^{-1}}.
\end{equation}

\section{Useful properties of the orthogonal polynomials}
Here, we show some useful properties of the Laguerre and Hermite polynomials used in the paper.

\subsection{Product of the orthogonal polynomials}
\paragraph{Product of Laguerre polynomials}
It is known that the product of the Laguerre polynomials of the same second parameter can be written by the linear combination of the Laguerre polynomials of the same type~\cite{Watson1938},
\begin{align}
& L^{(n)}_I(x)L^{(n)}_J(x) = \sum_{K=|I-J|}^{I+J} \overset{(n)}{\cX}{}^K_{I,J} L_K^{(n)}(x)
\end{align}
where the coefficients are given by
\begin{equation}
\overset{(n)}{\cX}{}^K_{I,J}  = \frac{(-2)^{I+J-K} K!}{(K-I)!(K-J)!(I+J-K)!} \hgfunc{3}{2}\left(\begin{array}{c}n+K+1,\fr{2}(K-I-J),\fr{2}(K-I-J+1)\\
 K-I+1,K-J+1\end{array};1\right).
\end{equation}
For $n=0$, the coefficient becomes symmetric in $(I,J,K)$, in which case we just write $\cX^K_{I,J}$.

\paragraph{Product of Hermite polynomials}
The decomposition of the product of the Hermite polynomials is also known 
\begin{equation}
 H_I(z) H_J(z) = \sum_{K=|I-J|}^{I+J} {\cal Q}^{K}_{I,J} H_{K}(z),
\end{equation}
where the coefficients have the non-zero value only if $I+J+K$ is even,
\begin{equation}
 {\cal Q}^K_{I,J} := \frac{2^\frac{I+J-K}{2}I!J! }{\left(\frac{I+K-J}{2}\right)!\left(\frac{J+K-I}{2}\right)!\left(\frac{I+J-K}{2}\right)!}. \label{eq:hermite-product-cf}
\end{equation}
It is worth noting that the coefficients in the above two formula become non-zero
 only if $(I,J,K)$ satisfy the trigonometric inequality: any of the three cannot exceed the sum of the rest two.

\subsubsection{Relation to Franel number}
\label{sec:franel}
Interestingly, the renormalization coefficient $\mu$ in eq.~(\ref{eq:axisym-mu1}), is related to the so called Franel number, which is known in combinatorics and number theory,
\begin{equation}
{\rm Fr}_N := \sum_{i=0}^N \binom{N}{i} ^3 = \hgfunc{3}{2}\left[\begin{array}{c}-N,-N,-N\\1,1\end{array};-1\right].
\end{equation}
Due to the identity,
\begin{equation}
\hgfunc{3}{2}\left[\begin{array}{c}-N,-N,-N\\1,1\end{array};-1\right] =    2^N \hgfunc{3}{2}\left[\begin{array}{c}N+1,-\frac{N}{2},-\frac{N-1}{2}\\
1,1\end{array};1\right],
\end{equation}
$\mu$ can be rewritten as
\begin{equation}
\mu = - \fr{4} \cX^N_{N,N} = - \fr{4}(-1)^N {\rm Fr}_N.
\end{equation}
Using the large $N$ approximation for the binomial coefficients, we can show the rapid growth in this number with respect to $N$, 
\begin{equation}
{\rm Fr}_N = \sum_{i=0}^N \binom{N}{i}^3 \sim \frac{2^{3N}}{\sqrt{N}} \int^\infty_{-\infty}
e^{-6Nx^2}dx \sim \frac{2^{3N}}{N}.
\end{equation}

\subsection{Integral of triple associated Laguerre polynomials}\label{sec:associated-laguerre-lauricella}
As found in~\cite{Erdelyi1936,Lee2000}, the triple integrals are given by
\begin{align}
&  \int_0^\infty z^\frac{i+j+k}{2} e^{-z} L^{(i)}_I(z) L^{(J)}_J(z) L^{(k)}_K(z) dz 
  = \frac{(i+I)!}{i!I!}\frac{(j+J)!}{j!J!}\frac{(k+K)!}{k!K!} \left(\frac{i+j+k}{2}\right)! \nonum
  & \qquad \times \quad F_A^{(3)} \left(\frac{i+j+k}{2}+1; -I,-J,-K;i+1,j+1,k+1;1,1,1\right)
\end{align}
where $F_A^{(3)}$ is one of the Lauricella's generalized hypergeometric functions defined by
\begin{align}
& F^{(n)}_A (a; b_1,\dots b_n; c_1,\dots,c_n;x_1,\dots,x_n) \nonum
& = \sum_{m_1=0}^\infty \cdots  \sum_{m_n=0}^\infty \frac{(a)_{m_1+\cdots+m_n} (b_1)_{m_1}
\cdots (b_n)_{m_n}}{(c_1)_{m_1} \cdots (c_n)_{m_n}m_1!\cdots m_n!}x_1^{m_1}\cdots x_n^{m_n}.
\end{align}
If $b_i$ is a negative integer, the summation with respect to $m_i$ stops at $|b_i|$.

%%%%%%%%%%%%%%%%%%%%%%%%%%%%%%%%%%%%%%%%%%%%%%%%%

\section{Derivative of Laguerre functions with respect to the parameter}
In this section, we study the infinitesimal parameter shift in the generalized Laguerre functions from
the Laguerre polynomials.

\subsection{Confluent hypergeometric equation}
We start from reviewing the confluent hypergeometric equation,
\begin{equation}
 z f''(z) + (b-z) f'(z) - a f(z) = 0.\label{eq:confluentHG}
\end{equation}
A solution is given by Kummer's confluent hypergeometric series
\begin{equation}
 \hgfunc{1}{1}(a,b,z) = \sum_{k=0}^\infty \frac{(a)_k}{(b)_k} \frac{z^k}{k!} \label{eq:hg-sol-1}
\end{equation}
where $(a)_k :=  \Gamma(a+k)/\Gamma(a)$ is the Pochhammer symbol.
If $b$ is not positive integer, the other solution is given by
\begin{equation}
 U(a,b,z) = \frac{\pi}{\sin \pi b} \left(\frac{\hgfunc{1}{1}(a,b,z)}{\Gamma(1+a-b)\Gamma(b)}-z^{1-b}\frac{\hgfunc{1}{1}(1+a-b,2-b,z)}{\Gamma(a)\Gamma(2-b)}\right).\label{eq:hg-sol-2}
\end{equation}
If $b$ is a positive integer, say $b=n+1\ (n = 0,1,2,\dots)$, the other solution is given by\footnote{The overall factor and the term proportional to the first solution $\hgfunc{1}{1}(a,n+1,z)$ are adjusted to give the valid formula for the non-positive integer value of $a$.}
\begin{align}
&U(a,n+1,z) = (-1)^n\frac{n!(n-1)!\Gamma(a-n)}{\Gamma(a)} z^{-n}  \sum_{k=0}^{n-1} \frac{n!(a-n)_k}{(1-n)_k k!}z^k \nonum
& \qquad - \sum_{k=0}^\infty \frac{(a)_k z^k}{(n+1)_k k! } [\psi(a+k)-\psi(1+k)-\psi(1+n+k)]\nonum
& \qquad -  \hgfunc{1}{1}(a,n+1,z)(\ln z+\pi \cot(\pi a))
\end{align}
where $\psi(z):=\Gamma'(z)/\Gamma(z)$ is the digamma function. For the negative value of $a$, it is convenient to rewrite this to
\begin{align}
&U(a,n+1,z) =  \sum_{k=1}^{n} \frac{n!(k-1)!\Gamma(1-a)}{\Gamma(k+1-a) (n-k)!}z^{-k} -  \hgfunc{1}{1}(a,n+1,z)\ln z\nonum
&  - \sum_{k=0}^\infty \frac{(a)_k z^k}{(n+1)_k k! } [\psi(1-a-k)-\psi(1+k)-\psi(1+n+k)] \label{eq:hg-sol-2-log}
\end{align}  
where we used the reflection formula for the gamma functions and digamma functions,
\begin{align}
\Gamma(z)\Gamma(1-z) = \frac{\pi}{\sin(\pi z)},\quad \psi(z) - \psi(1-z) = -\pi \cot(\pi z).
\end{align}

\subsection{Laguerre functions}
If $b$ is a positive integer, $b=n+1 \ (n=0,1,2,\dots)$, eq.~(\ref{eq:confluentHG}) is called the Laguerre equation, and the first solution~(\ref{eq:hg-sol-1}) is called the (generalized) Laguerre functions,
\begin{equation}
  \Phi(\alpha,n,z) := \hgfunc{1}{1}(-\alpha,n+1,z),
\end{equation}
or the following convention is more commonly used,
\begin{equation}
	L_{\alpha}^{(n)}(z) := \frac {\Gamma(\alpha+n+1)}{\Gamma(n+1)\Gamma(\alpha+1)} \hgfunc{1}{1}(-\alpha,n+1,z).
\end{equation}
These definitions are equal for $n=0$.
Throughout this section, we will use the former convention for the convenience.
In case of $\alpha = 0,1,2,\dots$, these functions reduce to the Laguerre polynomials. 

\paragraph{Laguerre functions of the second kind}
Recently, for $\alpha=N\  (N=0,1,2,\dots)$, the second solution
is found to be written in the closed form~\cite{ParkeMaximon2015,ParkeMaximon2016},
\begin{equation}
 \Psi(N,n,z) = \frac{n!}{(N+n)!}P(N,n,z) e^z z^{-n} - \Phi(N,n,z) {\rm Ei}(z) \label{eq:laguerre-2-PM}
\end{equation}
where ${\rm Ei}(z)$ is the exponential integral function.
The function $P(N,n,z)$ is given by
\begin{equation}
 P(N,n,z) = \sum_{m=0}^{n-1} \left[\frac{(N+m)!(n-m-1)!}{m!}\right]z^m + z^n \sum_{m=0}^{N-1} c(N,n,m) z^{m}
\end{equation}
where
\begin{equation}
c(N,n,m) = \frac{(-1)^{m+1}N!(N+n)!}{(N-m-1)!(m+n+1)!(m+1)!} \hgfunc{3}{2}\left(\begin{array}{c}1,1,-N+m+1\\2+m,2+m+n\end{array};1\right). \label{eq:laguerre-2-c}
\end{equation}
In ref.~\cite{ParkeMaximon2016}, eq.~(\ref{eq:laguerre-2-PM}) is shown to
coincide with the expression in eq.~(\ref{eq:hg-sol-2-log}),
\begin{equation}
\Psi(N,n,z) = U(-N,n+1,z).	
\end{equation}
Using the asymptotic expansion of ${\rm Ei}(z)$
, one can obtain the asymptotic behavior at the large $z$ as
\begin{align}
	\Psi(N,n,z) \simeq  (-1)^{N+1} N! n! z^{-N-n-1} e^z \left(1+\ord{z^{-1}}\right)  .\label{eq:Psi-as}
\end{align}
Close to $z=0$, we obtain
\begin{equation}
 \Psi(N,n,z) \simeq \sum_{k=1}^n \frac{N!n!(k-1)!}{(N+k)!(n-k)!}z^{-k}  - \log z -\gamma + H_{n}-H_N+\ord{z}. \label{eq:Psi-zero}
\end{equation}

\subsection{Derivative with respect to the parameter}
Here, we evaluate $\alpha$-derivative of $\Phi(\alpha,n,z)$ on a non-negative integer. It turns out, $\partial_\alpha \Phi(N,n,z)$ can be expressed in terms of $\Phi(N,n,z)$, $\Psi(N,n,z)$, $\log z$ and polynomials,
\begin{align}
&\partial_\alpha \Phi(N,n,z) = \Psi(N,n,z)+ (\gamma-H_N+\log z)\Phi(N,n,z)+\sum_{k=0}^{N-1} \frac{2}{N-k} \Phi(k,n,z)\nonum
&\hspace{2.5cm}-\sum_{k=1}^n \frac{N!n!(k-1)!}{(k+N)!(n-k)!} z^{-k}
 - \sum_{k=1}^n \fr{k} \hgfunc{2}{2}\left[\begin{array}{c}-N,k\\n+1,k+1\end{array};z\right].
\label{eq:dPhi}
\end{align}
\paragraph{Proof}
The above formula can be obtained through the expression in eq~.(\ref{eq:hg-sol-2-log}).
Using the reflection formula for the gamma function, we have
\begin{align}
\partial_\alpha \Phi(\alpha,n,z) = -\sum_{k=0}^\infty (\psi(\alpha+1-k)-\psi(\alpha+1)) \frac{(-\alpha)_k}{(n+1)_k k!}z^k.
\end{align}
Using eq.~(\ref{eq:hg-sol-2-log}), one can rewrite the above equation to
\begin{align}
&\partial_\alpha \Phi(\alpha,n,z) = U(-\alpha,n+1,z)+ (\psi(\alpha+1)+2\gamma+\log z)\Phi(\alpha,n,z)\nonum
&-\sum_{k=1}^n \frac{n!(k-1)!\Gamma(1+\alpha)}{\Gamma(k+1+\alpha) (n-k)!} z^{-k}-\sum_{k=0}^\infty (H_{k}+H_{k+n}) \frac{(-\alpha)_k}{(n+1)_k k!}z^k,
\end{align}
where $H_n$ is the harmonic number and we used $\psi(n) = H_{n-1}-\gamma$.
By setting $\alpha=N,\ (N=0,1,2,\dots)$, the last sum in the second line reduces to the $N$-th order polynomial, and then, we have the following expression, 
\begin{align}
&\partial_\alpha \Phi(N,n,z) = \Psi(N,n,z)+ (H_N+\gamma+\log z)\Phi(N,n,z)\nonum
&-\sum_{k=1}^n \frac{N!n!(k-1)!}{(k+N)!(n-k)!} z^{-k}-\sum_{k=0}^N (H_{k}+H_{k+n}) \frac{(-N)_k}{(n+1)_k k!}z^k.
\label{eq:dPhi-0}
\end{align}
The last summation can be simplified as
\begin{align}
& \sum_{k=0}^N (H_{k}+H_{k+n}) \frac{(-N)_k}{(n+1)_k k!}z^k  =
2\sum_{k=1}^N H_{k} \frac{(-N)_k}{(n+1)_k k!}z^k +  \sum_{k=0}^N \sum_{\ell=1}^n \fr{k+\ell} \frac{(-N)_k}{(n+1)_k k!}z^k\nonum
& =2 H_N \Phi(N,n,z) -\sum_{k=0}^{N-1} \frac{2}{N-k} \Phi(k,n,z) + \sum_{k=1}^n \fr{k} \hgfunc{2}{2}\left[\begin{array}{c}-N,k\\n+1,k+1\end{array};z\right].
\end{align}
This leads to eq.~(\ref{eq:dPhi}).

\newpage

\bibliographystyle{JHEP}
\bibliography{LargeDBumpyHolesReferences}

\end{document}